\documentclass[aps,preprint,preprintnumbers,notitlepage,superscriptaddress,amsmath,amssymb,noeprint]{revtex4-1}

\usepackage{hyperref} % hyperreferences
\usepackage{graphicx}
\usepackage{float}
\usepackage{amsmath}
\usepackage{mathrsfs}
\usepackage{amssymb}
\usepackage{siunitx}
\usepackage{braket}
\usepackage{bm}
\usepackage{xcolor}
\usepackage[normalem]{ulem}
\usepackage{setspace}
\usepackage{amsfonts}

\newcommand{\xx}{{x}}

\newcommand{\grad}{{\nabla}}
\newcommand{\avg}[1]{\left\langle  #1 \right \rangle}

\DeclareGraphicsExtensions{.png,.jpg,.eps}

\begin{document}

\title{Observation of KPZ universal scaling in a one-dimensional polariton condensate}

\author{Q. Fontaine}
\affiliation{Universit\'{e} Paris-Saclay, CNRS, Centre de Nanosciences et de Nanotechnologies (C2N), 91120, Palaiseau, France}

\author{D. Squizzato}
\affiliation{Univ. Grenoble Alpes and CNRS, Laboratoire de Physique et Mod\'elisation des Milieux Condens\'es, 38000 Grenoble, France.}
\affiliation{Dipartimento di Fisica, Universit\`a La Sapienza - 00185 Rome, Italy.}
\affiliation{Istituto Sistemi Complessi, Consiglio Nazionale delle Ricerche, Universit\`a La Sapienza - 00185 Rome, Italy.}

\author{F. Baboux}
\affiliation{Universit\'{e} Paris-Saclay, CNRS, Centre de Nanosciences et de Nanotechnologies (C2N), 91120, Palaiseau, France.}
\affiliation{Laboratoire Mat\'eriaux et Ph\'enom\`enes Quantiques, Universit\'e de Paris, CNRS-UMR 7162, Paris 75013, France.}

\author{I. Amelio}
\affiliation{INO-CNR BEC Center and Dipartimento di Fisica, Universit\`a di Trento, 38123 Povo, Italy.}

\author{A.~Lema\^{i}tre}
\affiliation{Universit\'{e} Paris-Saclay, CNRS, Centre de Nanosciences et de Nanotechnologies (C2N), 91120, Palaiseau, France}

\author{M. Morassi}
\affiliation{Universit\'{e} Paris-Saclay, CNRS, Centre de Nanosciences et de Nanotechnologies (C2N), 91120, Palaiseau, France}

\author{I.~Sagnes}
\affiliation{Universit\'{e} Paris-Saclay, CNRS, Centre de Nanosciences et de Nanotechnologies (C2N), 91120, Palaiseau, France}

\author{L.~Le~Gratiet}
\affiliation{Universit\'{e} Paris-Saclay, CNRS, Centre de Nanosciences et de Nanotechnologies (C2N), 91120, Palaiseau, France}

\author{A.~Harouri}
\affiliation{Universit\'{e} Paris-Saclay, CNRS, Centre de Nanosciences et de Nanotechnologies (C2N), 91120, Palaiseau, France}

\author{M. Wouters}
\affiliation{TQC, Universiteit Antwerpen, Universiteitsplein 1, B-2610 Antwerpen, Belgium.}

\author{I. Carusotto}
\affiliation{INO-CNR BEC Center and Dipartimento di Fisica, Universit\`a di Trento, 38123 Povo, Italy.}

\author{A.~Amo}
\affiliation{Univ. Lille, CNRS, UMR 8523 – PhLAM – Physique des Lasers Atomes et Molécules, F-59000 Lille, France}

\author{M. Richard}
\affiliation{Univ. Grenoble Alpes, CNRS, Grenoble INP, Institut N\'eel, 38000 Grenoble, France.}

\author{A. Minguzzi}
\affiliation{Univ. Grenoble Alpes and CNRS, Laboratoire de Physique et Mod\'elisation des Milieux Condens\'es, 38000 Grenoble, France.}

\author{L. Canet}
\affiliation{Univ. Grenoble Alpes and CNRS, Laboratoire de Physique et Mod\'elisation des Milieux Condens\'es, 38000 Grenoble, France.}

\author{S.~Ravets}
\affiliation{Universit\'{e} Paris-Saclay, CNRS, Centre de Nanosciences et de Nanotechnologies (C2N), 91120, Palaiseau, France}

\author{J.~Bloch}
\affiliation{Universit\'{e} Paris-Saclay, CNRS, Centre de Nanosciences et de Nanotechnologies (C2N), 91120, Palaiseau, France}

\date{\today}

\begin{abstract}
\vskip 2em

\textbf{Revealing universal behaviors is a hallmark of statistical physics.
Phenomena such as the stochastic growth of crystalline surfaces~\cite{krug1990}, of interfaces in bacterial colonies~\cite{wakita1997}, and spin transport in quantum magnets~\cite{5ljubotina2017, 6ljubotina2019,7scheie2021, 8wei2021} all belong to the same universality class, despite the great plurality of physical mechanisms they involve at the microscopic level.
This universality stems from a common underlying effective dynamics governed by the non-linear stochastic Kardar-Parisi-Zhang (KPZ) equation~\cite{1kardar1986}.
Recent theoretical works suggest that this dynamics also emerges in the phase of out-of-equilibrium systems displaying macroscopic spontaneous coherence~\cite{9altman2015, 10ji2015, 11he2015, 12zamora2017, 13comaron2018, 14squizzato2018, 15amelio2020, 16ferrier2020, 17deligiannis2021, 18mei2021}.
Here, we experimentally demonstrate that the evolution of the phase in a driven-dissipative one-dimensional polariton condensate falls in the KPZ universality class.
Our demonstration relies on a direct measurement of KPZ space-time scaling laws~\cite{38family1985, 2halpin1995}, combined with a theoretical analysis that reveals other key signatures of this universality class.
Our results highlight fundamental physical differences between out-of-equilibrium condensates and their equilibrium counterparts, and open a paradigm for exploring universal behaviors in driven open systems.}
\end{abstract}

{
\let\clearpage\relax
\maketitle
}

\begin{figure}[h!]
    \centering
    \includegraphics[scale=0.65]{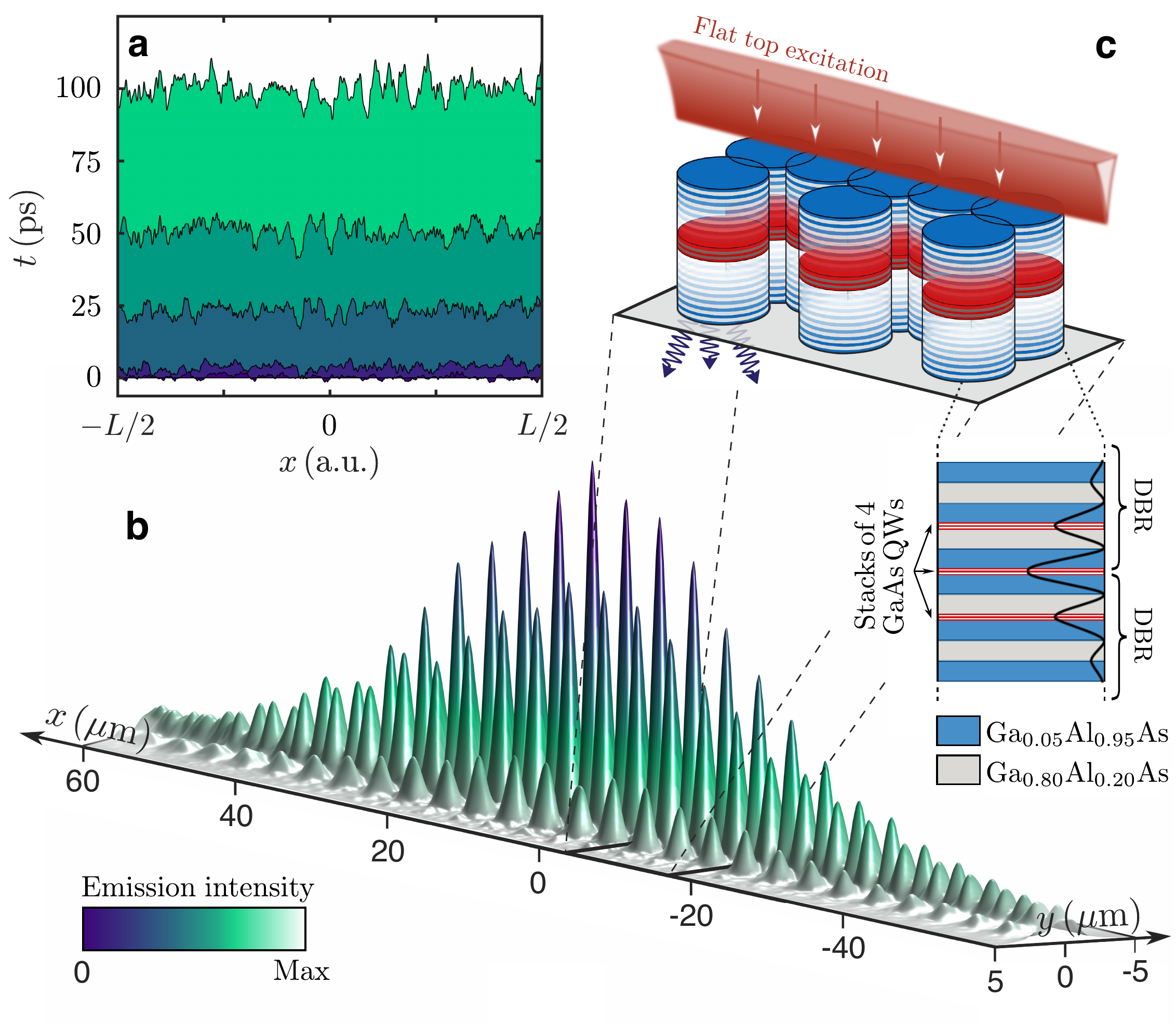}
    \caption{
    \textbf{KPZ physics in the phase dynamics of a 1D polariton condensate.}
    \textbf{a} Snapshots of the condensate phase evolution obtained by numerically solving Eqs.~\eqref{eq:GPE}-\eqref{eq:reservoir}.
    The unwrapped phase evolves in time as a growing interface.
    \textbf{b} Intensity distribution of the condensate emission measured for $P/P_{\mathrm{th}} \!=\! 1.1$. \textbf{c} Sketch of the lattice together with the excitation scheme.
    The lattice is excited using an elongated flat top beam.
    Inset: Sketch of a micropillar inner structure. A semiconductor optical microcavity is enclosed by two Bragg mirrors. Quantum wells (QWs, red layers) are distributed at the anti-nodes of the cavity mode (black line).
    }
    \label{fig:intro}
\end{figure}

\begin{figure}[h!]
    \centering
    \includegraphics[scale=0.60]{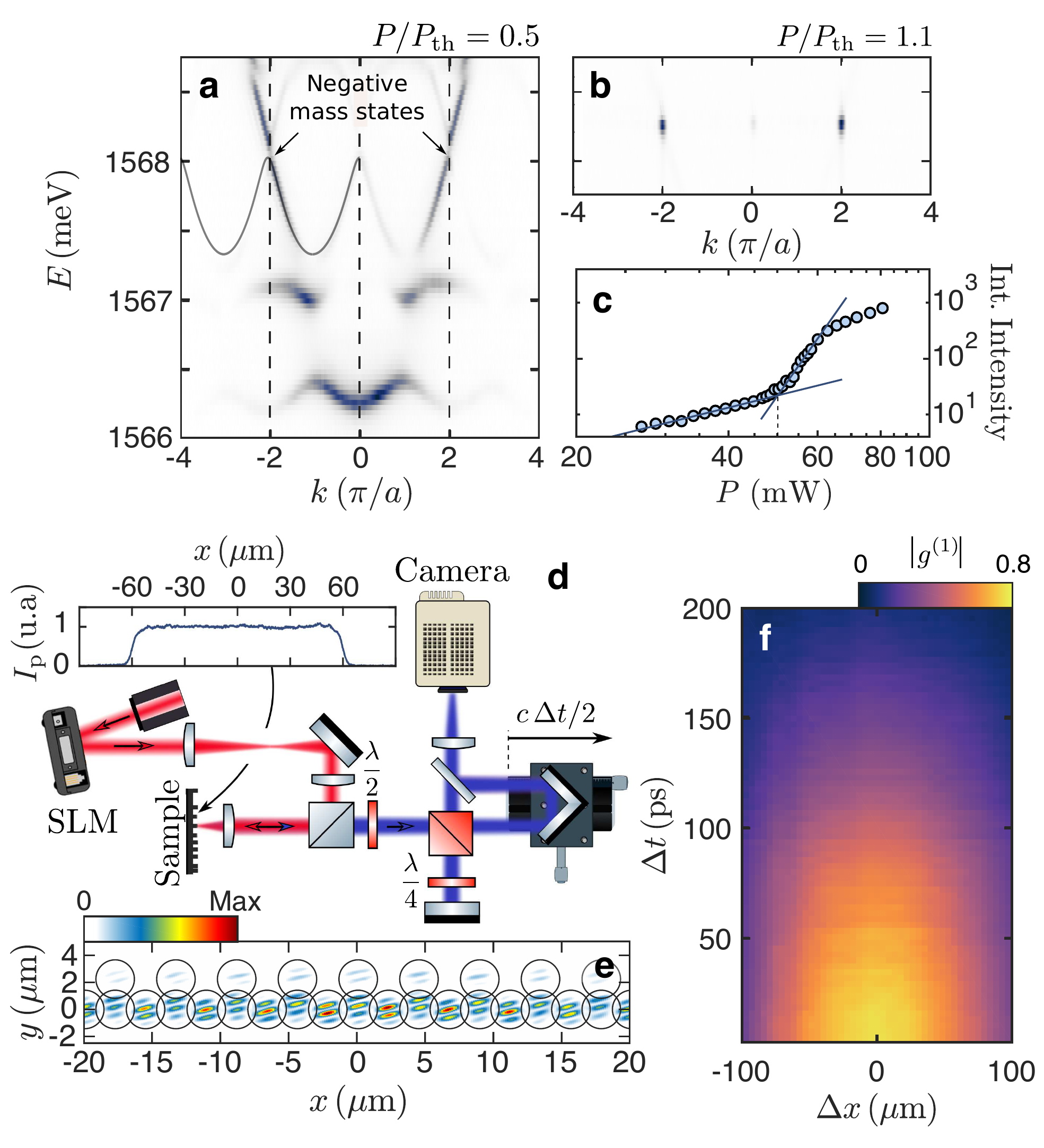}
    \caption{
    \textbf{Probing the coherence of 1D polariton condensates.}
    \textbf{a-b} Momentum-resolved emission spectra captured \textbf{(a)} below ($P/P_{\mathrm{th}} = 0.5$) and \textbf{(b)} above ($P/P_{\mathrm{th}} = 1.1$) condensation threshold.
    Above threshold, the emission is peaked at the top of the highest energy S-band, evidencing polariton condensation in a negative mass state.
    \textbf{c} Integrated emission intensity as a function of the excitation power.
    The onset of the intensity nonlinear increase is observed for $P_{\mathrm{th}} \approx 50 \, \mathrm{mW}$.
    \textbf{d} Sketch of the experimental setup.
    Inset: Flat top intensity profile of the excitation spot.
    \textbf{e} Interference pattern obtained for $P/P_{\mathrm{th}} = 1.13$ by overlapping the condensate field $\mathcal{E}(x, t_{0})$ with its mirror symmetric $\mathcal{E}(-x, t_{0})$ ($\Delta t = 0$).
    The black circles locate the lattice pillars.
    \textbf{f} Coherence map showing the value of $|g^{(1)}|$, retrieved from the fringe visibility using Eq.~\eqref{eq:g1retr}, as a function of $\Delta x$ and $\Delta t$.
    }
    \label{fig:exp}
\end{figure}

\begin{figure}[h!]
    \centering
    \includegraphics[scale=0.60]{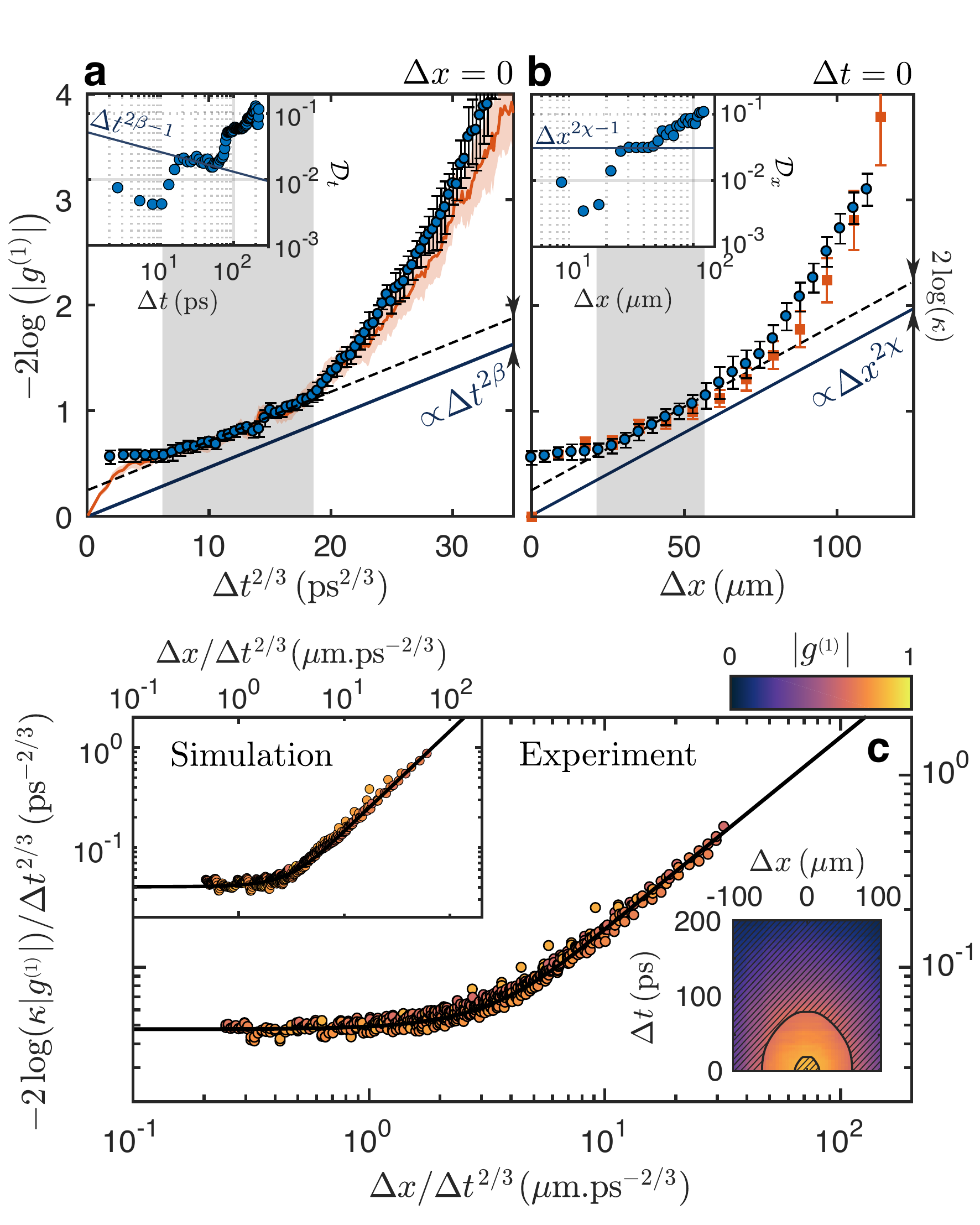}
    \caption{
    \textbf{KPZ scaling in the coherence decay of a 1D polariton condensate.}
    \textbf{a} Measured values of $-2 \, \mathrm{log} \left(| g^{(1)}(0, \Delta t) |\right)$ as a function of $\Delta t^{2/3}$.
    Inset: Temporal derivative of $-2 \, \mathrm{log} \left(| g^{(1)}(0, \Delta t) |\right)$, computed from the experimental data, as a function of $\Delta t$.
    \textbf{b} Measured values of $-2 \, \mathrm{log} \left(| g^{(1)}(\Delta x, 0) |\right)$ as a function of $\Delta x$.
    Inset: Spatial derivative of $-2 \, \mathrm{log} \left(| g^{(1)}(\Delta x, 0) |\right)$, computed from the experimental data, as a function of $\Delta x$. On both panels, the grey area delimits the KPZ temporal window.
    In panels (a) (resp. (b)), the simulated data are shown with an orange line (resp. squares).
    Errorbars on the experimental data points are calculated by performing a repeatability analysis on the numerical extraction of $g^{(1)}(\Delta x, \Delta t)$ from the interferograms.
    The orange-shaded area (resp. orange errorbars) on panel (a) (resp. (b)) show the $95\%$ confidence interval on the simulated data.
    \textbf{c} Measured values of $-2 \, \mathrm{log}(\kappa |g^{(1)}|)/\Delta t^{2/3}$ as a function of $y = \Delta x/\Delta t^{2/3}$, for points within the non-hatched region of the coherence map (bottom inset).
    Top inset: Same as in the main panel but for numerical data.
    On both graphs, the black line corresponds to the KPZ scaling function $F = C_{0} F_{\mathrm{KPZ}}(y/y_{0})$, adjusted to the experimental data by tuning the values of $C_{0}$ and $y_{0}$.
    Experimental and simulated $|g^{(1)}|$ datasets are normalized using $\kappa \!=\! 1.13$.
    In all panels, the numerical data are averaged over $10^{4}$ realizations of the noise.
    }
    \label{fig:scaling}
\end{figure}

\begin{figure}[h!]
    \centering
    \includegraphics[scale=0.65]{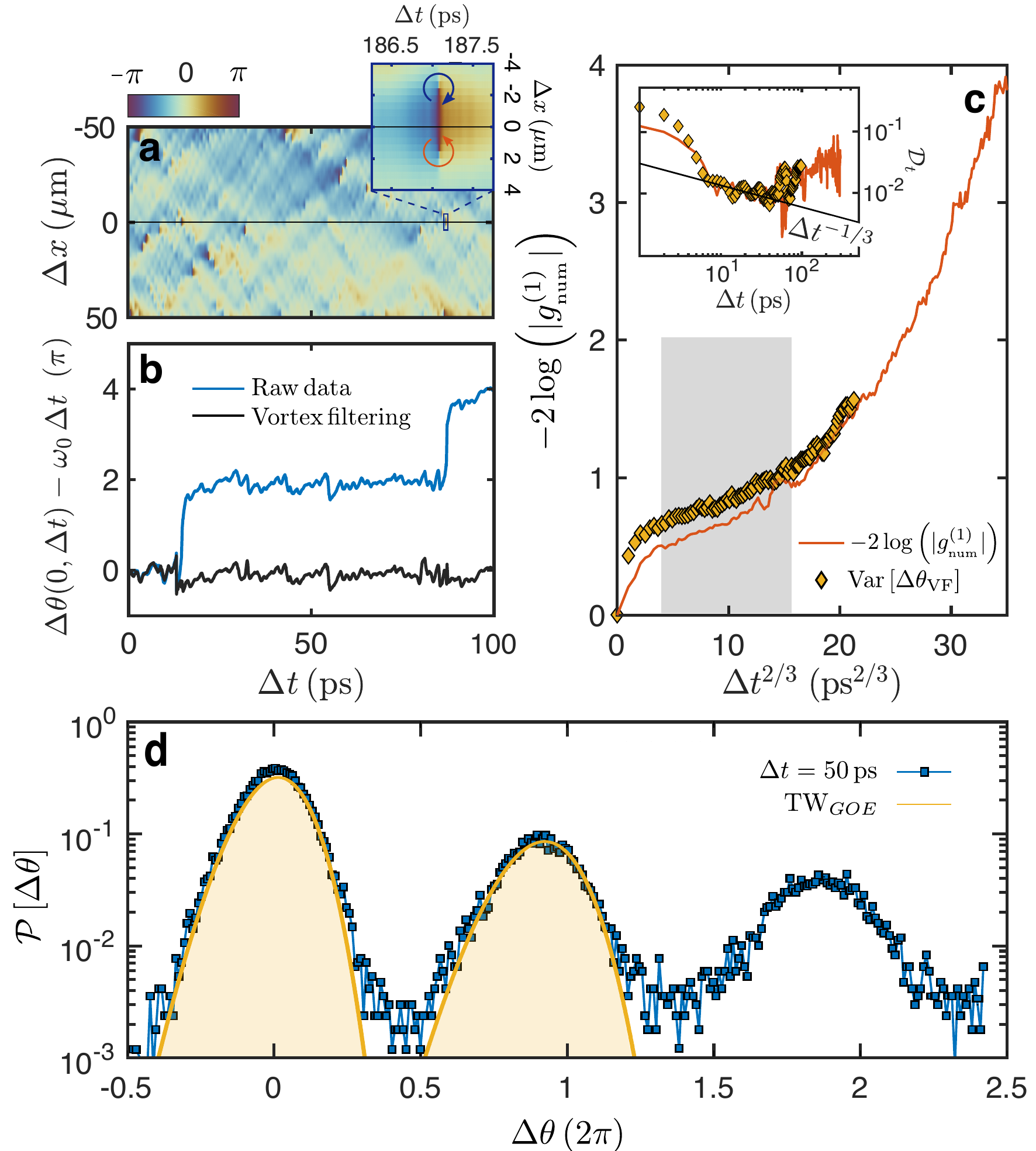}
    \caption{
    \textbf{Analysis of the simulated phase dynamics.}
    \textbf{a} Example of a phase map obtained after subtraction of the dynamical phase, $\omega_{0} \, \Delta t$.
    Inset: Zoom on a V-AV pair.
    \textbf{b} Time evolution of the unwrapped phase along $x = 0$ (blue line): two V-AV pairs are crossed, resulting in two phase jumps of approximately $2 \pi$.
    The black line of panel \textbf{(b)} is obtained by filtering all the vortices present in the phase map \textbf{(a)}.
    \textbf{c} $-2 \, \mathrm{log}\left( |g^{(1)}| \right)$ (red line) and $\mathrm{Var} \left[ \Delta \theta_{\mathrm{VF}}\right]$ (symbols) as a function of $\Delta t^{2/3}$ for $\Delta x = 0$, computed over $10^{4}$ realizations. The orange line is the same as in Fig.~\ref{fig:scaling}a. Inset: Temporal derivative of $-2 \, \mathrm{log} \left( |g^{(1)}| \right)$ (red line) and $\mathrm{Var} \left[ \Delta \theta_{\mathrm{VF}} \right]$ (symbols) as a function of $\Delta t$, showing similar behaviour in the KPZ window.
    \textbf{d} Probability distribution of phase fluctuations obtained at $\Delta t \!=\! 50 \, \mathrm{ps}$ (blue squares) from numerical simulations. Our computation involves $3 \times 10^{4}$ realizations of the noise. The first (resp. second, third, etc) peak from the left gathers the realizations for which no jump (resp one, two, etc) has occurred before $\Delta t = 50 \, \mathrm{ps}$. Yellow line: Fit of the numerical data by the Tracy-Widom GOE distribution.
    }
    \label{fig:vortex}
\end{figure}

\noindent Universality is a powerful concept in statistical physics that allows describing critical phenomena on the basis of a few fundamental ingredients. At thermal equilibrium, models like the Ising model are pivotal in understanding the critical properties of a wide class of physical systems.
Reversely, non-equilibrium systems lack a complete classification of their universal properties. In this context, the KPZ equation appears as a quintessential model to investigate non-equilibrium phenomena and phase transitions. Here, we provide the experimental demonstration that one-dimensional (1D) out-of-equilibrium condensates belong to the KPZ universality class.
\vspace{8pt}
\newline
The KPZ equation~\cite{1kardar1986} was originally proposed to describe the stochastic growth dynamics of an interface height $h(\boldsymbol{r}, t)$:
\begin{equation}
    \partial_{t} h(\boldsymbol{r}, t) = \nu \nabla^{2} h(\boldsymbol{r}, t) +\frac{\lambda}{2} \left[ \boldsymbol{\nabla}( h \left(\boldsymbol{r}, t \right)  \right]^{2} + \eta(\boldsymbol{r}, t).
\end{equation}
On the right, the first term corresponds to a smoothening diffusion, the second one to a nonlinear contribution leading to critical roughening, while the third one is a Gaussian white noise introducing stochasticity.
The spatial and temporal correlation functions of $h(\boldsymbol{r}, t)$ exhibit power law behaviors, with critical exponents specific to the KPZ universality class~\cite{1kardar1986}.
Currently available observations of KPZ dynamics have mainly focused on growing interfaces in classical systems~\cite{2halpin1995, 3krug1997, 4takeuchi2018}.
Interestingly, KPZ physics is also under investigation in quantum magnets~\cite{5ljubotina2017, 6ljubotina2019, 7scheie2021, 8wei2021}.
\vspace{8pt}
\newline
Recent theoretical works predicted that the spatio-temporal evolution of the phase of a polariton condensate falls into the KPZ universality class~\cite{9altman2015, 10ji2015, 11he2015, 12zamora2017, 13comaron2018, 14squizzato2018, 15amelio2020, 16ferrier2020, 17deligiannis2021, 18mei2021}.
However, unlike an actual interface height, the phase is defined periodically between $0$ and $2 \pi$.
This version of the KPZ equation is relevant for out-of-equilibrium systems developing macroscopic spontaneous coherence (lasers, arrays of coupled limit-cycles oscillators~\cite{19lauter2017}), and other systems like polar active smectic phases~\cite{20chen2013}.
The compactness of the phase field results in a rich phase diagram comprising not only the KPZ phase but also other regimes characterized by the proliferation of topological defects~\cite{19lauter2017, 20chen2013, 21he2017}.
In this article, we explore experimentally the spatio-temporal dynamics of the first order coherence in a 1D polariton condensate.
We observe the predicted coherence decay, and demonstrate the collapse of the data onto the universal KPZ scaling function.
Our theoretical analysis evidences how the observed 1D KPZ physics is resilient to the presence of vortex-antivortex (V-AV) pairs.
\vspace{8pt}
\newline
Cavity polaritons are hybrid quasi-particles emerging in semiconductor cavities from the strong coupling between electronic excitations (excitons) in a quantum well and cavity photons~\cite{22weisbuch1992, 23carusotto2013}.
Owing to their excitonic fraction, polaritons can be created from a gain medium of incoherent excitons (the excitonic reservoir) through bosonic stimulated scattering.
Due to photon leakage through the mirrors, the polariton dynamics is intrinsically out of equilibrium, its steady-state being the result of the balance between drive, relaxation and losses.
Finally, polaritons can be laterally confined in lattices~\cite{26schneider2016}, enabling band structure engineering.
\vspace{8pt}
\newline
The 1D polariton condensates at play consist in the macroscopic occupation of a given state obtained by incoherently pumping the exciton reservoir~\cite{24deng2002, 25kasprzak2006}. Above a critical density, exciton stimulated scattering from the reservoir into this state triggers a spontaneous $U(1)$ symmetry-breaking of the phase.
The condensate and reservoir dynamics are described by two coupled equations~\cite{23carusotto2013}:
\begin{align}
    i \hbar \frac{\partial}{\partial t}  \psi(x, t) &= \left[ E(\hat{k}) - \frac{i \hbar}{2} \gamma(\hat{k}) +g|\psi(x, t)|^{2} + 2 g_{R} n_{R}(x, t) +\frac{i \hbar}{2} R n_{R}(x, t) \right] \psi(x, t) +\xi(x, t)
    \label{eq:GPE} \\
    \frac{\partial}{\partial t} n_{R}(x, t) &= P(x)- \left( \gamma_{R} + R|\psi(x, t)|^{2} \right) n_{R}(x, t)
    \label{eq:reservoir}
\end{align}
Here, $\hat{k} = -i\hbar\partial/\partial x$ is the momentum operator, $\psi(x, t) = \sqrt{\rho(x, t)} \, e^{i \theta(x, t)}$ the polariton condensate field with density $\rho(x, t)$ and phase $\theta(x, t)$, $E(\hat{k})$ the polariton dispersion, $\gamma(\hat{k})$ the momentum-dependent decay rate, and $g$ the polariton-polariton interaction strength.
The exciton reservoir, with density $n_{R}(x, t)$, is pumped at rate $P(x)$.
Excitons either relax into the polariton condensate by stimulated scattering with rate $R$ or decay following other channels at rate $\gamma_{R}$.
The term $2 g_{R} n_{R}$ describes polariton repulsive interactions with reservoir excitons. It dominates the polariton blueshift close to threshold, and induces dephasing through inhomogeneous spectral broadening~\cite{28love2008}. Finally, $\xi(x, t)$ describes Gaussian noise induced by drive and loss.
\vspace{8pt}
\newline
Ignoring interactions with the reservoir ($g_{R} = 0$), previous theoretical studies have shown that the condensate phase $\theta(x, t)$ follows a KPZ equation~\cite{9altman2015, 10ji2015, 11he2015, 27grinstein1993}.
The condensate phase profile behaves as a classical interface (Fig.~\ref{fig:intro}a), and develops KPZ spatio-temporal correlations characterized by the phase variance $\mathrm{Var} \left( \Delta \theta(x, t) \right) = \langle \left( \Delta \theta(x, t) \!-\! \langle \Delta \theta(x, t) \rangle \right)^{2} \rangle$ (where $\langle . \rangle$ is the statistical averaging over different noise realizations, and $\Delta \theta = \theta(x, t)-\theta(x, t_{0})$, $t_{0}$ being a reference time).
Here,
we derive the mapping to the KPZ equation for $g_{R} \neq 0$ and obtain the KPZ parameters in terms of those entering Eqs.~\eqref{eq:GPE}-\eqref{eq:reservoir} (see~\cite{29SupMat}).
\vspace{8pt}
\newline
Experimentally probing KPZ correlations requires extended condensates to avoid finite size effects, a condition that was not fulfilled in early coherence measurements~\cite{30roumpos2012, 31fischer2014}.
This requirement is demanding due to the development of a modulation instability, which fragments the condensate into mutually incoherent micron-sized puddles~\cite{32bobrovska2014, 33daskalakis2015, estrecho2018, bobrovska2018}.
This instability originates from repulsive condensate-reservoir interactions, which result in effective attractive polariton-polariton interactions within the condensate and lead to its destabilization~\cite{34smirnov2014, 35liew2015}.
A solution to tame this instability is to spatially separate the excitonic reservoir from the condensate~\cite{36caputo2018}, or to use negative mass polaritons, obtained by band engineering in a lattice~\cite{37baboux2018}.
The negative mass changes the sign of the effective polariton-polariton interactions, thus restoring the condensate stability.
We use this negative mass technique to generate stable 1D polariton condensates extending over distances as large as $100 \, \mathrm{\mu m}$ (see Fig.~\ref{fig:intro}b).
\vspace{8pt}
\newline
The sample consists in a semiconductor microcavity embedding quantum wells (Fig.~\ref{fig:intro}c, and~\cite{29SupMat}).
We use nanotechnology processes to fabricate 1D asymmetric Lieb lattices of coupled micropillars containing three sites per unit cell (see Fig.~\ref{fig:intro}c, Methods Section, and~\cite{29SupMat}). We incoherently populate the excitonic reservoir using a blue-detuned cw laser focused on a single lattice, with an elongated flat-top intensity profile.
\vspace{8pt}
\newline
The polariton emission analysed in momentum space below condensation threshold (see Fig.~\ref{fig:exp}a) evidences the lattice band structure emerging from the hybridization of the discrete modes confined in each micropillar. Above a threshold $P_{\mathrm{th}} = 50 \, \mathrm{mW}$, the emission becomes peaked at the top of a negative mass band (Fig.~\ref{fig:exp}b).
This feature, together with the non-linear increase of the emission intensity (Fig.~\ref{fig:exp}c), indicates the onset of polariton condensation.
The condensate emission intensity in real space at $P = 1.1 \, P_{\mathrm{th}}$ reveals an extended and regular intensity profile envelope, as expected in absence of modulation instability (Fig.~\ref{fig:intro}b).
Although we use a flat-top intensity profile for the excitation beam, the condensate profile is rather peaked, which hints toward localization induced by disorder in the lattice.
\vspace{8pt}
\newline
We define the first order correlation evaluated between points separated in space by $\Delta x = 2x$ and delayed by $\Delta t$:
\begin{align}
     g^{(1)}(\Delta x, \Delta t) = \frac{\langle \psi^{*} (x, t_{0}) \psi(-x, t_{0}+\Delta t)\rangle}{\sqrt{\langle |\psi(x, t_{0})|^{2} \rangle} \sqrt{\langle|\psi(-x, t_{0}+\Delta t)|^{2}\rangle}} \label{Full_g1} \, .
\end{align}
Neglecting density-density and density-phase correlations, we show that $g^{(1)}(\Delta x, \Delta t) \approx \exp \left\{ -\mathrm{Var} \left[ \Delta \theta (\Delta x, \Delta t)\right]/2 \right\}$ (see~\cite{29SupMat}).
We thus expect KPZ universal scaling to show up as stretched exponentials in the coherence decay: $|g^{(1)}(\Delta x, \Delta t_{0})| \propto \exp{\left[ -(\Delta x/\lambda)^{2 \chi} /2\right]}$ (for fixed $\Delta t_{0}$) and $|g^{(1)}(\Delta x_{0}, \Delta t)| \propto \exp{\left[ -(\Delta t/\tau)^{2 \beta} /2\right]}$ (for fixed $\Delta x_{0}$), where $\chi$ and $\beta$ are the universal KPZ critical exponents while $\lambda$ and $\tau$ are two non-universal parameters.
In 1D, the “roughness” exponent $\chi$ is equal to $1/2$ and the “growth” exponent $\beta$ to $1/3$~\cite{2halpin1995, 38family1985}.
While $\chi = 1/2$ is common to several universality classes for 1D systems (such as linear systems described with Bogoliubov theory~\cite{39edwards1982, 40wouters2006}), $\beta = 1/3$ is an unambiguous signature of KPZ physics.
\vspace{8pt}
\newline
The condensate coherence $|g^{(1)}|$ is measured using Michelson interferometry (Fig.~\ref{fig:exp}d).
The field emitted by the condensate at time $t_{0}$, $\mathcal{E}(x, t_{0}) \propto \Psi(x, t_{0})$, and the one emitted at the mirror-symmetric point at time $t_{0}+\Delta t$, $\mathcal{E}(-x, t_{0} + \Delta t) \propto \Psi(-x, t_{0} + \Delta t)$, are overlapped ($\Delta t$ is the delay in the two interferometer arms).
The resulting intensity pattern exhibits well-contrasted interference fringes over the whole condensate (Fig.~\ref{fig:exp}e).
This indicates the emergence of extended spatial coherence as the fringe contrast gives a direct visualization of the degree of coherence between fields emitted at two points spatially separated by $\Delta x = 2x$ and delayed by $\Delta t$.
More specifically, $|g^{(1)}(\Delta x, \Delta t)|$ is determined from the fringe visibility $V(\Delta x, \Delta t)$ and from the intensity distributions $|\mathcal{E}(x, t_{0})|^{2}$ and $|\mathcal{E}(-x, t_{0}+\Delta t)|^{2}$ measured separately, using
\begin{equation}
    2|g^{(1)}(\Delta x, \Delta t)| \sqrt{|\mathcal{E}(x, t_{0})|^{2} |\mathcal{E}(-x, t_{0}+\Delta t)|^{2}} = V(\Delta x, \Delta t) \left[|\mathcal{E}(x, t_{0})|^{2} + |\mathcal{E}(-x, t_{0}+\Delta t)|^{2} \right].
    \label{eq:g1retr}
\end{equation}
The result is shown in Fig.~\ref{fig:exp}f.
\vspace{8pt}
\newline
We first focus on the temporal decay of the coherence.
To search for the growth exponent, we compute the temporal derivative $\mathcal{D}_{t} = -2 \, \partial \, \mathrm{log}\!\left(| g^{(1)}(0, \Delta t) |\right)\!/\partial \Delta t $ from our dataset.
According to KPZ theory, this derivative scales as a power law with exponent $2 \beta-1=-1/3$.
In the inset of Fig.~\ref{fig:scaling}a, we identify such scaling throughout the temporal window $15 \, \mathrm{ps} \leq \Delta t \leq 80 \, \mathrm{ps}$.
Equivalently, we observe in the main panel a linear increase of $-2 \, \mathrm{log} \! \left( |g^{(1)}(0, \Delta t)| \right)$ as a function of $\Delta t^{2/3}$ over the same window (grey shaded area), which demonstrates a key feature of KPZ dynamics.
At short timescales, the deviation from KPZ scaling and the saturation of $|g^{(1)}|$ are due to an incoherent background (spectrally broad photoluminescence from uncondensed states) that hides the onset of KPZ fluctuations.
For $\Delta t \geq 80 \, \mathrm{ps}$, $-2 \, \mathrm{log} \! \left( |g^{(1)}(0, \Delta t)| \right)$ also departs from KPZ scaling and follows a super-linear behavior that we attribute to slow reservoir population fluctuations~\cite{28love2008}.
We now perform a similar analysis in the spatial domain.
The results are shown in Fig.~\ref{fig:scaling}b.
The spatial derivative $\mathcal{D}_{x} =  -2 \, \partial \, \mathrm{log}\!\left(| g^{(1)}(\Delta x, 0) |\right)\!/\partial \Delta x $ exhibits a plateau within the spatial window $30 \, \mathrm{\mu m} \leq \Delta x \leq 60 \, \mathrm{\mu m}$, in agreement with $\chi = 1/2$ (inset).
The roughness exponent $\chi =1/2$ also shows up in the linear increase of $-2 \, \mathrm{log} \! \left( |g^{(1)}(\Delta x, 0)| \right)$ as a function of $\Delta x$, over the same window (grey shaded area in the main panel).
When approaching condensate edges ($\Delta x \geq 60 \, \mathrm{\mu m}$), the coherence decays faster as a consequence of enhanced fluctuations at smaller polariton density (see~\cite{41chiocchetta2013}).
Pushing further this data analysis, we fit the coherence decay curves with stretched exponentials and deduce experimental values for the scaling exponents: $\chi_{\mathrm{exp}} = 0.51 \pm 0.08$ and $\beta_{\mathrm{exp}} = 0.36 \pm 0.11$.
The uncertainty on $\beta_{\mathrm{exp}}$ allows us to discriminate between the different universality classes relevant for our system, as the KPZ value $\beta = 1/3$ remains the only one
lying within the $95\%$ confidence interval on $\beta_{\mathrm{exp}}$ (see~\cite{29SupMat}).
\vspace{8pt}
\newline
We now search for KPZ signatures over the whole space-time correlation map.
We select all data points where $0.57 \leq |g^{(1)}| \leq 0.75$, the range where we evidence KPZ scaling at $\Delta t = 0$. This space-time window is shown in the bottom inset of Fig.~\ref{fig:scaling}c.
For this subset of data points, we plot in Fig.~\ref{fig:scaling}c the value of $-2 \, \mathrm{log} \! \left( \kappa |g^{(1)}(\Delta x, \Delta t)| \right)/\Delta t^{2 \beta}$ as a function of the rescaled coordinate $y = \Delta x/\Delta t^{1/Z}$, where $Z = \beta/\chi = 2/3$ and $\kappa$ is a normalization factor (see~\cite{29SupMat}).
Strikingly, all these data points collapse onto the scaling function $F = C_{0} F_{\mathrm{KPZ}}(y/y_{0})$, where $F_{\mathrm{KPZ}}$ is the tabulated dimensionless KPZ universal scaling function~\cite{42prahofer2004, 43spohn2014}.
We use the non-universal constants $C_{0}$ and $y_{0}$ as fitting parameters in order to vertically and horizontally shift $F$ onto the collapsed data points.
This result demonstrates that 1D polariton condensates indeed belong to the KPZ universality class.
In order to reinforce the generality of this conclusion, we performed the same measurement and analysis on a different 1D lattice with four sites per unit cell.
We also found a KPZ space-time window where all data points collapse onto the universal scaling curve~\cite{29SupMat}.
To complete the picture, we carried out the same analysis at higher excitation powers, and found for both lattices that the spatio-temporal KPZ window shrinks for increasing $P/P_{\mathrm{th}}$ and eventually disappears when $P/P_{\mathrm{th}} > 1.2$~\cite{29SupMat}.
\vspace{8pt}
\newline
With the prospect of confronting these experimental data to simulations, we compute the phase evolution of the polariton condensate by numerically solving Eqs.~\eqref{eq:GPE}-\eqref{eq:reservoir}.
We obtain some microscopic parameters from the experiment, namely: the dispersion relation $E(k)$, the polariton blueshift $2 g_{R} \langle n_{R} \rangle$, the linewidth $\gamma(k)$, and the threshold-normalized laser power $P/P_{\mathrm{th}}$.
For the reservoir relaxation rate $R$ and decay rate $\gamma_{R}$, we choose values within some realistic range yielding the best agreement with the measured $|g^{(1)}|$.
We also adjust the pump size to make the condensate intensity profile $|\Psi(x)|^{2}$ comparable with the experimental one.
Finally, we neglect the polariton-polariton interaction energy $g|\Psi|^{2}$ and  take $g = 0$ in all simulations.
As such, this model also applies to spatially extended lasers in the weak coupling regime.
The calculated $|g_{_{\mathrm{num}}}^{(1)}|$ data are reported in Fig.~\ref{fig:scaling}a–b, showing excellent agreement with the experiment.
The short and long time behaviour is reproduced, together with the shrinking of the KPZ window when the excitation power is increased (see~\cite{29SupMat}).
\vspace{8pt}
\newline
We then perform on the numerical data the same analysis as on the experimental ones.
We plot $-2 \, \mathrm{log} \! \left( \kappa |g_{_{\mathrm{num}}}^{(1)}(\Delta x, \Delta t)| \right)/\Delta t^{2 \beta}$ as a function of $y$, selecting the points for which $0.57 \leq |g_{_{\mathrm{num}}}^{(1)}| \leq 0.75$.
The result is shown in Fig.~\ref{fig:scaling}c (top inset), together with the KPZ scaling function $F = C_{0} F_{\mathrm{KPZ}}(y/y_{0})$, using for $C_{0}$ and $y_{0}$ the same values as for the experimental data.
The simulated data align to the scaling function, thus fully validating our model.
\vspace{8pt}
\newline
To deepen our insight into the phase dynamics, we now analyze its stochastic behavior in the numerical simulations.
Fig.~\ref{fig:vortex}a shows an example of a phase map $\theta(x,\Delta t)$ corresponding to a given realization of the noise (others are shown in~\cite{29SupMat}).
For better visualization, we plot the wrapped phase after subtracting the dynamical phase $\omega_0 \Delta t$ ($\hbar \omega_0$ being the condensate energy).
We observe two kinds of phase variations: small amplitude fluctuations and fast (scarce) phase jumps.
These jumps are associated to pairs of close-by spatio-temporal vortices with opposite circulation (see inset), that we name V-AV pairs.
To analyse the effect of these V-AV pairs on the phase dynamics, we show in Fig.~\ref{fig:vortex}b the unwrapped phase temporal evolution at $x = 0$ (horizontal line in Fig.~\ref{fig:vortex}a).
The phase evolution exhibits plateaus with small amplitude phase fluctuations, separated by phase jumps of approximately $2 \pi$, occurring on a fast timescale ($\sim 1~{\rm ps}$) when passing through a V-AV pair.
Note that for the regime of parameters explored here, almost all vortices appear in V-AV pairs.
For higher noise or stronger interactions, activation of single vortices is expected and would lead to other dynamical regimes~\cite{21he2017}.
\vspace{8pt}
\newline
We now focus on showing that small amplitude phase fluctuations follow KPZ scaling laws by computing the phase variance $\mathrm{Var} \left[ \Delta \theta \right]$, and that the presence of V-AV only weakly affects $|g_{_{\mathrm{num}}}^{(1)}|$.
Since $\mathrm{Var} \left[ \Delta \theta \right]$ is extremely sensitive to phase jumps, we select for each trajectory a $100 \, \mathrm{ps}$ wide window where the phase undergoes the smallest amount of jumps.
When few phase jumps remain in the selected window, we filter them out by adding at every vortex location an other one of opposite charge (grey line in Fig.~\ref{fig:vortex}b).
Note that we discard $5\%$ of the realizations, where vortices proliferate (see~\cite{29SupMat}).
We then compute the phase variance $\mathrm{Var} \left[ \Delta \theta_{\mathrm{VF}} \right]$ over the set of vortex-free (VF) time windows.
The result is plotted in Fig.~\ref{fig:vortex}c together with the values of $-2 \mathrm{log}\left(|g_{_{\mathrm{num}}}^{(1)}|\right)$ (computed without filtering the V-AV pairs).
Both quantities exhibit the KPZ power-law scaling over the same time window (grey-shaded area), as further illustrated on Fig.~\ref{fig:vortex}c (inset) where we plot their time derivative.
This result reveals that the first order coherence is indeed a good observable to probe the KPZ dynamics of the condensate phase, even in presence of V-AV pairs.
\vspace{8pt}
\newline
Another striking signature of KPZ physics lies in the fact that phase fluctuations are governed by a probability distribution, which $-$ unlike the normal distribution $-$ is skewed and exhibits markedly different tails.
In Fig.~\ref{fig:vortex}d, we show the calculated probability distribution of the unwrapped $\Delta \theta (0, \Delta t)$, computed over all trajectories (thus including vortices) for $\Delta t = 50 \, \mathrm{ps}$, \textit{i.e.} in the center of the KPZ window.
All trajectories that have not crossed any V-AV pair contribute to the first peak in the distribution.
The second peak corresponds to trajectories which have crossed one V-AV pair before reaching $\Delta t = 50 \, \mathrm{ps}$ and have thus undergone one phase jump close to $2\pi$.
Strikingly the first two peaks are skewed and well reproduced by the Tracy-Widom Gaussian Orthogonal Ensemble (GOE) distribution associated to the flat subclass (see~\cite{29SupMat}).
Cumulating data for various $\Delta t$, we obtain an agreement with the Tracy-Widom GOE distribution over six decades (see~\cite{29SupMat} for details on the analysis).
The third peak corresponds to few realizations displaying two phase jumps.
The lack of statistics prevents precise analysis of its shape.
Our simulations highlight that V-AV pairs only modify the probability distribution by adding replicas of the main peak without significantly changing their shape.
Moreover, they confirm that in the regime where a low density of V-AV pairs stochastically shows up, KPZ dynamics is not destroyed but occurs piece-wise in the spatio-temporal domain.
\vspace{8pt}
\newline
To conclude, both our experimental and theoretical analysis prove that KPZ scaling laws are present in the decay of the first-order coherence of 1D driven-dissipative polariton condensates.
Our findings apply to any spatially extended driven open systems subject to gain and loss and characterized by a $U(1)$ symmetry breaking.
Our work opens many new challenges to be addressed in the future.
In 1D, while our results highlight the striking resilience of KPZ physics to V-AV pairs, regimes at higher noise strength or higher non-linearity remains to be explored~\cite{21he2017}.
Investigating different KPZ universality sub-classes, predicted for various geometries of the condensate environment~\cite{17deligiannis2021}, is now also within reach, when engineering the geometry of the condensate environment.
Beyond 1D, exciton-polariton lattices offer exciting perspectives for the exploration of KPZ physics in 2D, where an experimental realization is highly sought-after~\cite{12zamora2017, 13comaron2018, 14squizzato2018, 15amelio2020, 16ferrier2020, 17deligiannis2021}, and the role of topological defects still actively debated~\cite{12zamora2017}.
An experimental implementation involving polariton condensates would enable to test the different models and serve as a general analog simulator of complex physical systems belonging to KPZ universality class.

\newpage

\noindent{\Large \textbf{Method}}
\vspace{10pt}
\newline
The sample consists of a high quality factor ($Q \simeq 70000$) $\lambda/2$ Ga$_{0.05}$Al$_{0.95}$As microcavity surrounded by two Al$_{0.20}$Ga$_{0.80}$As/ Al$_{0.05}$Ga$_{0.95}$As distributed Bragg reflectors. Three stacks of four $8$ nm GaAs quantum wells are embedded in this microstructure, at the anti-nodes of the cavity mode electromagnetic field, resulting in a $15$ meV collective Rabi splitting. The as-grown planar cavity is patterned into 1D lattices of coupled micropillars ($3 \, \mathrm{\mu m}$ in diameter), using electron beam lithography and dry etching. In this work, we use a $200 \, \mathrm{\mu m}$ long asymmetric Lieb lattice, made of three pillars per unit cell with $2.2 \, \mathrm{\mu m}$ center-to-center separation distance. A close-cycle cryostation holds the sample at 4K.
\vspace{8pt}
\newline
We incoherently populate the excitonic reservoir using a non-resonant continuous-wave laser at $740 \, \mathrm{nm}$ (reflectivity minimum of the Bragg mirror). A spatial light modulator (SLM) enables shaping the excitation spot into a $125 \, \mu \mathrm{m}$ long flat-top beam in the lattice direction.
Note that our experiment is performed under truly continuous-wave excitation conditions, that is, without any chopper. The polariton emission leaking out through the cavity top mirror is analyzed in space, momentum (along the lattice direction $x$) and frequency with a monochromator coupled to a CCD camera.
\vspace{8pt}
\newline We retrieve the condensate first order coherence using Michelson intererometry. A 2-mirror retro-reflector mounted on a step-motorized translation stage in one of the interferometer arms enables us to overlap on a CCD camera the field $\mathcal{E}(x, t_{0})$ emitted by the condensate at time $t_{0}$ and position $x$, with $\mathcal{E}(-x, t_{0}+\Delta t)$, the field emitted at $t_{0}+\Delta t$ and position $-x$ ($\Delta t$ is the delay introduced between the interferometer arms by translating the retro-reflector). In order to probe the temporal scaling of the condensate coherence, we scan the retro-reflector position over $\Delta L = 5\, \mathrm{cm}$, corresponding to a maximum time delay of $\Delta t = 2 \Delta L/ c = 330 \, \mathrm{ps}$. During such a scan, we set the camera exposure time to $1\mathrm{s}$ and acquire a series of 250 images.

\newpage

\begin{singlespacing}

\newpage

\newpage

\newpage

\newpage

\end{singlespacing}

\newpage

\bibliographystyle{naturemag_noURL}

\let\oldaddcontentsline\addcontentsline% Store \addcontentsline
\renewcommand{\addcontentsline}[3]{}% Make \addcontentsline a no-op
\bibliography{Biblio}

\begin{thebibliography}{10}
\expandafter\ifx\csname url\endcsname\relax
  \def\url#1{\texttt{#1}}\fi
\expandafter\ifx\csname urlprefix\endcsname\relax\def\urlprefix{URL }\fi
\providecommand{\bibinfo}[2]{#2}
\providecommand{\eprint}[2][]{\url{#2}}

\bibitem{krug1990}
\bibinfo{author}{Krug, J.} \& \bibinfo{author}{Meakin, P.}
\newblock \bibinfo{title}{Universal finite-size effects in the rate of growth
  processes}.
\newblock \emph{\bibinfo{journal}{Journal of Physics A: Mathematical and
  General}} \textbf{\bibinfo{volume}{23}}, \bibinfo{pages}{L987}
  (\bibinfo{year}{1990}).

\bibitem{wakita1997}
\bibinfo{author}{Wakita, J.-i.}, \bibinfo{author}{Itoh, H.},
  \bibinfo{author}{Matsuyama, T.} \& \bibinfo{author}{Matsushita, M.}
\newblock \bibinfo{title}{Self-affinity for the growing interface of bacterial
  colonies}.
\newblock \emph{\bibinfo{journal}{Journal of the Physical Society of Japan}}
  \textbf{\bibinfo{volume}{66}}, \bibinfo{pages}{67--72}
  (\bibinfo{year}{1997}).

\bibitem{5ljubotina2017}
\bibinfo{author}{Ljubotina, M.}, \bibinfo{author}{{\v{Z}}nidari{\v{c}}, M.} \&
  \bibinfo{author}{Prosen, T.}
\newblock \bibinfo{title}{Spin diffusion from an inhomogeneous quench in an
  integrable system}.
\newblock \emph{\bibinfo{journal}{Nature communications}}
  \textbf{\bibinfo{volume}{8}}, \bibinfo{pages}{1--6} (\bibinfo{year}{2017}).

\bibitem{6ljubotina2019}
\bibinfo{author}{Ljubotina, M.}, \bibinfo{author}{{\v{Z}}nidari{\v{c}}, M.} \&
  \bibinfo{author}{Prosen, T.}
\newblock \bibinfo{title}{Kardar-parisi-zhang physics in the quantum heisenberg
  magnet}.
\newblock \emph{\bibinfo{journal}{Physical review letters}}
  \textbf{\bibinfo{volume}{122}}, \bibinfo{pages}{210602}
  (\bibinfo{year}{2019}).

\bibitem{7scheie2021}
\bibinfo{author}{Scheie, A.} \emph{et~al.}
\newblock \bibinfo{title}{Detection of kardar--parisi--zhang hydrodynamics in a
  quantum heisenberg spin-1/2 chain}.
\newblock \emph{\bibinfo{journal}{Nature Physics}}
  \textbf{\bibinfo{volume}{17}}, \bibinfo{pages}{726--730}
  (\bibinfo{year}{2021}).

\bibitem{8wei2021}
\bibinfo{author}{Wei, D.} \emph{et~al.}
\newblock \bibinfo{title}{Quantum gas microscopy of kardar-parisi-zhang
  superdiffusion}.
\newblock \emph{\bibinfo{journal}{arXiv preprint arXiv:2107.00038}}
  (\bibinfo{year}{2021}).

\bibitem{1kardar1986}
\bibinfo{author}{Kardar, M.}, \bibinfo{author}{Parisi, G.} \&
  \bibinfo{author}{Zhang, Y.-C.}
\newblock \bibinfo{title}{Dynamic scaling of growing interfaces}.
\newblock \emph{\bibinfo{journal}{Physical Review Letters}}
  \textbf{\bibinfo{volume}{56}}, \bibinfo{pages}{889} (\bibinfo{year}{1986}).

\bibitem{9altman2015}
\bibinfo{author}{Altman, E.}, \bibinfo{author}{Sieberer, L.~M.},
  \bibinfo{author}{Chen, L.}, \bibinfo{author}{Diehl, S.} \&
  \bibinfo{author}{Toner, J.}
\newblock \bibinfo{title}{Two-dimensional superfluidity of exciton polaritons
  requires strong anisotropy}.
\newblock \emph{\bibinfo{journal}{Physical Review X}}
  \textbf{\bibinfo{volume}{5}}, \bibinfo{pages}{011017} (\bibinfo{year}{2015}).

\bibitem{10ji2015}
\bibinfo{author}{Ji, K.}, \bibinfo{author}{Gladilin, V.~N.} \&
  \bibinfo{author}{Wouters, M.}
\newblock \bibinfo{title}{Temporal coherence of one-dimensional nonequilibrium
  quantum fluids}.
\newblock \emph{\bibinfo{journal}{Physical Review B}}
  \textbf{\bibinfo{volume}{91}}, \bibinfo{pages}{045301}
  (\bibinfo{year}{2015}).

\bibitem{11he2015}
\bibinfo{author}{He, L.}, \bibinfo{author}{Sieberer, L.~M.},
  \bibinfo{author}{Altman, E.} \& \bibinfo{author}{Diehl, S.}
\newblock \bibinfo{title}{Scaling properties of one-dimensional
  driven-dissipative condensates}.
\newblock \emph{\bibinfo{journal}{Physical Review B}}
  \textbf{\bibinfo{volume}{92}}, \bibinfo{pages}{155307}
  (\bibinfo{year}{2015}).

\bibitem{12zamora2017}
\bibinfo{author}{Zamora, A.}, \bibinfo{author}{Sieberer, L.},
  \bibinfo{author}{Dunnett, K.}, \bibinfo{author}{Diehl, S.} \&
  \bibinfo{author}{Szyma{\'n}ska, M.}
\newblock \bibinfo{title}{Tuning across universalities with a driven open
  condensate}.
\newblock \emph{\bibinfo{journal}{Physical Review X}}
  \textbf{\bibinfo{volume}{7}}, \bibinfo{pages}{041006} (\bibinfo{year}{2017}).

\bibitem{13comaron2018}
\bibinfo{author}{Comaron, P.} \emph{et~al.}
\newblock \bibinfo{title}{Dynamical critical exponents in driven-dissipative
  quantum systems}.
\newblock \emph{\bibinfo{journal}{Physical review letters}}
  \textbf{\bibinfo{volume}{121}}, \bibinfo{pages}{095302}
  (\bibinfo{year}{2018}).

\bibitem{14squizzato2018}
\bibinfo{author}{Squizzato, D.}, \bibinfo{author}{Canet, L.} \&
  \bibinfo{author}{Minguzzi, A.}
\newblock \bibinfo{title}{Kardar-parisi-zhang universality in the phase
  distributions of one-dimensional exciton-polaritons}.
\newblock \emph{\bibinfo{journal}{Physical Review B}}
  \textbf{\bibinfo{volume}{97}}, \bibinfo{pages}{195453}
  (\bibinfo{year}{2018}).

\bibitem{15amelio2020}
\bibinfo{author}{Amelio, I.} \& \bibinfo{author}{Carusotto, I.}
\newblock \bibinfo{title}{Theory of the coherence of topological lasers}.
\newblock \emph{\bibinfo{journal}{Physical Review X}}
  \textbf{\bibinfo{volume}{10}}, \bibinfo{pages}{041060}
  (\bibinfo{year}{2020}).

\bibitem{16ferrier2020}
\bibinfo{author}{Ferrier, A.}, \bibinfo{author}{Zamora, A.},
  \bibinfo{author}{Dagvadorj, G.} \& \bibinfo{author}{Szyma{\'n}ska, M.}
\newblock \bibinfo{title}{Searching for the kardar-parisi-zhang phase in
  microcavity polaritons}.
\newblock \emph{\bibinfo{journal}{arXiv preprint arXiv:2009.05177}}
  (\bibinfo{year}{2020}).

\bibitem{17deligiannis2021}
\bibinfo{author}{Deligiannis, K.}, \bibinfo{author}{Squizzato, D.},
  \bibinfo{author}{Minguzzi, A.} \& \bibinfo{author}{Canet, L.}
\newblock \bibinfo{title}{Accessing kardar-parisi-zhang universality
  sub-classes with exciton polaritons}.
\newblock \emph{\bibinfo{journal}{EPL (Europhysics Letters)}}
  \textbf{\bibinfo{volume}{132}}, \bibinfo{pages}{67004}
  (\bibinfo{year}{2021}).

\bibitem{18mei2021}
\bibinfo{author}{Mei, Q.}, \bibinfo{author}{Ji, K.} \&
  \bibinfo{author}{Wouters, M.}
\newblock \bibinfo{title}{Spatiotemporal scaling of two-dimensional
  nonequilibrium exciton-polariton systems with weak interactions}.
\newblock \emph{\bibinfo{journal}{Physical Review B}}
  \textbf{\bibinfo{volume}{103}}, \bibinfo{pages}{045302}
  (\bibinfo{year}{2021}).

\bibitem{38family1985}
\bibinfo{author}{Family, F.} \& \bibinfo{author}{Vicsek, T.}
\newblock \bibinfo{title}{Scaling of the active zone in the eden process on
  percolation networks and the ballistic deposition model}.
\newblock \emph{\bibinfo{journal}{Journal of Physics A: Mathematical and
  General}} \textbf{\bibinfo{volume}{18}}, \bibinfo{pages}{L75}
  (\bibinfo{year}{1985}).

\bibitem{2halpin1995}
\bibinfo{author}{Halpin-Healy, T.} \& \bibinfo{author}{Zhang, Y.-C.}
\newblock \bibinfo{title}{Kinetic roughening phenomena, stochastic growth,
  directed polymers and all that. aspects of multidisciplinary statistical
  mechanics}.
\newblock \emph{\bibinfo{journal}{Physics reports}}
  \textbf{\bibinfo{volume}{254}}, \bibinfo{pages}{215--414}
  (\bibinfo{year}{1995}).

\bibitem{3krug1997}
\bibinfo{author}{Krug, J.}
\newblock \bibinfo{title}{Origins of scale invariance in growth processes}.
\newblock \emph{\bibinfo{journal}{Advances in Physics}}
  \textbf{\bibinfo{volume}{46}}, \bibinfo{pages}{139--282}
  (\bibinfo{year}{1997}).

\bibitem{4takeuchi2018}
\bibinfo{author}{Takeuchi, K.~A.}
\newblock \bibinfo{title}{An appetizer to modern developments on the
  kardar--parisi--zhang universality class}.
\newblock \emph{\bibinfo{journal}{Physica A: Statistical Mechanics and its
  Applications}} \textbf{\bibinfo{volume}{504}}, \bibinfo{pages}{77--105}
  (\bibinfo{year}{2018}).

\bibitem{19lauter2017}
\bibinfo{author}{Lauter, R.}, \bibinfo{author}{Mitra, A.} \&
  \bibinfo{author}{Marquardt, F.}
\newblock \bibinfo{title}{From kardar-parisi-zhang scaling to explosive
  desynchronization in arrays of limit-cycle oscillators}.
\newblock \emph{\bibinfo{journal}{Physical Review E}}
  \textbf{\bibinfo{volume}{96}}, \bibinfo{pages}{012220}
  (\bibinfo{year}{2017}).

\bibitem{20chen2013}
\bibinfo{author}{Chen, L.}, \bibinfo{author}{Toner, J.} \emph{et~al.}
\newblock \bibinfo{title}{Universality for moving stripes: A hydrodynamic
  theory of polar active smectics}.
\newblock \emph{\bibinfo{journal}{Physical review letters}}
  \textbf{\bibinfo{volume}{111}}, \bibinfo{pages}{088701}
  (\bibinfo{year}{2013}).

\bibitem{21he2017}
\bibinfo{author}{He, L.}, \bibinfo{author}{Sieberer, L.~M.} \&
  \bibinfo{author}{Diehl, S.}
\newblock \bibinfo{title}{Space-time vortex driven crossover and vortex
  turbulence phase transition in one-dimensional driven open condensates}.
\newblock \emph{\bibinfo{journal}{Physical review letters}}
  \textbf{\bibinfo{volume}{118}}, \bibinfo{pages}{085301}
  (\bibinfo{year}{2017}).

\bibitem{22weisbuch1992}
\bibinfo{author}{Weisbuch, C.}, \bibinfo{author}{Nishioka, M.},
  \bibinfo{author}{Ishikawa, A.} \& \bibinfo{author}{Arakawa, Y.}
\newblock \bibinfo{title}{Observation of the coupled exciton-photon mode
  splitting in a semiconductor quantum microcavity}.
\newblock \emph{\bibinfo{journal}{Physical Review Letters}}
  \textbf{\bibinfo{volume}{69}}, \bibinfo{pages}{3314} (\bibinfo{year}{1992}).

\bibitem{23carusotto2013}
\bibinfo{author}{Carusotto, I.} \& \bibinfo{author}{Ciuti, C.}
\newblock \bibinfo{title}{Quantum fluids of light}.
\newblock \emph{\bibinfo{journal}{Reviews of Modern Physics}}
  \textbf{\bibinfo{volume}{85}}, \bibinfo{pages}{299} (\bibinfo{year}{2013}).

\bibitem{26schneider2016}
\bibinfo{author}{Schneider, C.} \emph{et~al.}
\newblock \bibinfo{title}{Exciton-polariton trapping and potential landscape
  engineering}.
\newblock \emph{\bibinfo{journal}{Reports on Progress in Physics}}
  \textbf{\bibinfo{volume}{80}}, \bibinfo{pages}{016503}
  (\bibinfo{year}{2016}).

\bibitem{24deng2002}
\bibinfo{author}{Deng, H.}, \bibinfo{author}{Weihs, G.},
  \bibinfo{author}{Santori, C.}, \bibinfo{author}{Bloch, J.} \&
  \bibinfo{author}{Yamamoto, Y.}
\newblock \bibinfo{title}{Condensation of semiconductor microcavity exciton
  polaritons}.
\newblock \emph{\bibinfo{journal}{Science}} \textbf{\bibinfo{volume}{298}},
  \bibinfo{pages}{199--202} (\bibinfo{year}{2002}).

\bibitem{25kasprzak2006}
\bibinfo{author}{Kasprzak, J.} \emph{et~al.}
\newblock \bibinfo{title}{Bose--einstein condensation of exciton polaritons}.
\newblock \emph{\bibinfo{journal}{Nature}} \textbf{\bibinfo{volume}{443}},
  \bibinfo{pages}{409--414} (\bibinfo{year}{2006}).

\bibitem{28love2008}
\bibinfo{author}{Love, A.} \emph{et~al.}
\newblock \bibinfo{title}{Intrinsic decoherence mechanisms in the microcavity
  polariton condensate}.
\newblock \emph{\bibinfo{journal}{Physical Review Letters}}
  \textbf{\bibinfo{volume}{101}}, \bibinfo{pages}{067404}
  (\bibinfo{year}{2008}).

\bibitem{27grinstein1993}
\bibinfo{author}{Grinstein, G.}, \bibinfo{author}{Mukamel, D.},
  \bibinfo{author}{Seidin, R.} \& \bibinfo{author}{Bennett, C.~H.}
\newblock \bibinfo{title}{Temporally periodic phases and kinetic roughening}.
\newblock \emph{\bibinfo{journal}{Physical review letters}}
  \textbf{\bibinfo{volume}{70}}, \bibinfo{pages}{3607} (\bibinfo{year}{1993}).

\bibitem{29SupMat}
\bibinfo{title}{See supplementary materials} .

\bibitem{30roumpos2012}
\bibinfo{author}{Roumpos, G.} \emph{et~al.}
\newblock \bibinfo{title}{Power-law decay of the spatial correlation function
  in exciton-polariton condensates}.
\newblock \emph{\bibinfo{journal}{Proceedings of the National Academy of
  Sciences}} \textbf{\bibinfo{volume}{109}}, \bibinfo{pages}{6467--6472}
  (\bibinfo{year}{2012}).

\bibitem{31fischer2014}
\bibinfo{author}{Fischer, J.} \emph{et~al.}
\newblock \bibinfo{title}{Spatial coherence properties of one dimensional
  exciton-polariton condensates}.
\newblock \emph{\bibinfo{journal}{Physical review letters}}
  \textbf{\bibinfo{volume}{113}}, \bibinfo{pages}{203902}
  (\bibinfo{year}{2014}).

\bibitem{32bobrovska2014}
\bibinfo{author}{Bobrovska, N.}, \bibinfo{author}{Ostrovskaya, E.~A.} \&
  \bibinfo{author}{Matuszewski, M.}
\newblock \bibinfo{title}{Stability and spatial coherence of nonresonantly
  pumped exciton-polariton condensates}.
\newblock \emph{\bibinfo{journal}{Physical Review B}}
  \textbf{\bibinfo{volume}{90}}, \bibinfo{pages}{205304}
  (\bibinfo{year}{2014}).

\bibitem{33daskalakis2015}
\bibinfo{author}{Daskalakis, K.~S.}, \bibinfo{author}{Maier, S.~A.} \&
  \bibinfo{author}{K{\'e}na-Cohen, S.}
\newblock \bibinfo{title}{Spatial coherence and stability in a disordered
  organic polariton condensate}.
\newblock \emph{\bibinfo{journal}{Physical review letters}}
  \textbf{\bibinfo{volume}{115}}, \bibinfo{pages}{035301}
  (\bibinfo{year}{2015}).

\bibitem{estrecho2018}
\bibinfo{author}{Estrecho, E.} \emph{et~al.}
\newblock \bibinfo{title}{Single-shot condensation of exciton polaritons and
  the hole burning effect}.
\newblock \emph{\bibinfo{journal}{Nature communications}}
  \textbf{\bibinfo{volume}{9}}, \bibinfo{pages}{1--9} (\bibinfo{year}{2018}).

\bibitem{bobrovska2018}
\bibinfo{author}{Bobrovska, N.}, \bibinfo{author}{Matuszewski, M.},
  \bibinfo{author}{Daskalakis, K.~S.}, \bibinfo{author}{Maier, S.~A.} \&
  \bibinfo{author}{K{\'e}na-Cohen, S.}
\newblock \bibinfo{title}{Dynamical instability of a nonequilibrium
  exciton-polariton condensate}.
\newblock \emph{\bibinfo{journal}{ACS Photonics}} \textbf{\bibinfo{volume}{5}},
  \bibinfo{pages}{111--118} (\bibinfo{year}{2018}).

\bibitem{34smirnov2014}
\bibinfo{author}{Smirnov, L.~A.}, \bibinfo{author}{Smirnova, D.~A.},
  \bibinfo{author}{Ostrovskaya, E.~A.} \& \bibinfo{author}{Kivshar, Y.~S.}
\newblock \bibinfo{title}{Dynamics and stability of dark solitons in
  exciton-polariton condensates}.
\newblock \emph{\bibinfo{journal}{Physical Review B}}
  \textbf{\bibinfo{volume}{89}}, \bibinfo{pages}{235310}
  (\bibinfo{year}{2014}).

\bibitem{35liew2015}
\bibinfo{author}{Liew, T. C.~H.} \emph{et~al.}
\newblock \bibinfo{title}{Instability-induced formation and nonequilibrium
  dynamics of phase defects in polariton condensates}.
\newblock \emph{\bibinfo{journal}{Physical Review B}}
  \textbf{\bibinfo{volume}{91}}, \bibinfo{pages}{085413}
  (\bibinfo{year}{2015}).

\bibitem{36caputo2018}
\bibinfo{author}{Caputo, D.} \emph{et~al.}
\newblock \bibinfo{title}{Topological order and thermal equilibrium in
  polariton condensates}.
\newblock \emph{\bibinfo{journal}{Nature materials}}
  \textbf{\bibinfo{volume}{17}}, \bibinfo{pages}{145--151}
  (\bibinfo{year}{2018}).

\bibitem{37baboux2018}
\bibinfo{author}{Baboux, F.} \emph{et~al.}
\newblock \bibinfo{title}{Unstable and stable regimes of polariton
  condensation}.
\newblock \emph{\bibinfo{journal}{Optica}} \textbf{\bibinfo{volume}{5}},
  \bibinfo{pages}{1163--1170} (\bibinfo{year}{2018}).

\bibitem{39edwards1982}
\bibinfo{author}{Edwards, S.~F.} \& \bibinfo{author}{Wilkinson, D.}
\newblock \bibinfo{title}{The surface statistics of a granular aggregate}.
\newblock \emph{\bibinfo{journal}{Proceedings of the Royal Society of London.
  A. Mathematical and Physical Sciences}} \textbf{\bibinfo{volume}{381}},
  \bibinfo{pages}{17--31} (\bibinfo{year}{1982}).

\bibitem{40wouters2006}
\bibinfo{author}{Wouters, M.} \& \bibinfo{author}{Carusotto, I.}
\newblock \bibinfo{title}{Absence of long-range coherence in the parametric
  emission of photonic wires}.
\newblock \emph{\bibinfo{journal}{Physical Review B}}
  \textbf{\bibinfo{volume}{74}}, \bibinfo{pages}{245316}
  (\bibinfo{year}{2006}).

\bibitem{41chiocchetta2013}
\bibinfo{author}{Chiocchetta, A.} \& \bibinfo{author}{Carusotto, I.}
\newblock \bibinfo{title}{Non-equilibrium quasi-condensates in reduced
  dimensions}.
\newblock \emph{\bibinfo{journal}{EPL (Europhysics Letters)}}
  \textbf{\bibinfo{volume}{102}}, \bibinfo{pages}{67007}
  (\bibinfo{year}{2013}).

\bibitem{42prahofer2004}
\bibinfo{author}{Pr{\"a}hofer, M.} \& \bibinfo{author}{Spohn, H.}
\newblock \bibinfo{title}{Exact scaling functions for one-dimensional
  stationary kpz growth}.
\newblock \emph{\bibinfo{journal}{Journal of statistical physics}}
  \textbf{\bibinfo{volume}{115}}, \bibinfo{pages}{255--279}
  (\bibinfo{year}{2004}).

\bibitem{43spohn2014}
\bibinfo{author}{Spohn, H.}
\newblock \bibinfo{title}{Nonlinear fluctuating hydrodynamics for anharmonic
  chains}.
\newblock \emph{\bibinfo{journal}{Journal of Statistical Physics}}
  \textbf{\bibinfo{volume}{154}}, \bibinfo{pages}{1191--1227}
  (\bibinfo{year}{2014}).

\bibitem{47porras2002}
\bibinfo{author}{Porras, D.}, \bibinfo{author}{Ciuti, C.},
  \bibinfo{author}{Baumberg, J.} \& \bibinfo{author}{Tejedor, C.}
\newblock \bibinfo{title}{Polariton dynamics and bose-einstein condensation in
  semiconductor microcavities}.
\newblock \emph{\bibinfo{journal}{Physical Review B}}
  \textbf{\bibinfo{volume}{66}}, \bibinfo{pages}{085304}
  (\bibinfo{year}{2002}).

\bibitem{48wouters2007}
\bibinfo{author}{Wouters, M.} \& \bibinfo{author}{Carusotto, I.}
\newblock \bibinfo{title}{Excitations in a nonequilibrium bose-einstein
  condensate of exciton polaritons}.
\newblock \emph{\bibinfo{journal}{Physical review letters}}
  \textbf{\bibinfo{volume}{99}}, \bibinfo{pages}{140402}
  (\bibinfo{year}{2007}).

\bibitem{49loirette2021}
\bibinfo{author}{Loirette-Pelous, A.}, \bibinfo{author}{Amelio, I.},
  \bibinfo{author}{Secl\`{\i}, M.} \& \bibinfo{author}{Carusotto, I.}
\newblock \bibinfo{title}{Linearized theory of the fluctuation dynamics in
  two-dimensional topological lasers}.
\newblock \emph{\bibinfo{journal}{Phys. Rev. A}}
  \textbf{\bibinfo{volume}{104}}, \bibinfo{pages}{053516}
  (\bibinfo{year}{2021}).

\bibitem{wouters2009}
\bibinfo{author}{Wouters, M.} \& \bibinfo{author}{Savona, V.}
\newblock \bibinfo{title}{Stochastic classical field model for polariton
  condensates}.
\newblock \emph{\bibinfo{journal}{Phys. Rev. B}} \textbf{\bibinfo{volume}{79}},
  \bibinfo{pages}{165302} (\bibinfo{year}{2009}).

\bibitem{53gladilin2014}
\bibinfo{author}{Gladilin, V.~N.}, \bibinfo{author}{Ji, K.} \&
  \bibinfo{author}{Wouters, M.}
\newblock \bibinfo{title}{Spatial coherence of weakly interacting
  one-dimensional nonequilibrium bosonic quantum fluids}.
\newblock \emph{\bibinfo{journal}{Physical Review A}}
  \textbf{\bibinfo{volume}{90}}, \bibinfo{pages}{023615}
  (\bibinfo{year}{2014}).

\bibitem{45kuhlmann2013}
\bibinfo{author}{Kuhlmann, A.~V.} \emph{et~al.}
\newblock \bibinfo{title}{Charge noise and spin noise in a semiconductor
  quantum device}.
\newblock \emph{\bibinfo{journal}{Nature Physics}}
  \textbf{\bibinfo{volume}{9}}, \bibinfo{pages}{570--575}
  (\bibinfo{year}{2013}).

\bibitem{46olivero1977}
\bibinfo{author}{Olivero, J.~J.} \& \bibinfo{author}{Longbothum, R.}
\newblock \bibinfo{title}{Empirical fits to the voigt line width: A brief
  review}.
\newblock \emph{\bibinfo{journal}{Journal of Quantitative Spectroscopy and
  Radiative Transfer}} \textbf{\bibinfo{volume}{17}}, \bibinfo{pages}{233--236}
  (\bibinfo{year}{1977}).

\bibitem{54werner1997}
\bibinfo{author}{Werner, M.} \& \bibinfo{author}{Drummond, P.}
\newblock \bibinfo{title}{Robust algorithms for solving stochastic partial
  differential equations}.
\newblock \emph{\bibinfo{journal}{Journal of computational physics}}
  \textbf{\bibinfo{volume}{132}}, \bibinfo{pages}{312--326}
  (\bibinfo{year}{1997}).

\bibitem{55dennis2013}
\bibinfo{author}{Dennis, G.~R.}, \bibinfo{author}{Hope, J.~J.} \&
  \bibinfo{author}{Johnsson, M.~T.}
\newblock \bibinfo{title}{Xmds2: Fast, scalable simulation of coupled
  stochastic partial differential equations}.
\newblock \emph{\bibinfo{journal}{Computer Physics Communications}}
  \textbf{\bibinfo{volume}{184}}, \bibinfo{pages}{201--208}
  (\bibinfo{year}{2013}).

\bibitem{Note1}
\bibinfo{note}{Note that for flat initial conditions, $v_{\infty }$ is equal to
  the microscopic KPZ non-linearity $\lambda $}.

\end{thebibliography}
\let\addcontentsline\oldaddcontentsline% Restore \addcontentsline

\noindent \textbf{Acknowledgments.} We thank Valentin Goblot, Daniel Vajner and Ateeb Toor for their assistance in the early development of the experiment.

\noindent \textbf{Funding.} This work was supported by the Paris Ile-de-France Région in the framework of DIM SIRTEQ, the French RENATECH network, the H2020-FETFLAG project PhoQus (820392), the QUANTERA project Interpol (ANR-QUAN-0003-05), the European Research Council via the project ARQADIA (949730), EmergenTopo (865151) and RG.BIO(785932), the French government through the Programme Investissement d’Avenir (I-SITE ULNE / ANR-16-IDEX-0004 ULNE) managed by the Agence Nationale de la Recherche, the Labex CEMPI (ANR-11-LABX-0007). L.C. acknowledges support from ANR (grant ANR-18-CE92-0019) and from Institut Universitaire de France.

\noindent \textbf{Author contributions.} Q.F. built the experimental setup, performed the experiments and analyzed the data. D.S. realized the theoretical calculations and numerical simulations. F.B. contributed to the design of the sample structure and initial characterization of the sample. A.L., M.M. grew the sample by molecular beam epitaxy. I.S, L.L.G., and A.H. fabricated the polariton lattices. Q.F., D.S., I.A., M.W., I.C., A.A., M.R., A.M., L.C., S.R., and J.B. participated to the scientific discussions about all aspects of the work. Q.F., A.M., L.C., S.R. and J.B. wrote the original draft of the paper. Q.F., D.S., I.A., M.W., I.C., A.A., M.R., A.M., L.C., S.R., and J.B. reviewed and edited the paper into its current form. A.M., L.C., S.R. and J.B. supervised the work.

\newpage

%%%%%%%%%%%%%%%%%%%%%%%%%%%%%%%%%%%%%%%%%%%%%%%%%%%%%%%%%%%%%%%%%%%%%%%%%%%%%%%%%%%%%%%%%%%%%%%%%%%%%%%%%%%%%%%%%%
%%%%%%%%%%%%%%%%%%%%%%%%%%%%%%%%%%%%%%%%%%%%%%%%%%%%%%%%%%%%%%%%%%%%%%%%%%%%%%%%%%%%%%%%%%%%%%%%%%%%%%%%%%%%%%%%%%
%%%%%%%%%%%%%%%%%%%%%%%%%%%%%%%%%%%%%%%%%%%%%%%%%%%%%%%%%%%%%%%%%%%%%%%%%%%%%%%%%%%%%%%%%%%%%%%%%%%%%%%%%%%%%%%%%%

\begin{center}
{\huge Supplementary Material}
\end{center}

\tableofcontents

\newpage

\section{Overview}

In this Supplemental Material, we provide  additional information on the experiments and on the numerical simulations, as well as  additional discussion of the results.
In Sec.~\ref{sec:theory}, we analytically derive the mapping from our two-coupled equation model for the dynamics of the condensate field and reservoir density to the KPZ equation for the phase dynamics. We precisely relate the $g^{(1)}$ first-order correlation function to the phase-phase correlations. 
In Sec.~\ref{sec:experiments}, we provide all the information on the experimental set-up and measurements, and we report complementary experimental results obtained on a symmetric Lieb lattice. 
In Sec.~\ref{sec:numerical-results} we perform an in-depth analysis of the phase dynamics and of the effect of space-time vortices. 

Beyond all the necessary discussion, let us emphasize below the main results reported in this material:
\begin{itemize}
    \item We establish the mapping to the KPZ equation for a more general and realistic model than previous studies in (Sec.~\ref{sec:model}).
    \item We consolidate the validity of our experimental findings by reproducing them in a different lattice featuring  condensation in a different type of bands (Sec.~\ref{sec:AddData}).
    \item We demonstrate that the measured scaling behavior of $g^{(1)}$ directly reflects the KPZ scaling of the phase (Sec.~\ref{sec:g1-contrib}),
    \item We analyze the effects of space-time vortices, and explain why KPZ dynamics can be resilient to their presence (Sec.~\ref{sec:distribution}).
\end{itemize}

\newpage

\section{The theoretical model: emergence of KPZ dynamics in incoherently pumped polaritons}
\label{sec:theory}

\subsection{The driven-dissipative Gross-Pitaevskii equation under incoherent pumping}\label{sec:model}

\vspace{-4pt}

We consider an out-of-equilibrium polariton condensate created in a one-dimensional lattice. Since the relevant dynamics occurs at low energy, we restrict the description to an effective single-band model, neglecting the contribution of the other lattice bands. We describe the polariton condensate wavefunction by the classical field  $\psi(x,t)$ at position $x$ and time $t$.
The excitation of the polariton condensate is modeled by introducing an external pump $P(x)$ filling an incoherent excitonic reservoir of density $n_R(x,t)$. The reservoir excitons either relax into the polariton condensate by stimulated scattering with rate $R$ or decay via other channels with total rate $\gamma_R$~\cite{47porras2002}.  We describe the polariton-polariton and exciton-polariton interactions as contact interactions of strength $g$ and $g_R$ respectively.
The coupled equations for the condensate and reservoir read~\cite{48wouters2007}:
\begin{equation}
\label{eq:cond+res}
\left\{
\begin{aligned}
&i\hbar\partial_t\psi=\left[ 
E (\hat k)
- \frac{i\hbar}{2}  \gamma(\hat k) 
+g|\psi|^2+2g_Rn_R+\frac{i\hbar}{2} R\,n_R  \right]\psi +  \hbar \xi \\
&\partial_t n_R=P-(\gamma_R+R|\psi|^2)n_R
\end{aligned}
\right. \,,
\end{equation}
where the first equation is a generalized, stochastic  Gross-Pitaevskii equation (gGPE) for the condensate and the second one is a rate equation for the reservoir.
In Eq.~(\ref{eq:cond+res}) $\hat k = -i\hbar\partial/\partial x$ is the momentum operator and $\xi$ is a white noise with correlations $\langle \xi(x, t) \xi^* (x', t') \rangle =2 \xi_0 \delta(x-x')\delta(t-t')$. 
In the vicinity of $k=0$, we approximate the lattice dispersion by a parabola  $E(k) = E_{0} + \hbar^2 k^2/2m$. 
The polariton linewidth $\gamma(k)$ is taken as momentum dependent, consistently with the experimental observations (see Sec.~\ref{sec:gamma2} below) and is well approximated by  $\gamma(k \simeq 0) \simeq \gamma_0 + \gamma_2 k^2$ near $k=0$.
The momentum-dependent linewidth plays a very important role in our model as it ensures the stability of the polariton condensate in our simulations~\cite{37baboux2018}. 
Note that it also has a crucial impact on the edge dynamics of topological lasers, where $k$-dependent losses naturally occur from the $k$-dependent confinement of the edge mode~\cite{15amelio2020, 49loirette2021}. 
For the variance of the noise used in Eq.~\eqref{eq:cond+res}, we take the value $\xi_0=\frac{R}{2}\,n_R\,,$ that represents quantum noise due to pumping within the truncated Wigner picture. 
In this way, the correlators of the quantum field can be extracted from the noise-averaged spatio-temporal correlators of $\psi(x,t)$~\cite{wouters2009, 23carusotto2013}.

\subsection{Mapping to the KPZ equation}
\label{sec:mapping}

In the following, we use the density-phase representation of $\psi(x,t)$ within the rotating frame of the condensate: $\psi(x, t)=\sqrt{\rho(x,t)} \, \mathrm{exp} \big[ i(\theta(x,t)-\overline{\omega_{0}} t) \big]$, where $\overline{\omega_{0}} =\frac{g_R}{\hbar R} \gamma_0 \big[ 2 +\frac{g}{g_R}\frac{\gamma_R}{\gamma_0}(p-1) \big]$ is the condensate emission frequency. 
We focus on the dynamics of the system at small momentum and frequency. For the sake of simplicity,
we therefore perform our analysis using the parabolic approximations of both $E(k)$ and $\gamma(k)$, introduced in section~\ref{sec:model}. 
In terms of the phase and density fields, the Laplacian and time-derivative operators read:
\begin{align}\label{eq:d1op}
\partial_t \psi=&\psi\left( \frac{1}{2}\rho^{-1}\partial_t\rho+i\partial_t\theta -i\overline{\omega_{0}} \right) \quad \,,
\end{align}
\begin{align}\label{eq:d2op}
\nabla^2\psi(x, t)=& \; \psi(x, t)\Big( -\frac{1}{4}\rho(x, t)^{-2} (\grad\rho(x, t))^2+\frac{1}{2}\rho(x, t)^{-1}\nabla^2\rho(x, t)-(\grad\theta(x, t))^2 \nonumber \\ 
& \hspace{1.3cm} + i\rho(x, t)^{-1}\grad\rho(x, t)\cdot\grad\theta(x, t)+i\nabla^2\theta(x,t) \Big) \nonumber\\
\equiv& \; \psi(x,t) \, \mathcal{D}[\rho,\theta ]\,,
\end{align}
where in the last line we have introduced the differential operator $\mathcal{D}[\rho,\theta ]$ for simplicity in the notation.
Equations~\eqref{eq:cond+res} turn into a set of three coupled equations for the real-valued fields $\theta(x, t)$, $\rho(x, t)$ and $n_R(x, t)$,
\begin{equation}\label{eq:phdensressys}
\left\{
\begin{aligned}
& \partial_t\theta=
\overline{\omega_{0}}  +
\frac{\hbar}{2m}\Re\{\mathcal{D}[\rho,\theta]\}+\frac{\gamma_2}{2}\Im\{\mathcal{D}[\rho,\theta]\}-\frac{g}{\hbar}\rho-2\frac{g_R}{\hbar}n_R+\Re\{\bar\xi\}\\
&\frac{1}{2\rho}\partial_t\rho=-\frac{\hbar}{2m}\Im\{\mathcal{D}[\rho,\theta]\}+\frac{\gamma_2}{2}\Re\{\mathcal{D}[\rho,\theta]\}+\frac{1}{2}\Big(R\,n_R-\gamma_0\Big)+\Im\{\bar\xi\}\\
&\partial_t n_R=P-\left(\gamma_R+R\,\rho\right)n_R
\end{aligned}
\right.
\end{equation}
where $\Re\{\cdot\}$, $\Im\{\cdot\}$ stand for the real- and imaginary-part and $\bar\xi=-i e^{-i\theta + i\overline{\omega_{0}} t}\rho^{-1/2}\xi$.

The existence of a Goldstone mode entails that phase fluctuations dominate in the long time, large-distance regime.  Conversely, the polariton and reservoir densities are subject to a restoring force that makes them relax, within a short timescale, to their stationary values $\rho_0$,  $n_{R,0}$.
It is thus convenient to perform the following decomposition: $\delta\rho = \rho(x, t)-\rho_0$ and $\delta n_R=n_R(x,t)-n_{R,0}$. 
We now assume that $\delta\rho/\rho_0 \ll 1$ and $\delta n_R/n_{R,0} \ll 1$, and that these fluctuations are stationary,
i.e. we neglect  $\partial_t \delta\rho$ and  $\partial_t \delta n_R$. This is justified when the time scales of the reservoir and condensate density fluctuations are well separated from the ones of the phase fluctuations. This is similar in spirit to the usual decoupling approximation \cite{9altman2015}, but on the two equations for the densities.
We also neglect the spatial dependence of the condensate density fluctuations, which leads to the following set of equations, describing the effective dynamics of the system:
\begin{equation}
\left\{
\begin{aligned}
& \partial_t\theta=-\frac{\hbar}{2m}(\grad\theta)^2+\frac{\gamma_2}{2}\nabla^2\theta-\frac{g}{\hbar}\delta\rho-2\frac{g_R}{\hbar}\delta n_R+\Re\{\bar\xi\}\\
&\delta n_R=\frac{2}{R} \left[ \frac{\hbar}{2m}\nabla^2\theta +\frac{\gamma_2}{2}(\grad\theta)^2-\Im\{\bar\xi\} \right]\\
&\delta\rho=-p\frac{\gamma_R}{\gamma_0}\delta n_R  \,.
\end{aligned} \,
\right.
\end{equation}
After a simple substitution, the equation governing the phase evolution becomes:
\begin{align}
\partial_t\theta&=\left[\frac{\gamma_2}{2}- u \frac{g_R}{\hbar R} \frac{\hbar}{m} \right]\nabla^2\theta-\left[\frac{\hbar}{2m}+ u \frac{g_R}{\hbar R} \gamma_2 \right](\grad\theta)^2+\eta\nonumber\\
&\equiv\nu \nabla^2\theta+\frac{\lambda}{2}(\grad\theta)^2+\eta
\label{eq:bareKPZ}
\end{align}
with
\begin{align}
\langle\eta(x, t)\eta(x', t') \rangle&=\frac{\xi_0}{\rho_0}\left[1+4 \left( u  \frac{ g_R}{\hbar R}\right)^2 \right]\delta(\xx-\xx')\delta(t-t')\nonumber\\
&\equiv 2D\delta(\xx-\xx')\delta(t-t')\, \label{eq:noiseKPZ},
\end{align}
and
\begin{equation}\label{eq:u}
    u=2-p\frac{g}{ g_R}\frac{\gamma_R}{\gamma_0} \,.
\end{equation}
Equation~\eqref{eq:bareKPZ} is the KPZ equation for the phase. 
The terms neglected during the derivation of this equation may slightly renormalize the KPZ parameters, but they do not drive the phase dynamics out of the KPZ universality class. The numerical results (see Sec.~\ref{sec:numerical-results} below) show that their effect is negligible in our case.

Equations~\eqref{eq:bareKPZ} and \eqref{eq:noiseKPZ} allow us to obtain the expression of the KPZ parameters $\nu$, $\lambda$ and $D$ in terms of the microscopic parameters entering Eq.~\eqref{eq:cond+res}: 
\begin{align}\label{eq:kpzcoeffs}
\nu=\frac{\gamma_2}{2}-& u  \frac{g_R}{\hbar R}  \frac{\hbar}{ m}\,,\quad 
\lambda=-2\left[\frac{\hbar}{2m }+ u  \frac{ g_R}{\hbar R} \gamma_2 \right]\,,\quad 
D=\frac{ R \,n_{R,0}}{4\rho_0}\left[1+4 \left( u  \frac{ g_R}{\hbar R}\right)^2  \right]\,,
\end{align}
with $u$ given by Eq.~\eqref{eq:u}.
Note that the effective diffusivity $\nu$ must be positive in order for the linear limit of the KPZ equation, \textit{i.e.} the Edwards-Wilkinson equation (EW), to be stable. Therefore, the expression of $\nu$ provides important insights into the role played by the non-trivial $k$-dependence of the linewidth, expressed by the parameter 
$\gamma_2$, and by the sign of the mass.
For the parameters used in our simulations, $u$ is positive (see section~\ref{sec:numerical-simulations}). 
The negative sign of the polariton mass is thus crucial in stabilizing the system~\cite{37baboux2018}.  In the case of a positive mass, a non-zero value of $\gamma_2$ is absolutely necessary in order for the phase to be stable: a vanishing value of $\gamma_2$ would yield a negative value for $\nu$, hence an instability of the KPZ equation. 

Finally, we would like to stress that our derivation  of the mapping 
between the generalized Gross-Pitaevskii equation for the polariton condensate and the KPZ equation for the phase is more general and realistic than those found in previous studies~\cite{9altman2015, 10ji2015, 11he2015, 12zamora2017, 13comaron2018, 14squizzato2018, 16ferrier2020, 17deligiannis2021, 18mei2021, 53gladilin2014}: 
we formulate the decoupling approximation for the three-equation system (\ref{eq:phdensressys}) and we include in the gGPE the reservoir-induced blue-shift term $2 g_{R} n_R$.
Both aspects are crucial to provide a faithful description of the experiment.

\subsection{Connection between the condensate first-order correlation and the two-point phase-phase correlations}
\label{sec:g1-var}

In this section, we detail the link between the condensate first-order correlation function $g^{(1)}$, which is measured experimentally, and the phase-phase correlation function, which displays 
universal spatio-temporal KPZ scaling. 
In particular, we derive the conditions required in order to ensure that the scaling behavior of $g^{(1)}$ reflects the underlying KPZ dynamics of the condensate phase.

The general definition of the first-order correlation reads:
\begin{align}
g^{(1)}(\Delta x, \Delta t)&=\frac{\langle \psi^*(x, t_{0})\psi(-x, t_0+\Delta t) \rangle}{\sqrt{\avg{\rho(x, t_{0})}}\sqrt{\avg{\rho(-x, t_{0} + \Delta t)}}} \nonumber\\
&= \frac{\avg{\sqrt{\rho(x, t_{0})\rho(-x, t_{0} + \Delta t)}e^{i\Delta\theta(\Delta x, \Delta t)}}}{\sqrt{\avg{\rho(x, t_0)}}\sqrt{\avg{\rho(-x, t_0+ \Delta t)}}}\,.
\label{eq:g1-def}
\end{align}
In the left-hand side, we omitted the dependence on $t_0$ due to the stationarity of the condensate dynamics. 
In our work, we are interested in accessing the phase dynamics from the field-field correlator. 
If we assume that the dynamics of the phase is decoupled from the one of the density, we get:
\begin{equation}
\label{eq:phase-density}
g^{(1)}(\Delta x, \Delta t)= \frac{\avg{\sqrt{\rho(x, t_0)\rho(-x, t_0+\Delta t)}}}{\sqrt{\avg{\rho(x, t_0)}}\sqrt{\avg{\rho(-x, t_0 + \Delta t)}}}\avg{e^{i\Delta\theta(\Delta x,\Delta t)}}\,.
\end{equation}
We then decompose the density field into a mean-field and a fluctuating contribution, $\rho(x, t)=\rho_0+\delta\rho(x, t)$, and assume that $\delta\rho(x, t)/\rho_0 \ll 1$. 
We expand both the numerator and the denominator in the right-hand side of Eq.~\eqref{eq:phase-density}, which become, to linear order in $\delta\rho/\rho_0$:
\begin{align}
\avg{\sqrt{\rho(x, t_0)\rho(-x, t_0+\Delta t)}}&\simeq \rho_0\avg{\left( 1+\frac{1}{2}\frac{\delta\rho(x, t_0)}{\rho_0} \right)\left( 1+\frac{1}{2}\frac{\delta\rho(-x, t_0+\Delta t)}{\rho_0} \right)}\nonumber\\ &\simeq \rho_0+\frac{1}{2}\left(\avg{ \delta\rho(x, t_0)}+\avg{\delta\rho(-x, t_0+\Delta t)}\right)\,,
\end{align}
and
\begin{align}
\sqrt{\avg{\rho(x, t_0)}\avg{\rho(-x, t_0+\Delta t)}}&\simeq \rho_0 \left( 1+\frac{1}{2}\frac{\avg{\delta\rho(x, t_0)}}{\rho_0} \right)\left( 1+\frac{1}{2}\frac{\avg{\delta\rho(-x, t_0+\Delta t)}}{\rho_0} \right)\nonumber\\ &\simeq \rho_0+\frac{1}{2}\left(\avg{ \delta\rho(x, t_0)}+\avg{\delta\rho(-x, t_0+\Delta t)}\right)\,.
\end{align}
In this limit, the density terms in Eq.~\eqref{eq:phase-density} simplify and one hence gets
\begin{equation}
\label{eq:g1-exp}
g^{(1)}(\Delta x, \Delta t) = \Big\langle\exp\left[ i\Delta\theta(\Delta x, \Delta t) \right]\Big\rangle.
\end{equation}
Furthermore, for small fluctuations of the phase, we can use the cumulant expansion to get:
\begin{equation}\label{eq:g1cumuls}
|g^{(1)}(\Delta x, \Delta t)|^2 \simeq \exp(-\langle \Delta\theta(\Delta x, \Delta t)^2 \rangle + \langle \Delta\theta(\Delta x, \Delta t) \rangle^2)\equiv \exp(-\textrm{Var} \left[ \Delta \theta(\Delta x, \Delta t) \right])\,,
\end{equation}
and hence
\begin{equation}\label{eq:g1delta}
-2 \log( |g^{(1)}(\Delta x, \Delta t)| ) \simeq \textrm{Var} \left[ \Delta \theta(\Delta x, \Delta t) \right].
\end{equation}
It is instructive to study the validity of approximation \eqref{eq:g1cumuls}. At $\mathcal{O}(\Delta\theta^4)$ we have

\begin{align}
|g^{(1)}(\Delta x, \Delta t)|^2\;=\;&1- \langle \Delta\theta(\Delta x, \Delta t)^2 \rangle + \langle \Delta\theta(\Delta x, \Delta t) \rangle^2 +\frac{1}{12}\langle \Delta\theta(\Delta x, \Delta t)^4 \rangle \nonumber \\
&\hspace{0.2cm}-\frac{1}{3} \langle \Delta\theta(\Delta x, \Delta t) \rangle\langle \Delta\theta(\Delta x, \Delta t)^3 \rangle 
+\frac{1}{4} \langle \Delta\theta(\Delta x, \Delta t)^2 \rangle^2  + \mathcal{O}(\Delta\theta(\Delta x, \Delta t)^6)
\end{align}
and
\begin{align}
 \exp(-\textrm{Var} \Delta \theta(\Delta x, \Delta t))\;=\;&1- \langle \Delta\theta(\Delta x, \Delta t)^2 \rangle + \langle \Delta\theta(\Delta x, \Delta t) \rangle^2 \nonumber\\ 
 &\hspace{0.2cm}+\frac{1}{2}  \langle \Delta\theta(\Delta x, \Delta t)^2 \rangle^2  
 -\langle \Delta\theta(\Delta x, \Delta t)^2 \rangle  \langle \Delta\theta(\Delta x, \Delta t) \rangle^2\rangle \nonumber\\ 
 &\hspace{0.2cm}+\frac{1}{2} \langle \Delta\theta(\Delta x, \Delta t) \rangle^4 + \mathcal{O}(\Delta\theta(\Delta x, \Delta t)^6)\,.
\end{align}
We thus expect the two quantities to differ significantly when $\mathcal{O}(\Delta\theta^4)$ fluctuations become comparable to $1-\textrm{Var} \left[\Delta \theta \right]$.
The effect of the density-density and density-phase correlations are studied in Sec.~\ref{sec:g1-contrib}. 
We show that they do not affect the KPZ scaling for all time delays within the time window where KPZ scaling is observed, thus supporting the corresponding assumption in the derivation of the relation~\eqref{eq:g1-exp}. The validity of Eq.~\eqref{eq:g1delta} is also discussed in the same section.

\newpage

\section{Experiments: additional information and data}
\label{sec:experiments}

\subsection{Sample description}\label{subsec:sample}

\begin{figure}[h!]
    \centering
    \includegraphics[scale=0.40]{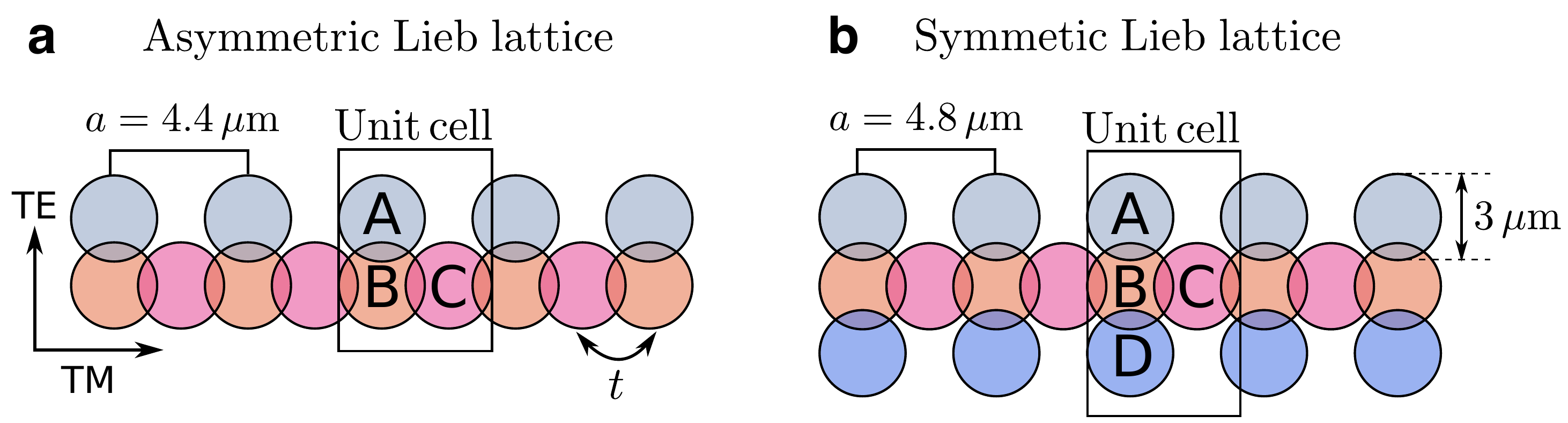}
    \caption{
    \textbf{Sketches of the lattices.} 
    \textbf{a} Asymmetric and \textbf{b} symmetric Lieb lattices.
    }
    \label{fig:SUP_MAT_Lattices}
\end{figure}

The sample $-$ grown by molecular beam epitaxy $-$ consists of a $\lambda/2$ Ga$_{0.05}$Al$_{0.95}$As microcavity surrounded by two Al$_{0.20}$Ga$_{0.80}$As/Al$_{0.05}$Ga$_{0.95}$As distributed Bragg reflectors with 28 (resp. 40) pairs in the top (resp. bottom) mirror, yielding a nominal quality factor of $Q=70000$. 
Three stacks of four 7nm GaAs quantum wells are embedded in the microstructure, resulting in a $15$ meV Rabi splitting.
The first stack lies at the center of the cavity spacer and the other two at the first anti-nodes of the electromagnetic field in each mirror (inset of Fig.~\ref{fig:SUP_MAT_Lattices}a). 
\vspace{4pt}
\newline
The planar cavity is patterned into 200 $\mu$m long 1D lattices of coupled micropillars, using electron beam lithography and dry etching. 
We choose to work on two different Lieb lattices, where we found experimental conditions that give rise to condensation in negative mass states. 
The first one $-$ namely, the asymmetric Lieb lattice $-$ exhibits three micropillars of 3 $\mu$m diameter per unit cell, with a lattice period $a = 4.4 \, \mu$m (Fig.~\ref{fig:SUP_MAT_Lattices}a). 
The second one, referred to as the symmetric Lieb lattice, exhibits four micropillars per unit cell, with $a = 4.8 \, \mu$m (Fig.~\ref{fig:SUP_MAT_Lattices}b). 
The cavity-exciton detuning (\textit{i.e.} the difference between the lowest energy cavity mode and the exciton line) is about $-12\,\mathrm{meV}$ (resp. $-15\,\mathrm{meV}$) for the  asymmetric (resp. the symmetric) Lieb lattice.

\subsection{From a single micropillar to lattices}\label{subsec:BandEngineering}

Micropillars constitute the elementary building block of the lattices we use. In such a structure, the electric field is confined in all directions: longitudinally by the cavity mirrors, transversely by the large refractive index mismatch between AlGaAs and vacuum. 
In the transverse plane, polaritons are thus confined through their photonic component in a quasi-infinite circular potential. 
This confinement yields discrete energy modes whose spatial shape are similar to the hydrogen atomic orbitals. 
The lowest energy mode has a single bright lobe and thus corresponds to a S-state; the next two modes correspond to P-states; and so on. 
\vspace{4pt}
\newline As mentioned in section~\ref{subsec:sample}, the unit cell of a 1D asymmetric Lieb lattice contains three sites (labeled A, B and C in Fig.~\ref{fig:SUP_MAT_Lattices}a), linked by the coupling constant $t$.
In the quasi-continuum limit where several unit cells are arranged along a 1D lattice, this coupling between sites yields the hybridization of the pillar S-orbitals into three dispersive S-bands, gapped one from the other.
The same reasoning enables describing the appearance of six higher energy P-bands, resulting from the hybridization of the pillar P-orbitals.
\vspace{4pt}
\newline The 1D symmetric Lieb lattice contains four sites (labeled A, B, C and D in Fig.~\ref{fig:SUP_MAT_Lattices}b), and thus presents four dispersive S-bands, and eight dispersive P-bands.

\subsection{Asymmetric lattice characterization - Microscopic parameters}
\label{sec:gamma2}

\subsubsection{Low-power photoluminescence spectrum}

\begin{figure}[h!]
    \centering
    \includegraphics[scale=0.60]{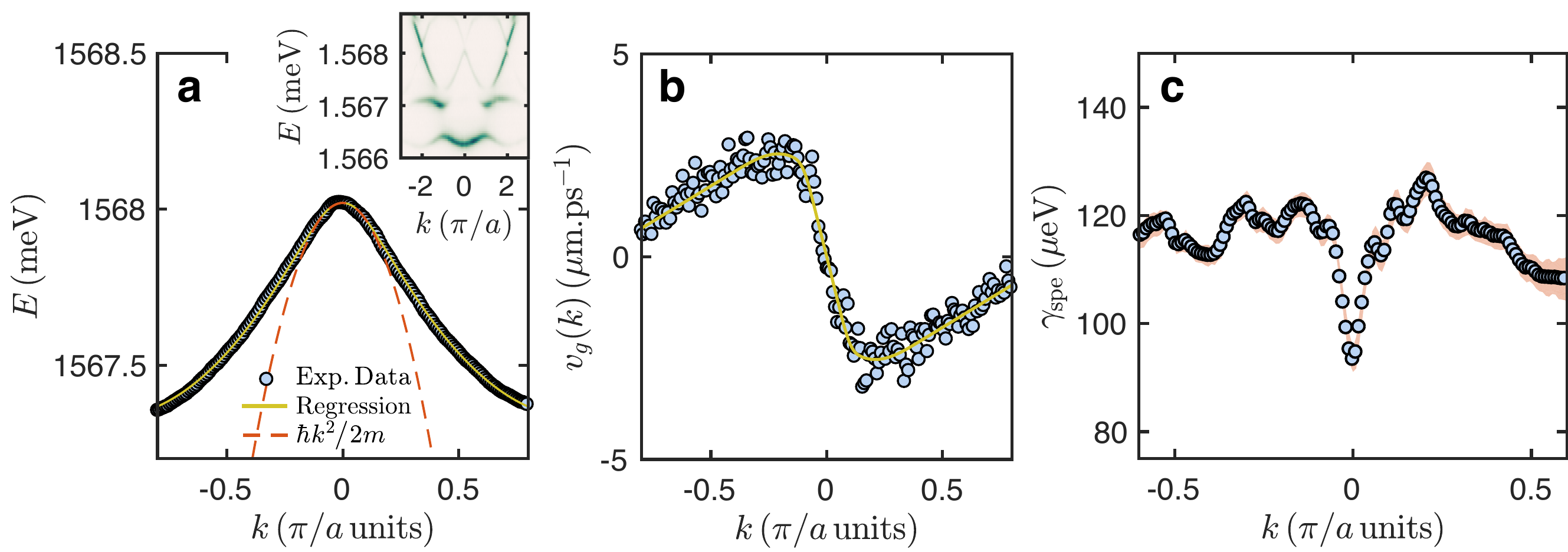}
    \caption{\textbf{a} Dispersion of the upper S-band obtained by fitting the spectrum (inset) by a lorenztian lineshape for all wave-vectors lying in the first Brillouin zone. 
    Red dotted line: parabolic fit of the data points in the vicinity of $k=0$.
    Yellow line: fit of the dispersion using a nonlinear regression model.
    \textbf{b} Group velocity as a function of $k$, computed from the data in \textbf{(a)}.
    Yellow line: derivative of the fit function in \textbf{(a)}. \textbf{c} Spectral linewidth $\gamma_{\mathrm{spe}}$ as a function of $k$.   
    }
    \label{fig:SUP_MAT_Carac_1}
\end{figure}

Linear spectroscopy enables visualizing the band structure of our lattices, from which we can extract some of the parameters entering our numerical simulations. 
The inset of Fig.~\ref{fig:SUP_MAT_Carac_1}a shows the far-field emission (in TM polarization, parallel to the lattice axis) of the asymmetric Lieb lattice at low excitation power ($P/P_{th} \approx 0.5$).
The bottom three bands visible on this image correspond to the three lattice S-bands. 
Above, we also see the first P-band, separated by a small gap from the upper S-band, at the top of which condensation takes place. 
Fitting the latter with a Lorentzian lineshape for all wave-vectors lying in the first Brillouin zone enables us to retrieve all at once: 
\vspace{-6pt}
\begin{itemize}
    \item[-] the polariton dispersion $E(k)$ (see Fig.~\ref{fig:SUP_MAT_Carac_1}a), from which we extract the polariton mass $\boldsymbol{m \!=\! -3.3 \times 10^{-6} \, m_{e}}$ (where $m_{e}$ is the electron mass);
    \vspace{-6pt}
    \item[-] the polariton group velocity $v_{g}(k)$ (see Fig.~\ref{fig:SUP_MAT_Carac_1}b), obtained by differentiating the dispersion;
    \vspace{-6pt}
    \item[-] the spectral linewidth $\gamma_{\mathrm{spe}}(k)$ (light blue points in Fig.~\ref{fig:SUP_MAT_Carac_1}c), from which we obtain an estimate of the polariton linewidth at $k=0$: $\boldsymbol{\gamma_{\mathrm{spe}}(0) \approx 80 \,\mu\mathrm{eV}}$.
\end{itemize}
\noindent The $k = 0$ value of the measured spectral linewidth appears to be relatively large compared to the $22 \, \mu\mathrm{eV}$ nominal linewidth expected for this structure.
This is most probably due to electrostatic fluctuations in the sample during the integration time ($\sim 60 \, \mathrm{s}$), which induce a spectral wandering of the emission energy through the polariton excitonic component~\cite{45kuhlmann2013}. 
This leads, in turn, to an inhomogeneous broadening of the polariton linewidth.
 
\subsubsection{Propagation measurement}

\begin{figure}[h!]
    \centering
    \includegraphics[scale=0.65]{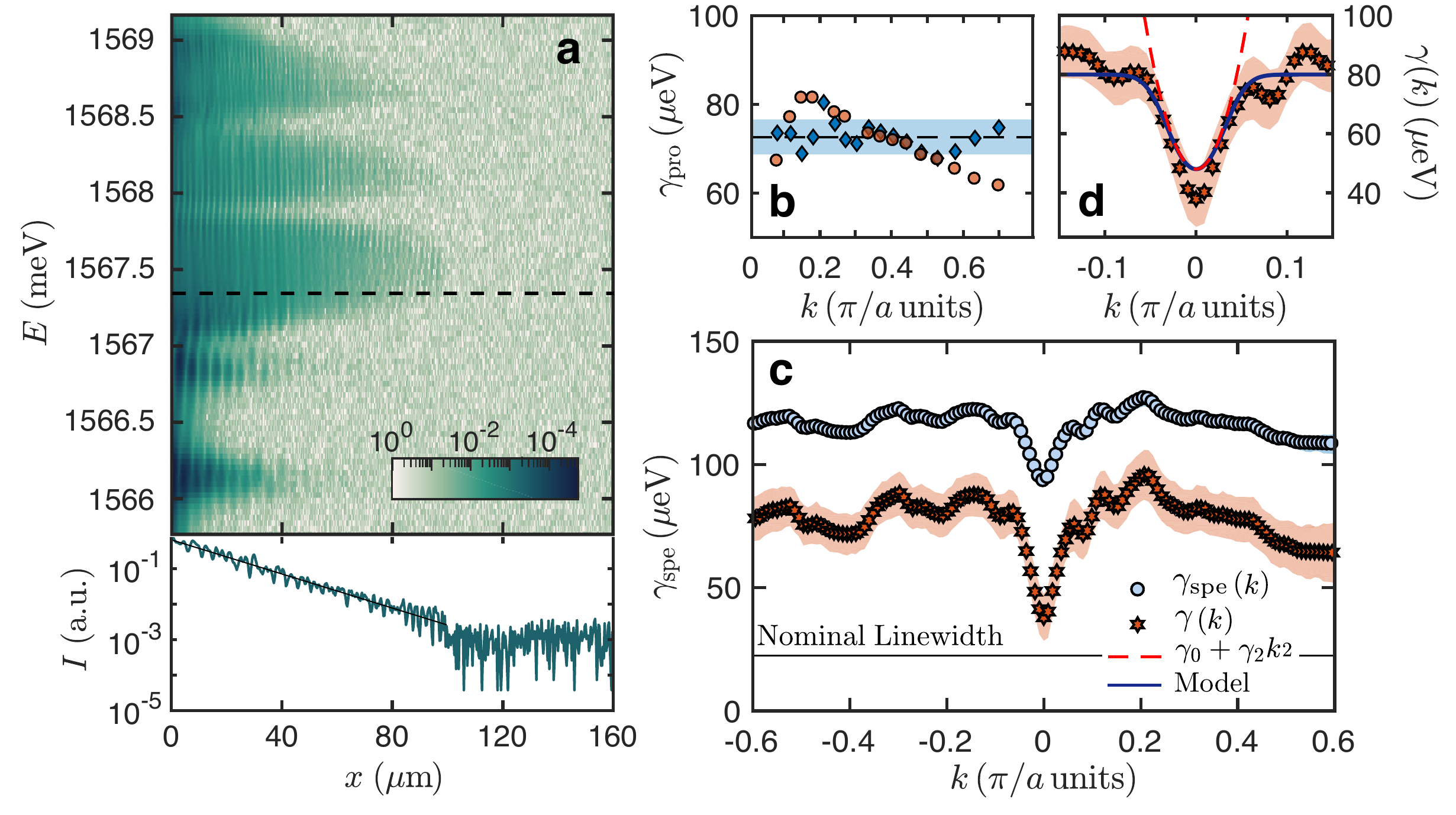}
    \caption{\textbf{a} Energy-resolved photoluminescence in real space (log scale), obtained under local non-resonant excitation in the vicinity of $x = 0$. 
    Bottom inset: Exponential decay of the luminescence intensity at $E = 1.5673$ meV (log scale). 
    Fitting this decay (black solid line) enables to retrieve the Lorentzian contribution to $\gamma_{\mathrm{spe}}$. 
    \textbf{b} Lorentzian ($\gamma_{\mathrm{pro}}$, red diamonds) and Gaussian contribution ($\gamma_{g}$, blue dots) to $\gamma_{\mathrm{spe}}$. 
    \textbf{c} Spectral linewidth $\gamma_{\mathrm{spe}}$ (light blue dots) and polariton linewidth $\gamma$ (red stars) as a function of $k$. 
    The red dots are obtained from the blue ones by removing the inhomogeneous Gaussian broadening in $\gamma_{\mathrm{spe}}$.
    \textbf{d} Polariton linewidth $\gamma$ together with our theoretical model (blue line). 
    The dashed red line shows the quadratic expansion of the model in the vicinity of $k=0$. 
    }
    \label{fig:SUP_MAT_Carac_2}
\end{figure}

We can get a better estimate of the polariton linewidth $\gamma(k)$ by probing in real space the energy resolved propagation of polaritons along the lattice, under localized excitation. 
Depending on their wave-vector, polaritons travel away from the excitation spot, with a constant group velocity $|v_{g}(k)|$.
Because of their finite lifetime, this propagation results in an exponential decrease of the photoluminescence intensity along the lattice direction ($I(x) \!\propto\! \exp \! \left\{ -|x|/L_{x} \right\}$), as shown in Fig.~\ref{fig:SUP_MAT_Carac_2}a. 
Fitting this decay at different energies $E(k)$ allows us to retrieve the polariton linewidth $\gamma_{\mathrm{pro}} = v_{g}(k)/L_{x}$ as function of $k$ (red dots in Fig.~\ref{fig:SUP_MAT_Carac_2}b). 
\vspace{4pt}
\newline This method has the advantage of being less sensitive to charge fluctuations. 
Indeed, polaritons leave the pumping area with a given initial energy $E_{0}$, setting the group velocity at which they travel. 
This group velocity remains constant over the whole propagation as charge fluctuations (i) mainly affect the reservoir energy locally (under the pump spot) and (ii) occur on a time scale much larger than the polariton lifetime. 
Therefore, the propagation length $L_{x}$ only depends on $E_{0}$ regardless of the exciton energy at the time at which the polariton was emitted.
Consequently, we assume that the measurement of $\gamma_{\mathrm{pro}}\;$is less affected by the wandering of the exciton energy and, moreover, that it almost corresponds to the Lorentzian contribution to $\gamma_{\mathrm{spe}}$.
\vspace{4pt}
\newline 
The Gaussian contribution $\gamma_{\mathrm{g}}$ to the spectral linewidth $-$ arising from the inhomogeneous broadening $-$ can then be retrieved using the following approximation~\cite{46olivero1977}: 
\begin{equation}
\gamma_{\mathrm{spe}} = 0.535 \, \gamma_{\mathrm{pro}} + \sqrt{0.217 \, \gamma_{\mathrm{pro}}^{2} + \gamma_{\mathrm{g}}^{2}}.
\end{equation}

\newpage

\noindent We notice that $\gamma_{g}$ (diamonds in Fig.~\ref{fig:SUP_MAT_Carac_2}b) is nearly constant and equal to $73 \, \mu\mathrm{eV}$ in the range  $0.1 \, \pi/a \!<\! k \!<\! 0.7 \, \pi/a$ where the propagation measurement is reliable (outside of this range, $v_{g}$ is too small to properly extract $\gamma_{\mathrm{pro}}$). 
Assuming that $\gamma_{g}$ remains constant over the Brillouin zone, we can finally remove the contribution of the inhomogeneous broadening to the spectral linewidth and obtain a better estimate of the polariton linewidth $\gamma$ (red stars in Fig.~\ref{fig:SUP_MAT_Carac_2}c). We find $\boldsymbol{\gamma (0) = 40 \pm 10 \, \mu\mathrm{eV}}$, which is reasonable compared to the nominal linewidth given earlier. 
The large errorbars on the data points (red shaded area) mainly come from the uncertainty on the measurement of $\gamma_{g}$. 
We also show in Fig.~\ref{fig:SUP_MAT_Carac_2}d a comparison between the experimental data and the fit of the linewidth behavior used in the numerical simulations (see section~\ref{sec:numerical-simulations}), and its parabolic approximation around $k=0$ ($\gamma = \gamma_{0}+\gamma_{2} k^{2}$), used in the derivation of the mapping in Sec. \ref{sec:mapping} (red dotted line). 
The optimal parameters found in our simulations are given by $\boldsymbol{\gamma_{0,\mathrm{th}} = 48.5 \, \mu\mathrm{eV}}$ and $\boldsymbol{\gamma_{2,\mathrm{th}} = 1.6 \times 10^{4} \, \mu\mathrm{eV}.\mu\mathrm{m}^{2}}$.

\subsection{Optical setup and data analysis}

\subsubsection{Optical setup}

The sketch of the optical setup is shown on Fig. 2d of the main text. 
In our experiment, polaritons are excited using a non-resonant continuous-wave laser of wavelength $740 \, \mathrm{nm}$ (where the cavity mirror reflectivity exhibits a minimum). 
A spatial light modulator (SLM) enables shaping the excitation spot into a $125 \, \mu \mathrm{m}$ long flat-top beam in the lattice direction, and a Gaussian with a $3.5 \, \mu \mathrm{m}$ FWHM in the transverse direction.
The light emitted by the sample is collimated by the excitation lens, passes through a polarizer (selecting the TM polarization), and is sent through an interferometer. 
A polarized beam splitter, combined with a half wave-plate, enables splitting the incoming light into two beams while controlling their power ratio. 
The first beam reflects on a plane mirror, making a round trip through a quarter-wave plate which turns its polarization by 90$^{\mathrm{o}}$. 
The second beam reflects on a retroreflector. 
The latter is mounted on a motorized translation stage, allowing for a variation of the path length difference between the interferometer arms. 
Both beams are finally recombined in a non-polarized beam splitter before being imaged onto a CCD camera. 
This arrangement of the interferometer enables us to tune the interfringe spacing of the resulting interference pattern, as it allows to control the incident wave-vector of both beams before the last lens as well as their relative spacing. 
In order to probe the temporal scaling of the condensate first-order correlation function, we typically scan the retroreflector position over a distance of $\Delta L = 5\, \mathrm{cm}$, corresponding to a maximum time delay of $\Delta t = 2 \Delta L/ c = 330 \, \mathrm{ps}$. During such a scan, we set the camera exposure time to $1\,\mathrm{s}$ and acquire a serie of 250 images.
The zero delay position ($\Delta L = 0$) has been calibrated beforehand by sending white light through the interferometer.

\subsubsection{Data analysis procedure} \label{subsubsec:Analysis}

\begin{figure}[h!]
    \centering
    \includegraphics[scale=0.68]{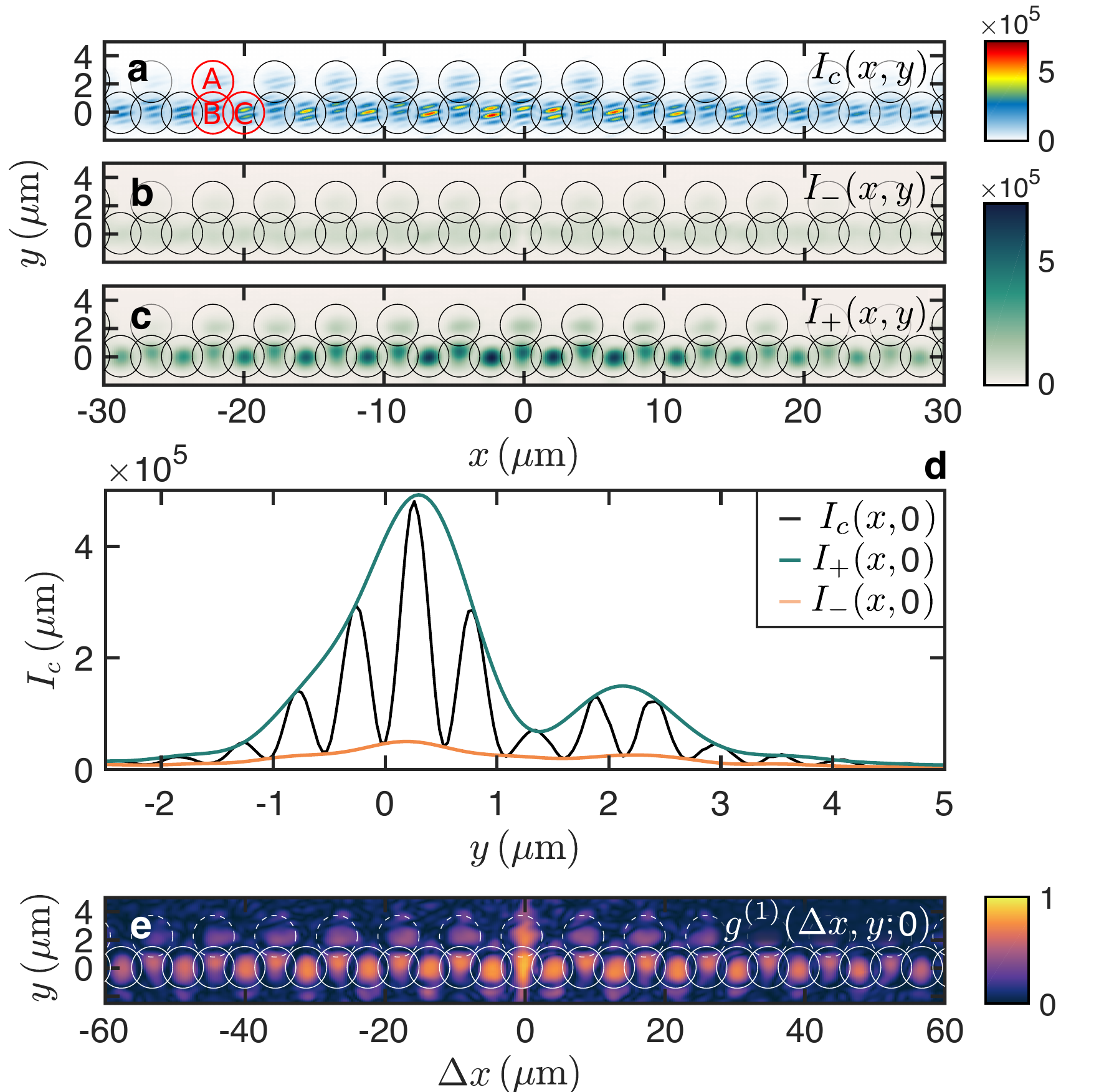}
    \caption{\textbf{Retrieving the first order coherence of polariton condensates.} \textbf{a} Interference pattern captured by the CCD for $\Delta t = 0$. \textbf{b-c} lower and upper envelopes of the interferogram. \textbf{d} Cut of the first three maps along $x=0$. Coherence retrieved from Eq.~\eqref{eq:Visibility}. The photoluminescence background strongly interferes with itself at $(\Delta x = 0, \Delta t = 0)$, creating a peak in $|g^{(1)}|$ at $\Delta x = 0$. This peak disappears after one scan step only, as the photoluminescence background light is fully incoherent.}
    \label{fig:SUP_MAT_DataAnalysis}
\end{figure}

In our experimental setup, the condensate image (reference arm) is overlapped with image at the mirror-symmetric point with respect to a plane orthogonal to the lattice (retroreflector arm). 
The resulting interference pattern (at $\Delta t =0$) is shown in Fig.~\ref{fig:SUP_MAT_DataAnalysis}a and Fig.~\ref{fig:SUP_MAT_Sym}d for the asymmetric and symmetric Lieb lattices respectively.
At each point $\boldsymbol{r} = (x, y)$ of the image plane, the intensity $I_{c}(x, y)$ is given by the interference between the fields emitted at $x$ and $-x$ in the sample plane. Dropping the $y$ coordinate, we thus expect that: 
\begin{equation} \label{eq:intrfint}
    \begin{aligned}
        I_{c}(x, \Delta t) = \frac{1}{4} \left[I(x) + I(-x) +2 \sqrt{I(x) I(-x)} \, |g^{(1)}(\Delta x, \Delta t)|  \, \cos\left(\Delta \Phi \right)\right],
    \end{aligned}
\end{equation}
\noindent where $\Delta x =2x$, $\Delta \Phi = \boldsymbol{\delta q \cdot r}$ stands for the relative geometrical phase between the condensate field and its mirror symmetric (originating from the non-zero relative transverse wave-vector $\boldsymbol{\delta q}$ between them) and $I(x) = \langle |\mathcal{E}(x, t)|^{2} \rangle_{\tau}$ for the time-averaged intensity distribution of the sample emission at position $x$. 
Here, $\langle \cdots \rangle_{\tau}$ is a time averaging over $\tau$ arising from the fact that the camera integration time $\tau = 1 \, \mathrm{s}$ is much longer than all time scales involved in the condensate dynamics.
In the main text, we implicitly assume that the ergodic hypothesis is valid, which implies that averaging physical observables over long time (as done experimentally) or over a large set of different noise realizations (as in simulations) is equivalent.
In what follows, $\langle \cdots \rangle$ indistinctly denotes temporal or statistical averaging.

\noindent The first order correlation function $g^{(1)}$ in Eq.~\eqref{eq:intrfint} is defined by: 
\begin{equation} \label{eq:g1}
    \begin{aligned}
        g^{(1)}(\Delta x, \Delta t) = \frac{\langle \mathcal{E}^{\star}(x, t_{0})\mathcal{E}(-x, t_{0}+\Delta t) \rangle}{\sqrt{\langle|\mathcal{E}(x, t_{0})|^{2}\rangle \, \langle|\mathcal{E}(-x, t_{0} + \Delta t)|^{2}\rangle }}.
    \end{aligned}
\end{equation}
\noindent 

\newpage

\noindent Experimentally, we retrieve the correlation function~\eqref{eq:g1} by measuring the fringe visibility: $V = (I_{\mathrm{+}}-I_{\mathrm{-}})/(I_{\mathrm{+}}+\mathrm{I_{-}})$, where $I_{\mathrm{+}}$ and $I_{\mathrm{-}}$ stand respectively for the upper and lower envelopes of $I_{c}$. 
At every time delay $\Delta t$, $I_{\mathrm{+}}$ and $I_{\mathrm{-}}$ are extracted using Fourier analysis on the interferogram. 
As an example, Fig.~\ref{fig:SUP_MAT_DataAnalysis}b and Fig.~\ref{fig:SUP_MAT_DataAnalysis}c respectively show the lower and upper envelopes associated to the interference pattern in Fig.~\ref{fig:SUP_MAT_DataAnalysis}a. 
A cut of those three intensity maps along $x=0\,$yields the graph in Fig.~\ref{fig:SUP_MAT_DataAnalysis}d. 
The visibility $V$ is finally related to the first-order coherence through:
\begin{equation} \label{eq:Visibility}
    \begin{aligned}
        V(\Delta x, \Delta t) = \frac{2 \sqrt{I(x) I(-x)}}{I(x)+I(-x)} \, |g^{(1)}(\Delta x, \Delta t)| = K(\Delta x) \, |g^{(1)}(\Delta x, \Delta t)|,
    \end{aligned}
\end{equation}
\noindent where $K(\Delta x)$ is a normalization factor taking into account potential imbalance between $I(x)$ and $I(-x)$. In our case, this factor remains close to 1.
Using Eq.~\eqref{eq:Visibility}, we finally retrieve $|g^{(1)}(\Delta x, y; \Delta t)|$ (see Fig.~\ref{fig:SUP_MAT_DataAnalysis}e). 
The coherence map shown in Fig. 2e of the main text is obtained by keeping only the maximum value of $|g^{(1)}(\Delta x, y; \Delta t)| \,$over the pillars identified through white solid circles in Fig.~\ref{fig:SUP_MAT_DataAnalysis}e. 

\subsubsection{Normalization of \texorpdfstring{$|g^{(1)}|$}{TEXT}}
\label{sec:normalization}

After having retrieved $|g^{(1)}(\Delta x, \Delta t)|$ from the data analysis detailed in the previous section, we search for KPZ scalings in the spatio-temporal variations of $-2 \mathrm{log}\left( |g^{(1)}| \right)$. 
In particular, we show in Fig. 3c of the main text the collapse onto the universal KPZ scaling function of the $|g^{(1)}|$ data points within a certain spatio-temporal window. 
In order to do so, we plot in log-log scale $-2 \mathrm{log}\left( \kappa |g^{(1)}| \right) / \Delta t^{2/3}$ as function of the rescaled coordinate $y = \Delta x/\Delta t^{2/3}$, where $\kappa$ is a normalization factor that needs to be properly set. 
Indeed, representing the data in such a way implicitly requires that the temporal KPZ scaling extends all the way to $\Delta t = 0$, where $|g^{(1)}(0,0)|$ is expected to be 1. 
Experimentally, we observe a transient regime at short time delays, preceding the establishment of the $\Delta t^{2\beta}$ power law behavior of $-2 \mathrm{log}\left(|g^{(1)}| \right)$. 
We thus need to ensure that the extrapolation of this power law passes through 0 at $\Delta t = 0$ in order for the chosen graphic representation to be meaningful. 
This amounts to shifting downward the data points shown in Fig.3 a-b until they match the blue solid line, which, in turn, translates into multiplying the whole $|g^{(1)}|$ data set by a factor $\kappa$. 
Note that this normalization does not change the coherence decay, which remains a stretched exponential.

\subsection{Additional results} \label{sec:AddData} 

\subsubsection{Variations of \texorpdfstring{$|g^{(1)}|$}{TEXT} in linear scale}

\begin{figure}[h!]
    \centering
    \includegraphics[scale=0.54]{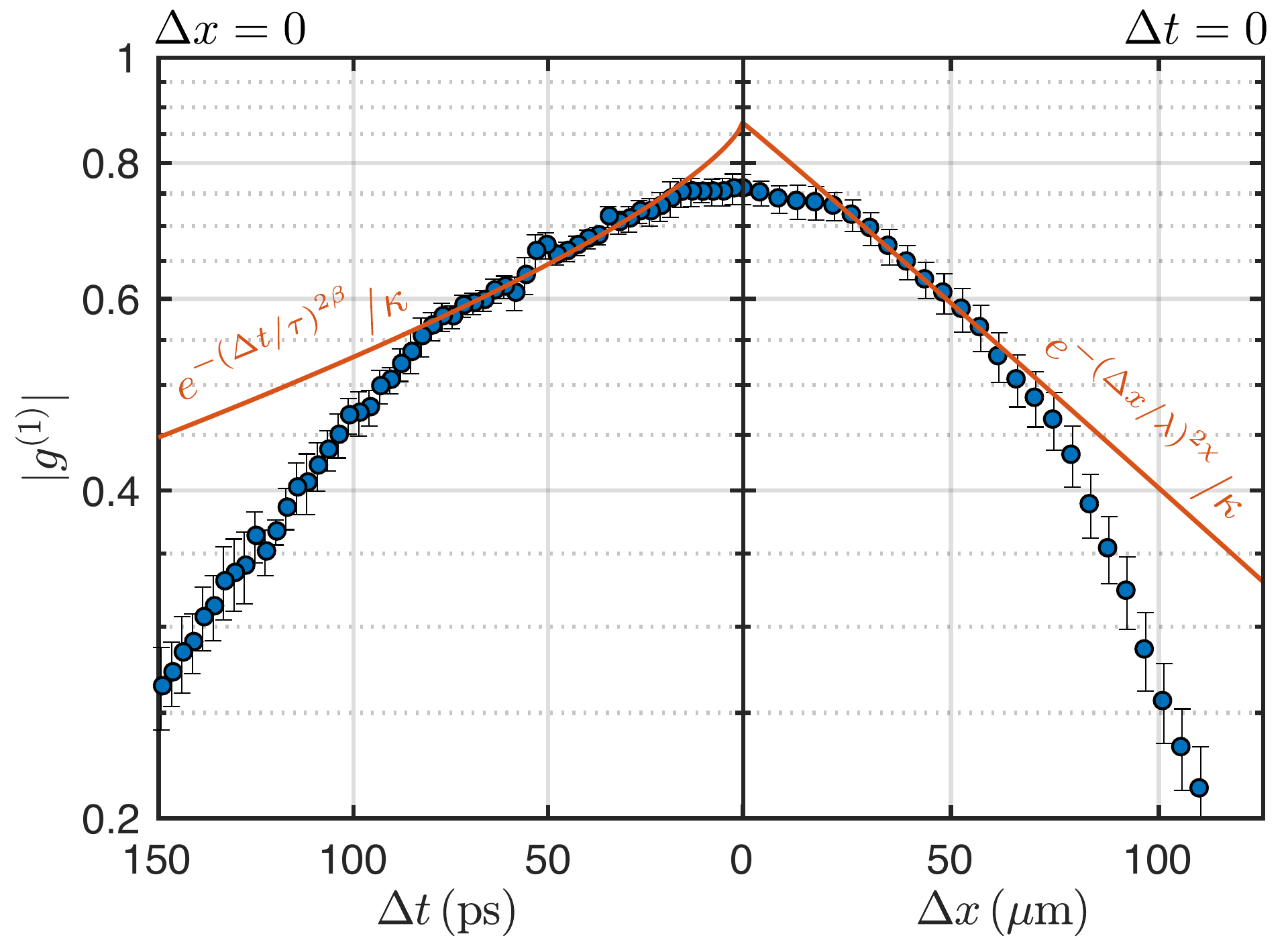}
    \caption{\textbf{Variations of $|g^{(1)}|$ in linear scale.} 
    Errorbars are calculated by performing a repeatability analysis on the numerical extraction of $g^{(1)}$ from the interferograms.
    We report on this plot the result of the fitting procedure describe in~\ref{sec:FittingProcedure} (red lines). 
    }
    \label{fig:SUP_MAT_LinScale}
\end{figure}

In order for the reader to be readily able to compare our results with with other works in the literature where data are reported in a different way, we show in figure~\ref{fig:SUP_MAT_LinScale} the variations of $\boldsymbol{|g^{(1)}|}$ as a function of $\Delta t$ (for $\Delta x = 0$) and $\Delta x$ (for $\Delta t = 0$).
We also report on this graph the fit of the data points by streched-exponential decays over the spatial and temporal KPZ window. Further details on the fitting procedure can be found in section~\ref{sec:FittingProcedure}.

\subsubsection{Estimation of the scaling exponents \texorpdfstring{$\beta$}{TEXT} and \texorpdfstring{$\chi$}{TEXT}}
\label{sec:FittingProcedure}

\vspace{-4pt}

\begin{figure}[h!]
    \centering
    \includegraphics[scale=0.60]{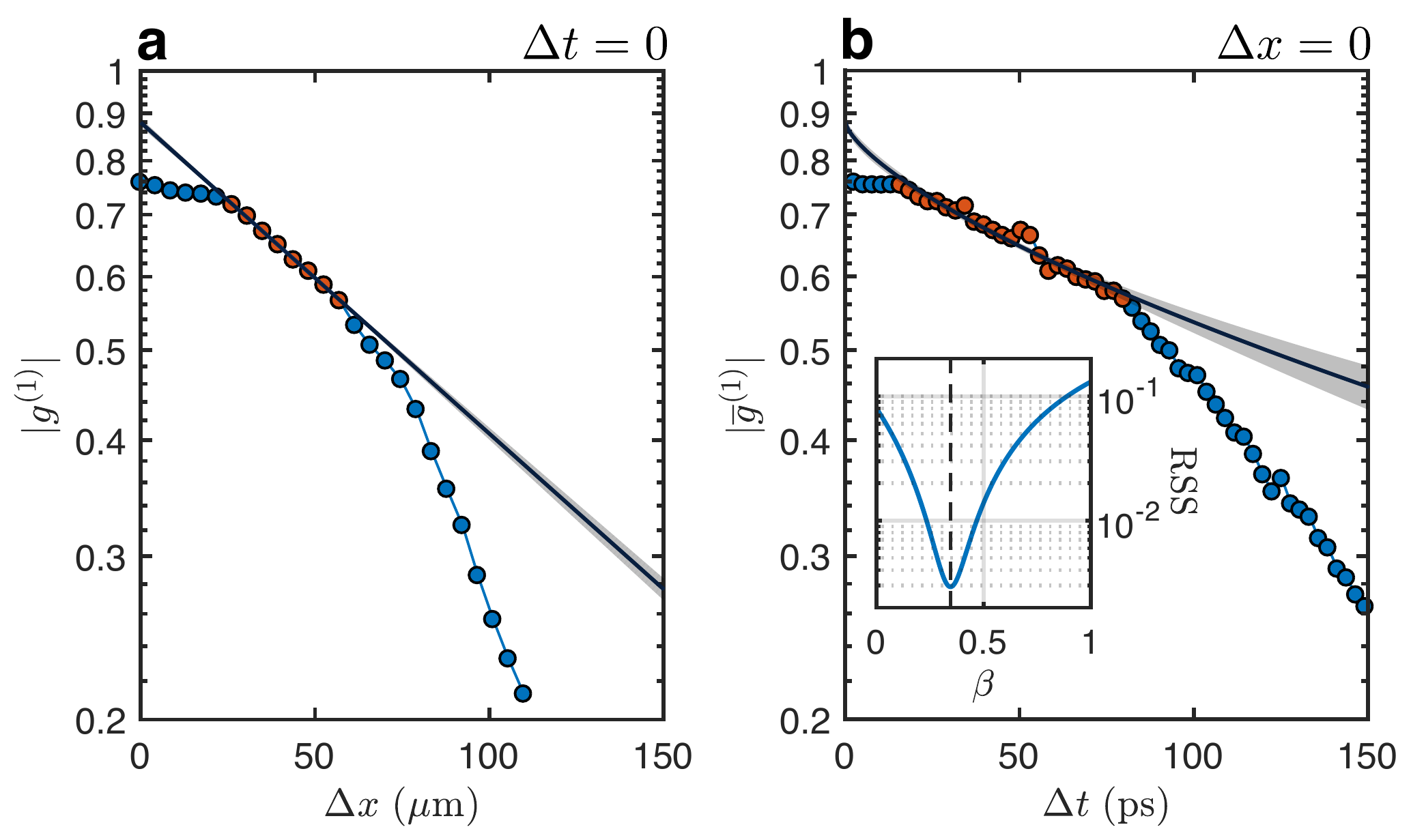}
    \caption{\textbf{Fit of the coherence spatial and temporal decay setting $\boldsymbol{\chi = 1/2}$.} 
    \textbf{a} Fit of the spatial decay of $|g^{(1)}|$ at $\Delta t = 0$ by the stretched exponential function $f_{x}$. 
    \textbf{b} Fit of the temporal decay of $|\overline{g}^{(1)}| = \kappa |g^{(1)}|$ at $\Delta x = 0$ by the stretched exponential function $f_{t}$.
    Errobars are estimated by propagating the uncertainty on $\kappa$, as explained in the text.
    They are smaller than the points diameter.
    In both graphs, the red dots indicate where the fits are performed.
    The grey-shaded area gives the $95\%$ confidence interval on those fits. \textbf{Inset:} Residual Sum of Squares (RSS) as a function of $\beta$, reaching a minimum at $0.35$. 
    }
    \label{fig:SUP_MAT_FitFixChi}
\end{figure}

In the main text, we show that the temporal ($\Delta x = 0$, Fig 3.a) and spatial variations ($\Delta t = 0$, Fig 3.b) of $|g^{(1)}|$ qualitatively agree with the stretched exponential scaling predicted by KPZ theory. 
In this section, we present a more quantitative analysis of the experimental data, based on curve fitting, which aims at measuring the universal scaling exponent $\chi$ and $\beta$ within a $95\%$ confidence interval.
\vspace{4pt}
\newline
As mentioned in the main text, the theoretical value of the roughness exponent $\chi = 1/2$ is not characteristic to the KPZ universality class  but rather shared among three different classes: Edward-Wilkinson $\left(\chi = 1/2, \, \beta = 1/4 \right)$, KPZ $\left(\chi = 1/2, \, \beta = 1/3 \right)$ and the class $\left(\chi = 1/2, \, \beta = 1/2 \right)$ to which linear systems described by Bogoliubov theory pertain. 
As all the classes to which our system could belong share the same value for $\chi$, we first set $\chi = 1/2$ and fit the spatial decay of $|g^{(1)}\left(\Delta x, \Delta t = 0 \right)|$ with the stretched exponential function $f_{x}(\Delta x) = \exp \left( - \Delta x/\lambda\right)/\kappa$ (see Fig.~\ref{fig:SUP_MAT_FitFixChi}a). 
The normalization factor $\kappa$ and the non-universal space-scale $\lambda$ are two fitting parameters.
From this fit, we obtain: $\kappa = 1.14 \pm 0.01 $ (uncertainties are estimated from the $95\%$ confidence interval on the fit parameters).
We then focus on the temporal decay of $|g^{(1)}\left(\Delta x=0, \Delta t \right)|$. 
In order to properly propagate the error on $\kappa$, we first renormalize the data points by defining $\overline{g}^{(1)} = \kappa g^{(1)}$. 
If we omit the uncertainty on the experimental data and only consider the uncertainties originating from the fitting procedure, the error on $\overline{g}^{(1)}$ can simply be expressed as: $\delta \overline{g}^{(1)} = \delta \kappa \, g^{(1)}$.
The renormalized data points are shown on Fig.~\ref{fig:SUP_MAT_FitFixChi}b (on this graph, errorbars are smaller than the points diameter).
We finally fit the temporal decay of $|\overline{g}^{(1)}\left(\Delta x=0, \Delta t \right)|$ with an other stretched exponential function $f_{t}(\Delta t) = \exp \left[ - (\Delta t/\tau)^{2 \beta}\right]$, using a weighted nonlinear least squares algorithm to take the error on $\overline{g}^{(1)}$ into account. 
From this fitting procedure, we obtain: $\beta = 0.35 \pm 0.02$. The fitted value of $\beta$ is in close agreement with the theoretical prediction for the KPZ universality class where $\beta_{\mathrm{th}} = 1/3$.

\vspace{-16pt}

\begin{figure}[h!]
    \centering
    \includegraphics[scale=0.60]{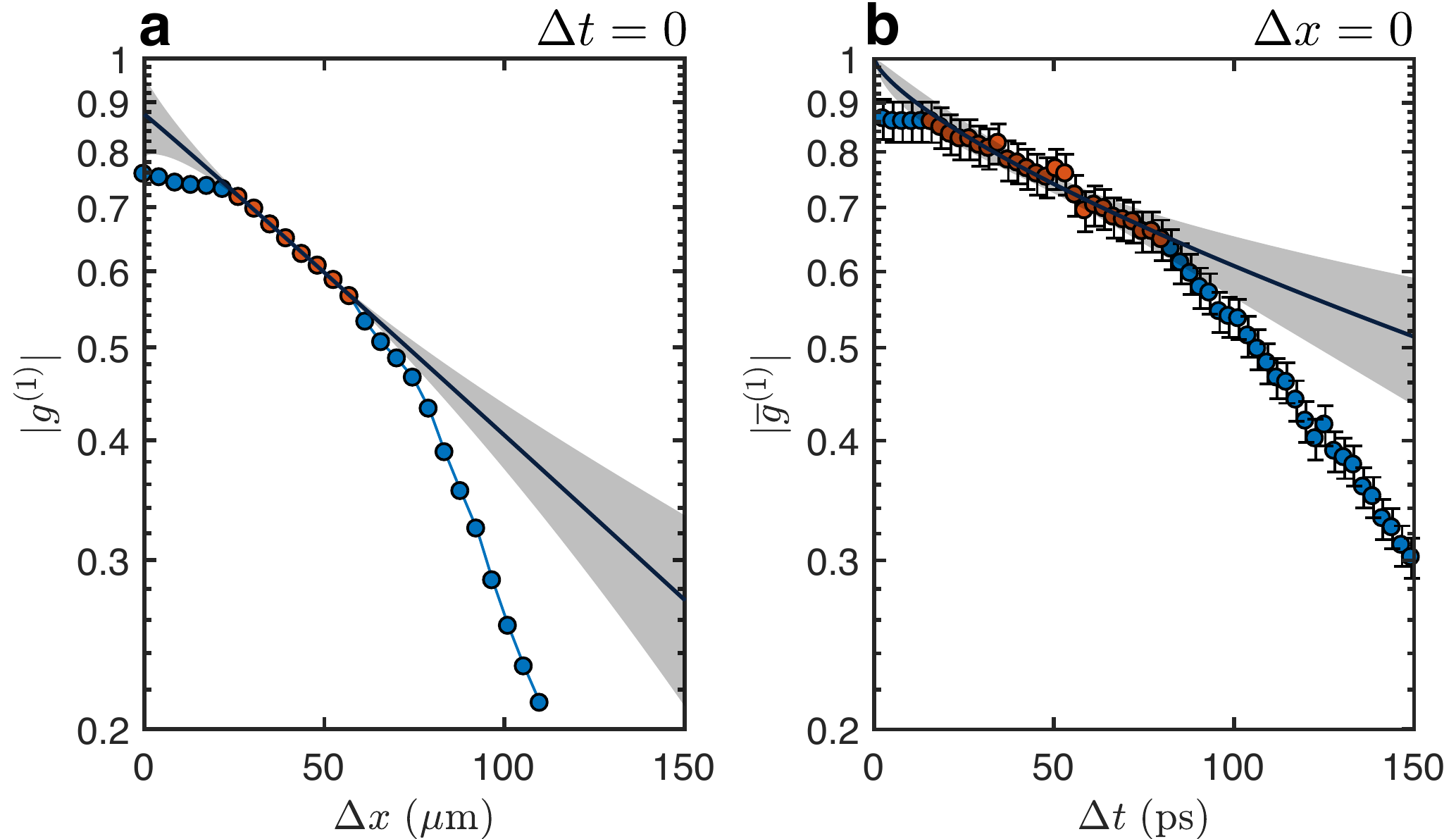}
    \caption{\textbf{Fit of the coherence spatial and temporal decay setting $\boldsymbol{\chi}$ as a fitting parameter.}
    \textbf{a} Fit of the spatial decay of $|g^{(1)}|$ at $\Delta t = 0$ by the stretched exponential function $\tilde{f}_{x}$.
    \textbf{b} Fit of the temporal decay of $|\overline{g}^{(1)}| = \kappa |g^{(1)}|$ at $\Delta x = 0$ by the stretched exponential function $f_{t}$.
    Errobars are estimated by propagating the uncertainty on $\kappa$, as explained in the text.
    In both graphs, the red dots indicate where the fits are performed.
    The grey-shaded area gives the $95\%$ confidence interval on those fits.
    }
    \label{fig:SUP_MAT_FitAll}
\end{figure}

We can push our analysis further relaxing the constraint on $\chi$. We fit the spatial decay of $|g^{(1)}\left(\Delta x, \Delta t = 0 \right)|$ with the stretched exponential function $\tilde{f}_{x}(\Delta x) = \exp \left[ - (\Delta x/\lambda)^{2 \chi}\right]/\kappa$, where $\chi$ is now a third fitting parameter (see Fig.~\ref{fig:SUP_MAT_FitAll}a). We obtain: $\kappa = 1.14 \pm 0.06$ and $\chi = 0.51 \pm 0.08$. Propagating the error on $\kappa$ in the same way as before and fitting the renormalized data points by $f_{t}$ (see Fig.~\ref{fig:SUP_MAT_FitAll}b), we get: $\beta = 0.36 \pm 0.11$. As expected, the uncertainty on $\beta$ is now larger but it still allows us to discriminate between the different universality classes, as the KPZ value $\beta_{\mathrm{th}} = 1/3$ remains the only one lying within the $95\%$ confidence interval on $\beta$.

\subsubsection{Symmetric Lieb lattice}
\label{sec:sym-Lieb}
\begin{figure}[h!]
    \centering
    \includegraphics[scale=0.63]{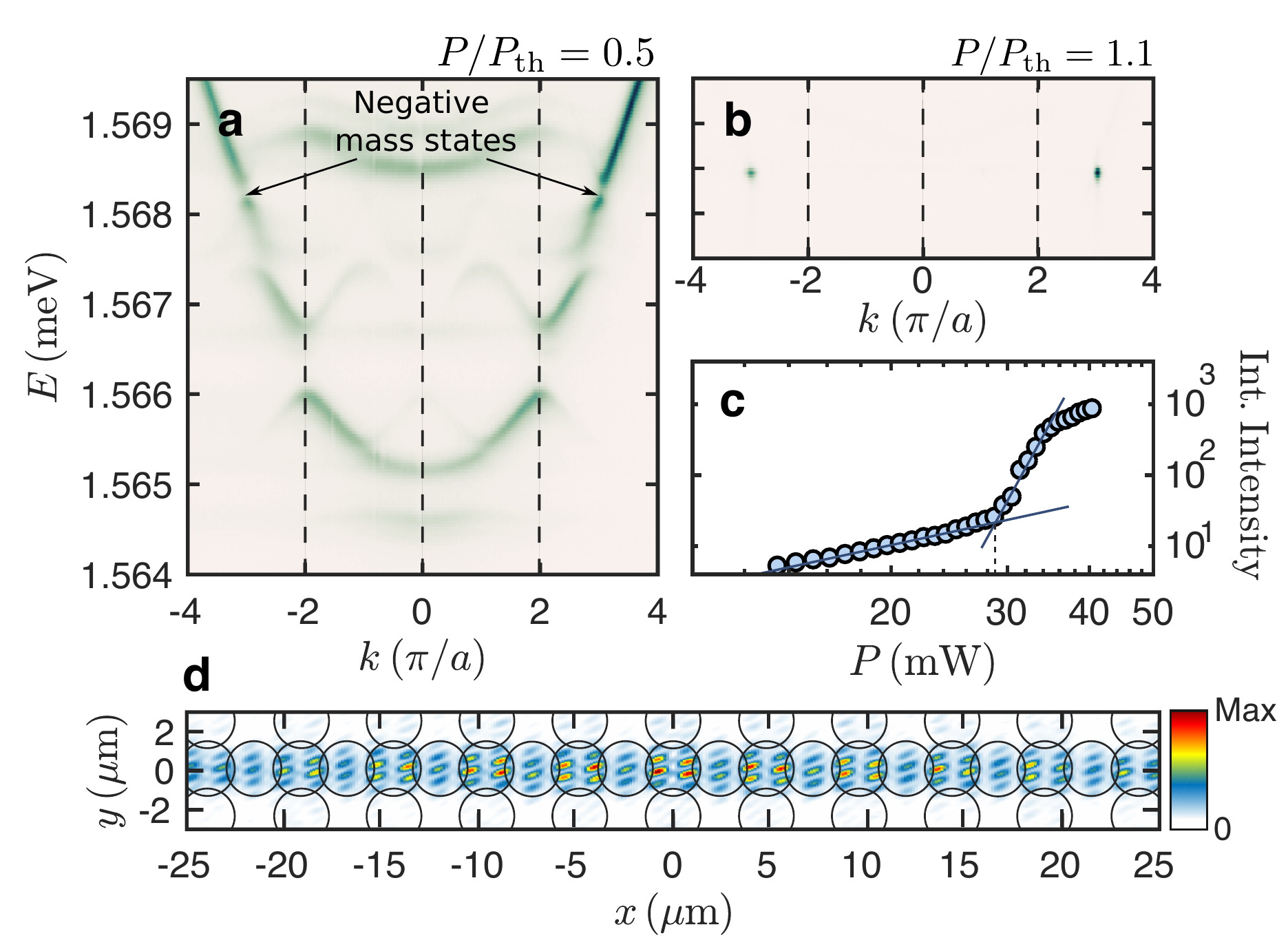}
    \caption{\textbf{Condensation in the symmetric Lieb lattice.} \textbf{a}-\textbf{b} Far-field energy-resolved photoluminescence (TM polarization) below \textbf{(a)} and above \textbf{(b)} condensation threshold. The condensate forms at the edge of the second Brillouin zone, in a negative mass state lying in the P-bands of the Lieb lattice. \textbf{c} Integrated emission intensity as function of the incident pump power, showing a condensation threshold at $P_{\mathrm{th}} \approx 28 \, \mathrm{mW}$. \textbf{d} Interference pattern at $\Delta t = 0$. The black circles show where the pillars are located.}
    \label{fig:SUP_MAT_Sym}
\end{figure}

As mentioned in section~\ref{subsec:sample}, the cavity-exciton detuning of the symmetric lattice is $3 \, \mathrm{meV}$ larger (in absolute value) than the asymmetric lattice one. 
As a consequence, the interplay between gain and dissipation gives rise to condensation in the P-bands of the symmetric lattice, at an energy close to the one at which condensation was observed in the asymmetric one.
The black arrows in the low-power far-field photoluminescence (see Fig.~\ref{fig:SUP_MAT_Sym}a) indicate the top of the P-band in which the polariton condensate forms. 
The polariton mass is negative there, thus preventing the formation of modulation instability. 
Note the distinctive spatial distribution of the condensate (visible on the interferogram in Fig.~\ref{fig:SUP_MAT_Sym}d), that exhibits two lobes on each pillar, confirming the fact that condensation occurs in P-bands.  

\begin{figure}[h!]
    \centering
    \includegraphics[scale=0.70]{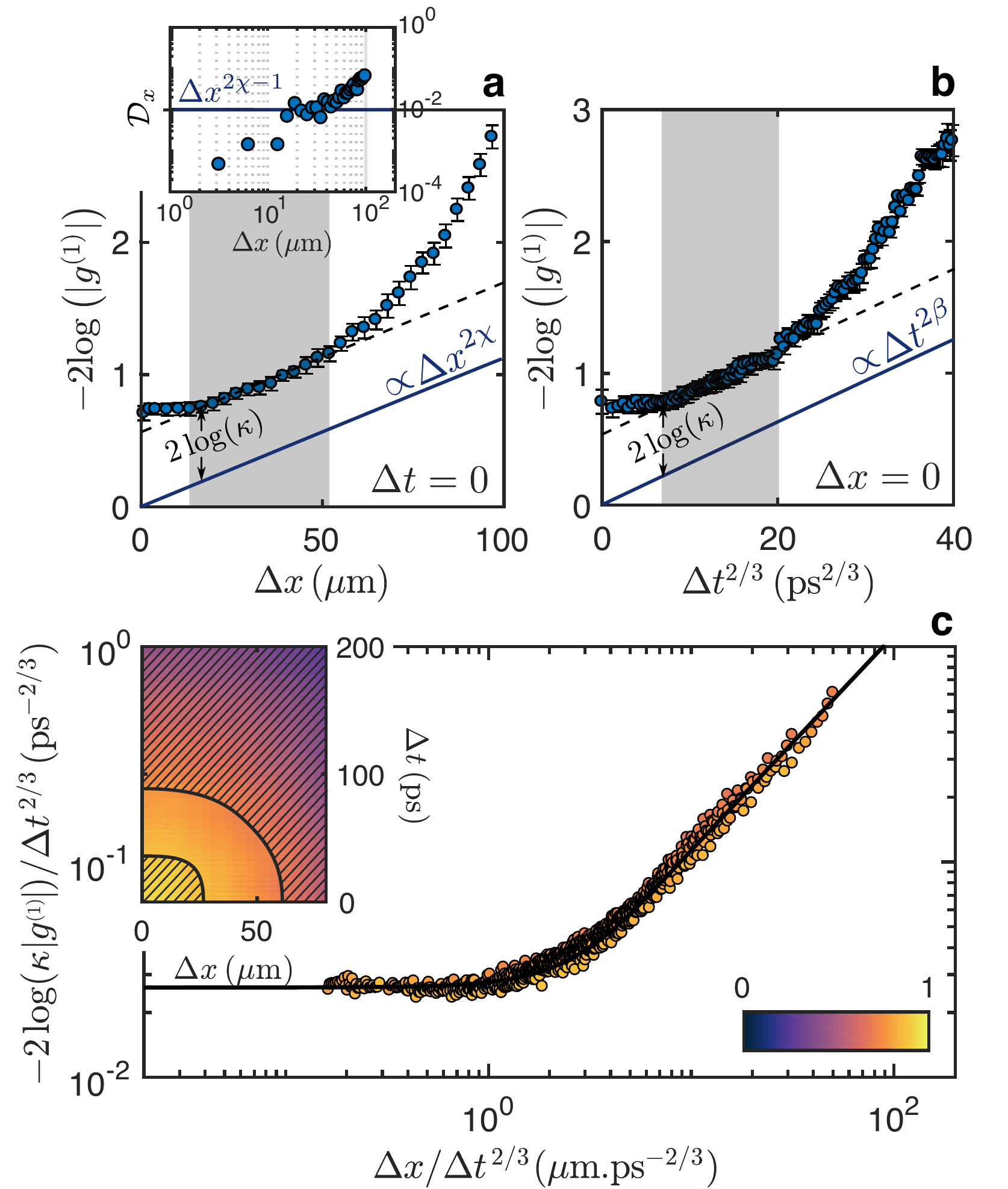}
    \caption{\textbf{KPZ scaling in the coherence decay of a polariton condensate forming in the symmetric Lieb lattice.} 
    \textbf{a} Measured values of $-2 \, \mathrm{log} (| g^{(1)} (\Delta x, 0) |)$ as a function of $\Delta x$. 
    The grey shaded area highlights the temporal window where KPZ scaling settles in.
    Inset: Variations of $\mathcal{D}_{x}$ computed from the experimental data.
    The solid line shows the expected scaling of $\mathcal{D}_{x}$. 
    \textbf{b} Measured values of $-2 \, \mathrm{log} (|g^{(1)} (0, \Delta t) |)$ as a function of $\Delta t^{2/3}$. 
    The grey shaded area shows the temporal window inside which KPZ scaling is observed.  
    Errorbars on the experimental data points are calculated by performing a repeatability analysis on the numerical extraction of $g^{(1)}(\Delta x, \Delta t)$ from the interferograms.
    \textbf{c} Measured values of $-2 \, \mathrm{log} ( \kappa | g^{(1)} |)$ as a function of the rescaled coordinates $y = \Delta x/\Delta t^{2/3}$.
    We observe the collapse of the data points within the non hatched region of the coherence map (inset) onto the KPZ universal scaling function (black solid line).
    The normalization factor is $\kappa = 1.32$.
    }
    \label{fig:SUP_MAT_Scaling}
\end{figure}

Following the procedure described in section~\ref{subsubsec:Analysis}, we can retrieve $|g^{(1)} (\Delta x, y)|$ for any time delay $\Delta t$ and then study the spatio-temporal scaling of $-2 \mathrm{log} \left(  |g^{(1)}| \right)$ for the symmetric lattice. 
The experimental data obtained for $P/P_{\mathrm{th}} \approx 1.12$ are shown in Fig.~\ref{fig:SUP_MAT_Scaling}. 
The variations of $-2 \mathrm{log} \left(  |g^{(1)}| \right)$ as a function of $\Delta x$ for $\Delta t =0$ (Fig.~\ref{fig:SUP_MAT_Scaling}a) exhibit a linear trend over the spatial window $15 \, \mathrm{\mu m} \!<\! \Delta x \!<\! 50 \, \mathrm{\mu m}$ (grey shaded area), in agreement with KPZ predictions. 

This result is supported by the observation of a plateau in $\mathcal{D}_{x} = -2\partial \mathrm{log} ( |g^{(1)}(\Delta x, 0)|) /\partial \Delta x $ (inset) in the same range of $\Delta x$. 
The variations of $-2 \mathrm{log} \left(  |g^{(1)}| \right)\;$as a function of $\Delta t^{2/3}$ for $\Delta x =0$ (Fig.~\ref{fig:SUP_MAT_Scaling}b) clearly show a linear increase over the temporal window $25 \, \mathrm{ps} \!<\! \Delta t \!<\! 90 \, \mathrm{ps}$ (grey area), indicating that this quantity scales as a $\Delta t^{2 \beta}$ power law, with $\beta = 1/3$.
Finally, we observe in Fig.~\ref{fig:SUP_MAT_Scaling}c the collapse onto a single curve of all the data points lying within the non-hatched region of $|g^{(1)}|$ (see inset). 
This curve can be reproduced with remarkable agreement using the KPZ universal scaling function (black solid line), that has been shifted horizontally and vertically to fit the data points. 
These results highlight the fact that our experimental findings apply to different lattices and different types of bands.

\newpage

\newpage

\subsubsection{Effect of the pumping power on the KPZ window}\label{subsubsec:Power}

\vspace{-4pt}

\begin{figure}[h!]
    \centering
    \includegraphics[scale=0.55]{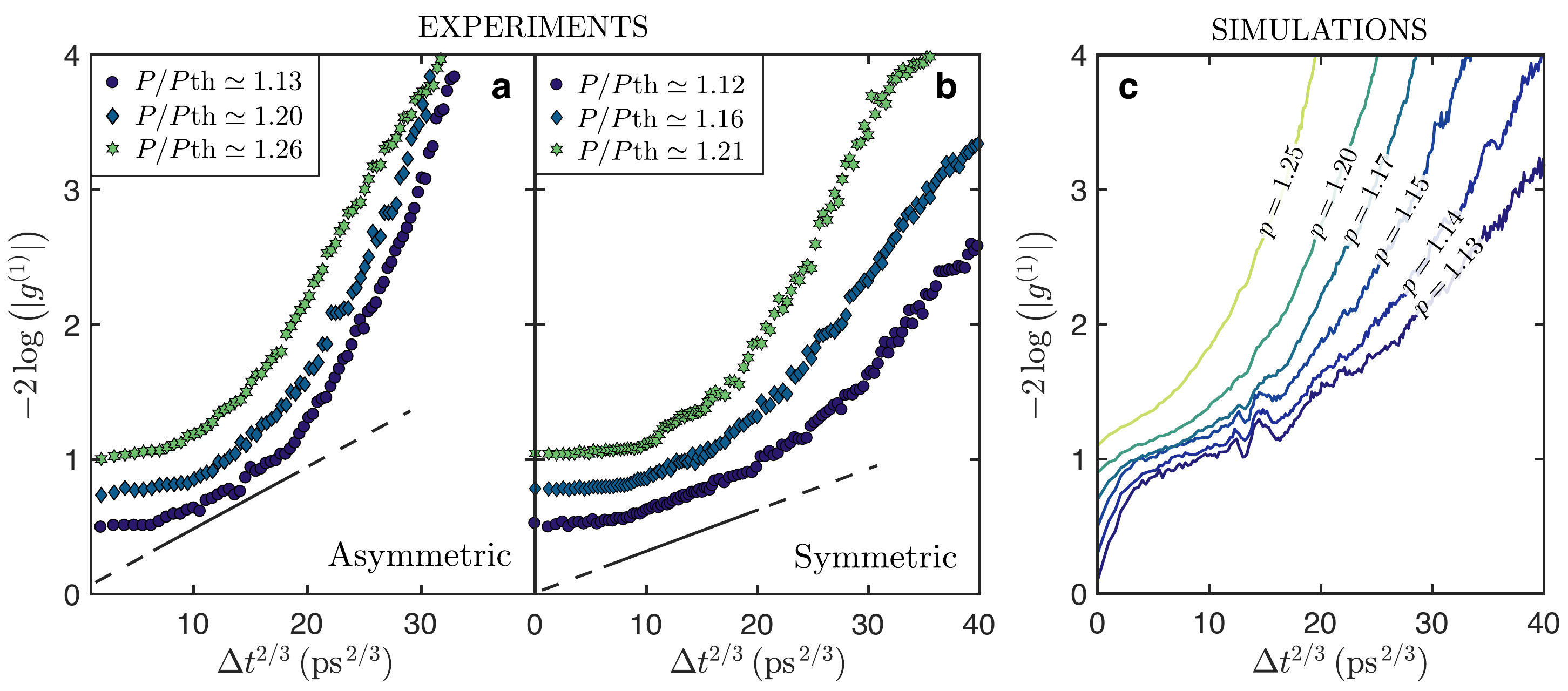}
    \caption{ 
    \textbf{a-b} Measured values of $-2 \mathrm{log} (|g^{(1)}(0, \Delta t)|)$ as a function of $\Delta t^{2/3}$ for various excitation powers in \textbf{(a)} the asymmetric and \textbf{(b)} the symmetric Lieb lattice. 
    For $P/P_{\mathrm{th}} < 1.15$, the data exhibits a linear increase over a given time window characteristic of KPZ scaling.
    This window shortens as the excitation power increases before vanishing completely.
    \textbf{c} Computed values of $-2 \mathrm{log} (|g^{(1)}(0, \Delta t)|)$ as a function of $\Delta t^{2/3}$, for various excitation powers ($p = P/P_{\mathrm{th}}$ varies from 1.13 to 1.25, as indicated on the curves).
    Our simulations qualitatively reproduce the change observed experimentally in the scaling of $-2 \mathrm{log} (|g^{(1)}(0, \Delta t)|)$ when increasing the excitation power.
    The parameters used in our numerical analysis are detailed in section~\ref{sec:numerical-simulations}.
    }
    \label{fig:SUP_MAT_ComPower}
\end{figure}

\begin{figure}[h!]
    \centering
    \includegraphics[scale=0.80]{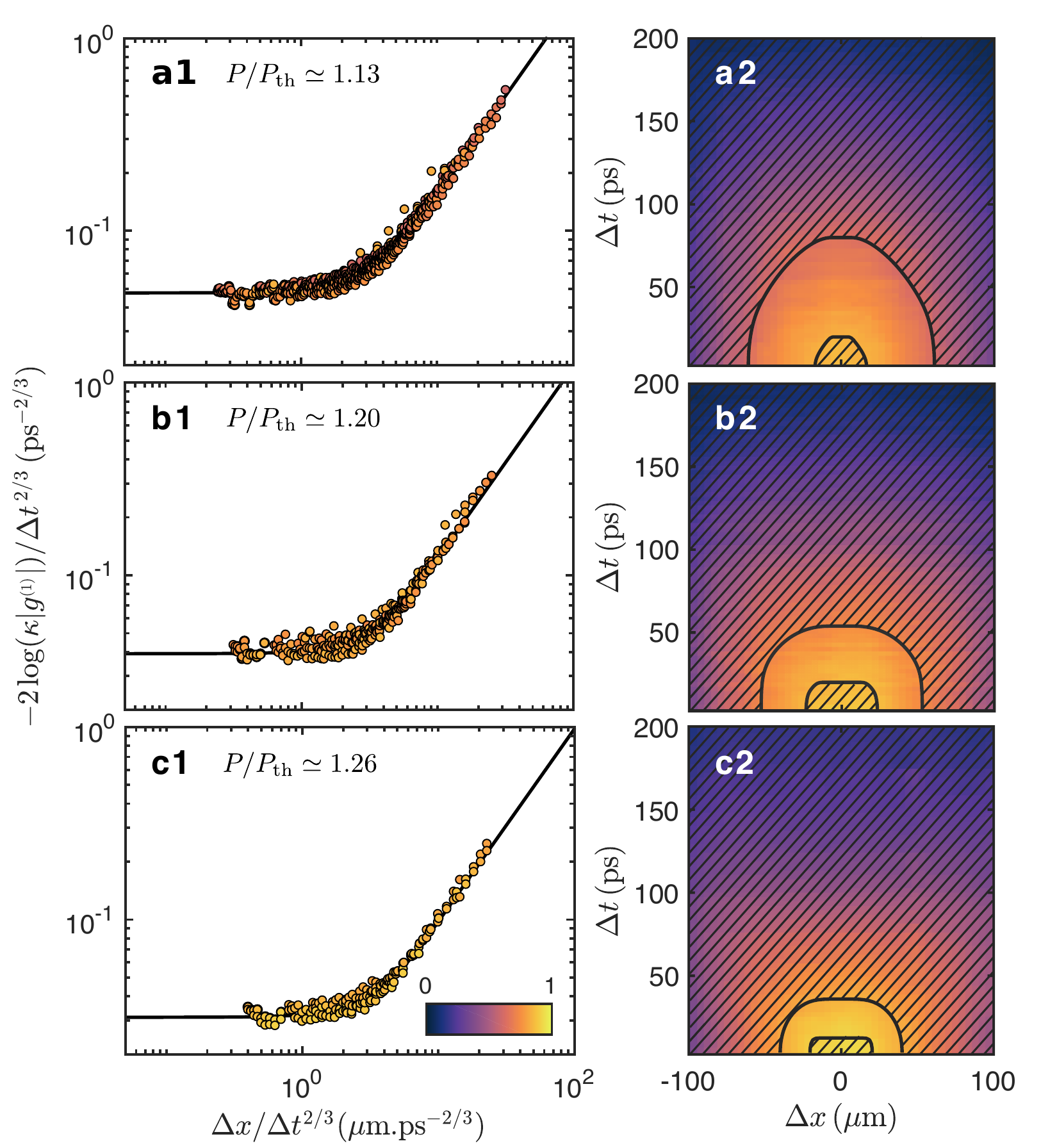}
    \caption{
    \textbf{Shrinking of the spatio-temporal KPZ window when increasing the excitation power.} \textbf{a} $P/P_{th} = 1.13$. \textbf{b} $P/P_{th} = 1.20$. \textbf{c} $P/P_{th} = 1.26$. The normalization factor $\kappa$ is equal to 1.13 in (a-b) and to 1.11 in \textbf{(c)}. 
    }
    \label{fig:SUP_MAT_Shrink}
\end{figure}

We briefly discuss in this section the impact of the excitation power on the coherence decay of the polariton condensate. 
Fig.~\ref{fig:SUP_MAT_ComPower} shows, for $\Delta x = 0$, the variations of $-2 \mathrm{log} ( |g^{(1)}| )$ as a function of $\Delta t^{2/3}$ for different excitation powers, both on the asymmetric (left) and symmetric (right) Lieb lattice. 
The linear trend observed at low excitation power ($P/P_{\mathrm{th}} < 1.15$), emphasized by the blue lines on both panels, becomes less and less visible when $P/P_{\mathrm{th}}$ increases. 
Moreover, the extension of the temporal KPZ window over which this linear trend occurs shrinks progressively as the power increases. Simulations presented in Fig.~\ref{fig:SUP_MAT_ComPower}c reproduces the observed features and are discussed in Sec.~\ref{sec:cond-linewidth}.
The power dependent behavior is further supported by the results shown in Fig.~\ref{fig:SUP_MAT_Shrink}, where the collapse of the data onto the KPZ scaling function is plotted for three values of the power ($P/P_{\mathrm{th}} = 1.13$, 1.20 and 1.26) used to illuminate the asymmetric Lieb lattice. 
The spatio-temporal KPZ window (non-hatched region in Fig.~\ref{fig:SUP_MAT_Shrink}a2, Fig.~\ref{fig:SUP_MAT_Shrink}b2 and Fig.~\ref{fig:SUP_MAT_Shrink}c2), defined by the data points in the $|g^{(1)}|$ data set that collapse onto $F_{\mathrm{KPZ}}$ (black solid line), shrinks as we increase the power. 

\newpage

\newpage

\section{Numerical simulations: method and parameters}
\label{sec:numerical-simulations}

The numerical integration of Eq.~\eqref{eq:cond+res} is performed using the interaction picture method~\cite{54werner1997, 55dennis2013}.
The idea behind this integration scheme is similar to the interaction picture in quantum mechanics. 
We first split Eq.~\eqref{eq:cond+res} into a linear, exactly solvable part and a remaining nonlinear part.
We then solve the linear component in Fourier space and transform it back to real space. 
We transform Eq.~\eqref{eq:cond+res} by moving into the interaction picture and integrate the resulting nonlinear equation using semi-implicit Runge-Kutta method, with an adaptive time-step.
We take as initial condition $\psi(x, t=0)=0$, and let the condensate grow under the action of the pumped reservoir. 
The sampling starts at $t_0$, long after the condensate has reached its stationary density profile.
Usually, we perform our simulations using $t_0=10$ ns. 

As mentioned in section~\ref{sec:gamma2}, some of the parameters entering the numerical simulations are known experimentally.  
For instance, we use a lattice spacing equal to the experimental lattice period $a=4.4\, \mu \mathrm{m}$. The measurement of the polariton dispersion relation $E(k)$
shown  in Fig.~\ref{fig:SUP_MAT_Carac_1}a provides us with a good estimate of the polariton mass ${m = -3.3 \!\times\! 10^{-6} m_{e}}$. 
The $k$-dependent polariton linewidth $\gamma(k)$ is modelled by the function
\begin{equation}
    \label{gamma_k}
    \gamma(k) = \gamma_{0} + \left(\gamma_{\infty} - \gamma_{0} \right) \left[ 1 - \mathrm{exp} \left( -\frac{\gamma_{2}k^{2}}{\gamma_{\infty}-\gamma_{0}} \right) \right]
\end{equation}
which is compared to the experimental data in the inset of Fig.~\ref{fig:SUP_MAT_Carac_2}d (blue solid line). The parameter values are $\gamma_{0, \mathrm{sim}} \equiv \gamma(k=0)= 48.5 \mu \mathrm{ eV}$, 
$\gamma_{2, \mathrm{sim}} \equiv 1/2 (\partial^2 \gamma(k)/\partial k^2) |_{k=0}=1.6 \times 10^{4} \, \mathrm{\mu eV.\mu m^{2}}$  and $\gamma_{\infty, \mathrm{sim}} \equiv \underset{k\rightarrow \infty}{\lim} \gamma(k)=77\mu$eV.
Furthermore, the energy blue-shift at threshold is known with good accuracy:
\begin{equation}    
2 \, g_R \, n_R \Big|_{P=P_{\mathrm{th}}} \approx %\boldsymbol
{0.6\, m \text{eV} \equiv \mu_{\mathrm{th}}} \, ,
\end{equation}
\noindent where $n_{R}|_{_{P=P_{\mathrm{th}}}} = P_{\mathrm{th}}/\gamma_{R} = \gamma_{0}/R$ stands for the reservoir density at condensation threshold and $P_{\mathrm{th}} = \gamma_{0} \gamma_{R}/R$ for the threshold power. 
Assuming that $g_{R}$ does not depend on the pumping power $P$, we find:
\begin{equation}
g_{R}=\frac{1}{2}\frac{\mu_{\mathrm{th}}R}{\gamma_0 } \, .
\end{equation}
Since the reservoir-induced blueshift $2 g_{R} n_{R}$ is two orders of magnitude larger than the polariton-induced blueshift $g |\psi|^{2}$, we take $g = 0$ in our simulations.   
In order to qualitatively reproduce the spatial density profile of the condensate in our experiments, we use a spatially-dependent pump profile  modelled by
\begin{equation}
P(x) = P \, \frac{\left[ 1+\tanh((L_0+x)/\sigma) \right] \left[ 1+\tanh((L_0-x)/\sigma) \right]}{\left[ 1+\tanh(L_0/\sigma) \right]^2} \end{equation}
with $L_0 = 80 \, \mu$m and $\sigma = 9.7 \, \mu$m.
The remaining free parameters in our numerical simulations are thus the scattering rate $R$ of excitons into the condensate and the reservoir decay rate $\gamma_{R}$. We adjust them so as to find the best possible agreement with the experimental data.
All the simulations presented in this article were performed using ${\hbar}{R = 8.8 \times 10^{-4} \, \mathrm{\mu m\, ps^{-1}}}$ and ${\gamma_{R} = 0.45 \gamma_{0}}$. 

In connection with the derivation of the mapping from the gGPE equation for the condensate field to the effective KPZ equation for the phase field presented in Sec.~\ref{sec:mapping}, let us comment on the involved time scales. It is not straightforward to determine the  dynamical time scales for density and phase fluctuations which are in general due to many-body non-linear effects. However, one can assume that they are typically set by the one-body relaxation rates $\gamma_R$ and $\gamma_0$ for the densities, and by the mean field frequency $\overline{\omega_{0}} = 2 g_R n_R + g \rho_0$  for the phase. Thus one has $\overline{\omega_{0}} \simeq 12 \gamma_0 \simeq 6 \gamma_R$,  which suggests that the decoupling of time scales assumed in the mapping to the KPZ equation is verified. In our system, the phase dynamics is faster than the density dynamics of both the condensate and the reservoir. Moreover, the numerical  simulations confirm   the emergence of KPZ dynamics for our set of parameters. 

This mapping  allows us to evaluate the values of the parameters of the KPZ equation.  Using the previous values for the gGPE parameters, we find:
\begin{equation}\label{eq:kpzcoefnum}
    {\lambda=-5.7\times 10^2 \,\mu\textrm m^2\,p\textrm s^{-1} \,,\quad \nu=3.8\times 10^2\, \mu\textrm m^2\,p\textrm s ^{-1} \,,\quad D=2\,\mu\textrm m} \,.
\end{equation}
This allows us in the following to compute the theoretical values for the non-universal normalization constants entering the universal scaling function and distribution, and compare them with the results from direct fits of the experimental and numerical data.

\newpage

\section{Numerical simulations: discussion}
\label{sec:numerical-results}

\vspace{-6pt}

\subsection{Deducing the universality subclass from the collapse of numerical data}
\label{sec:univ-subclass}

\vspace{-6pt}

\begin{figure}[h!]
\includegraphics[scale=0.70]{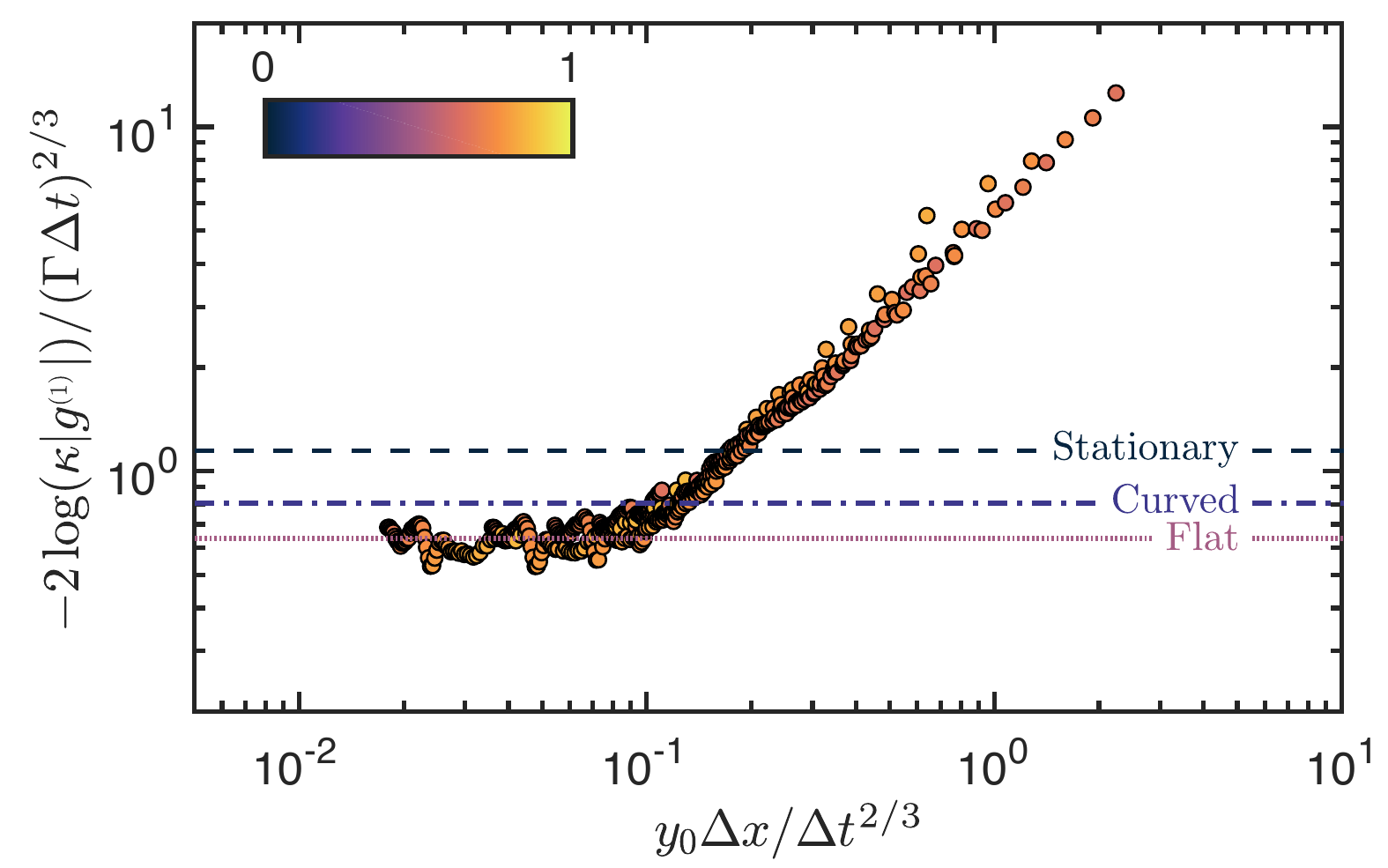}
\caption{
\textbf{Collapse of the simulated data.} 
Collapse of the numerical data for $|g^{(1)}(\Delta x,\Delta t)|$ plotted in dimensionless units, using the normalization constants $y_0$ and $\Gamma$ determined from the microscopic coefficients of Eq.~\eqref{eq:nonunivparams}. 
The dashed lines show the exact values of horizontal asymptotes expected for the flat, curved and stationary universality subclasses. 
}
\label{fig:collthnum}
\end{figure}

In this section, we explain how the horizontal asymptote of the curve onto which the numerical data points collapse provide information about the KPZ universality subclass the system belongs to.
The KPZ universal scaling function associated with the correlation function $\mathrm{Var} \left[ \Delta \theta (\Delta x,\Delta t)\right]$ in 1D is defined by:
\begin{equation}\label{eq:gyth}
F_{\mathrm{KPZ}}
\left(
y=y_0 \frac{\Delta x}{\Delta t^{2/3}}
\right)=\frac{\mathrm{Var}\left[ \Delta \theta (\Delta x,\Delta t)\right]}{ \left(|\Gamma| \Delta t \right)^{2/3}} \simeq \frac{-2 \, \mathrm{log}\left( |g^{(1)} (\Delta x,\Delta t)|\right)}{ \left(|\Gamma| \Delta t \right)^{2/3}} \,,
\end{equation}
where $y_0$ and $\Gamma$ are non-universal normalization constants which can be directly expressed in terms of the KPZ parameters $\lambda$, $\nu$ and $D$ as follows
\begin{equation}
y_0=(2 D \lambda^2 / \nu)^{-1/3}\,,\quad \Gamma=\lambda D^{2} /2 \nu^{2} \,.
\label{eq:gamth}
\end{equation}
Using the parameters given in section~\ref{sec:numerical-simulations}, we obtain:
\begin{equation}\label{eq:nonunivparams}
y_0=6.6\times  10^{-2} \, \mu \textrm m^{-1} \, \mathrm{p}\textrm s^{2/3} \, ,\quad \Gamma=-8.1\times 10^{-3} \, \mathrm{p}\textrm s^{-1}\,.
\end{equation}
In Fig.~\ref{fig:collthnum}, we report the same data points as shown in the inset of Fig.~3c  but using dimensionless coordinates, that we calculate based on the values of $\Gamma$ and $y_{0}$ given above.
We recognize the typical features of the KPZ scaling function, showing a plateau at small $y$ and a linear growth at large $y$~\cite{17deligiannis2021}. It is worth mentioning that the horizontal asymptote  $F_{\mathrm{KPZ}}(0)$ of the universal scaling function is a universal constant that depends on the universality subclass the system falls into. It is known exactly for the flat, stationary and curved initial conditions (see Sec.~\ref{sec:distribution} for details). In Fig.~\ref{fig:collthnum}, we indicate by horizontal dashed lines the values of $F_{\mathrm{KPZ}}(0)$ expected in the three different KPZ subclasses.
We observe that the plateau reached by the simulated data at small $y$ is in close agreement with the exact theoretical value for the flat subclass $F_{\mathrm{\mathrm{KPZ},flat}}(0)\simeq 0.64$ (which is the one expected in our case, as we will see Sec.~\ref{sec:distribution}). This agreement is remarkable given that $\Gamma$ and $y_{0}$ are non-universal quantities (\textit{i.e.} sensitive to the microscopic details of the model). \textbf{It provides even at a quantitative level a strong support for the validity of our mapping between the gGPE model and the KPZ equation for the phase dynamics} (see Sec.~\ref{sec:mapping}). \\

Note that in the main text, the black curve that we adjust to the data points is the KPZ universal scaling function for the stationary case, which is the only one that is known exactly~\cite{42prahofer2004}.

\subsection{Different contributions to \texorpdfstring{$\mathbf{g^{(1)}}$}{TEXT} and influence of space-time vortices}
\label{sec:g1-contrib}

In the previous sections, we have demonstrated that both the experimental and the numerical data for the first-order coherence $g^{(1)}$ exhibit the KPZ scaling in space and time, yielding a clear collapse of all data points within the KPZ window onto the universal scaling function. 
The crucial question which arises is to what extent the behavior of $g^{(1)}$ properly reflects the properties of the phase itself.
This is all the more important that, as shown in the main text, typical space-time phase maps exhibit the formation of space-time vortices. 
\textbf{In this section, we analyze the effect of these vortices and show that \textit{(i)} they do not spoil the KPZ regime as long as almost all vortices appear as pairs of close-by vortex and anti-vortex and the density of single vortices remains low enough, and \textit{(ii)} the scaling behavior of $g^{(1)}$ is indeed inherited from the scaling behavior of the phase-phase correlations.}

\subsubsection{Analysis of the effect of density-density and density-phase correlations}

\begin{figure}[h!]
    \centering
    \includegraphics[scale=0.90]{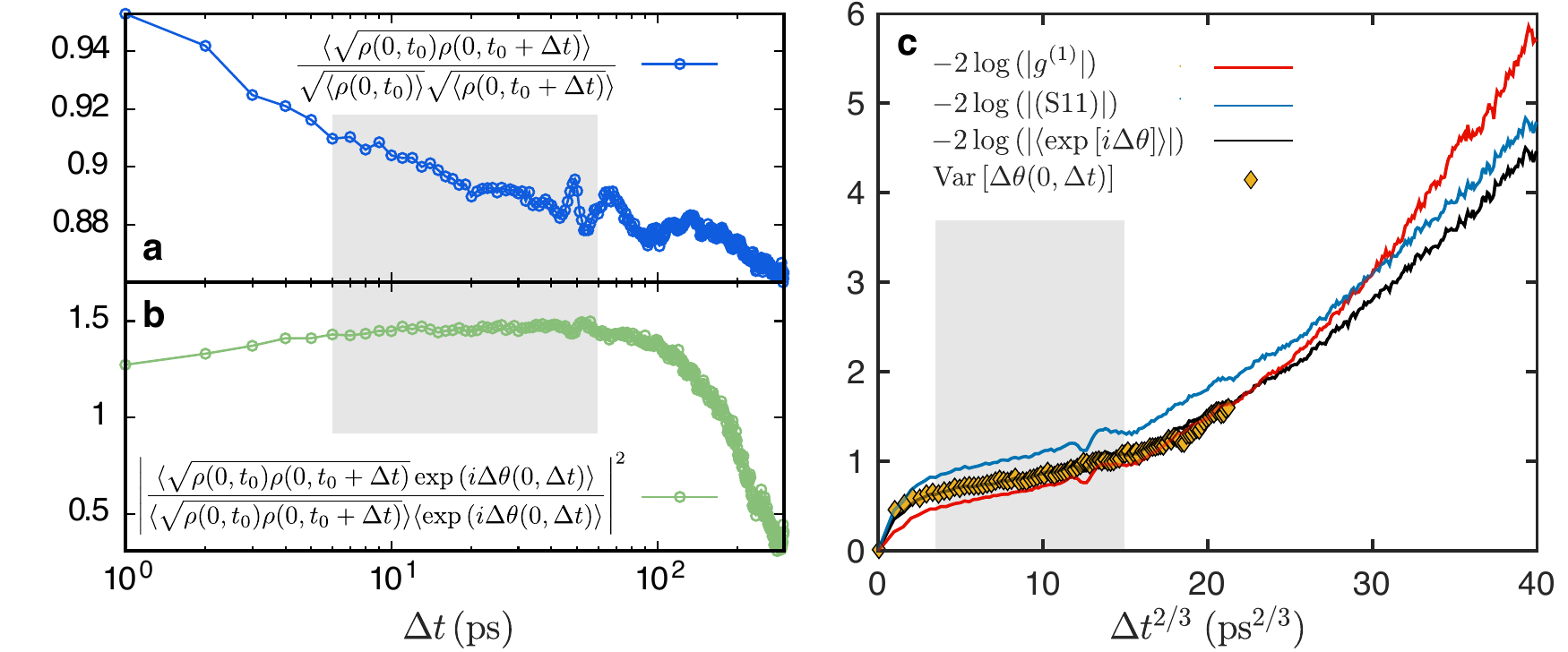}
    \caption{\textbf{a} Density-density correlations. \textbf{b} Density-phase correlations. \textbf{c} Successive approximations of $-2\log(|g^{(1)}|)$. The red line is the full reconstruction of $-2\log(|g^{(1)}|)$, taking into account phase-phase, density-density and density-phase correlations. The blue one is obtained from Eq.~\eqref{eq:phase-density}, neglecting density-phase correlations in $g^{(1)}$. The black line shows the variations of $-2\log(|\langle\exp(i\Delta \theta)\rangle|)$ where density-density and density phase correlations are neglected. The grey shaded region in all panels indicates the KPZ window.}
    \label{fig:g1varcomp}
\end{figure}

We first examine the different contributions to $g^{(1)}$ in order to test the assumptions made in Sec.~\ref{sec:g1-var} to relate the first-order coherence of the condensate field to the phase-phase correlations. 
The assumptions required to derive Eq.~\eqref{eq:g1-exp} from Eq.~\eqref{eq:g1-def} are to neglect both the density-phase and density-density correlations. 
The calculated temporal variations of these correlations are displayed in Figs.~\ref{fig:g1varcomp}a and b, for $\Delta x = 0$. 
We observe that although such correlations are present in the system, they remain approximately constant over the KPZ window.
Furthermore, Fig.~\ref{fig:g1varcomp}c shows the effect of each of these contributions, by comparing:
\begin{itemize}
    \item [(1)] The quantity $-2 \, \mathrm{log} \left( | \langle \mathrm{exp(i \Delta \theta)} \rangle | \right)$ defined in Eq.~\eqref{eq:g1-exp} when only considering phase-phase correlations in $g^{(1)}$ (black line),
    \item [(2)] The right hand side of Eq.~\eqref{eq:phase-density}, obtained by considering phase-phase and density-density correlations in $g^{(1)}$ (blue line),
    \item [(3)] $-2 \, \mathrm{log} \left( |g^{(1)}| \right)$, taking into account all correlations (red line).
\end{itemize}

Comparing the black and blue curves, one observes that except at very short time delays ($\Delta t < 3$ ps), density-density correlations only lead to a global shift, and thus do not affect the scaling behavior. When including density-phase correlations (red curve) the main effect we observe is a faster decoherence at long time delays ($\Delta t > 80$ ps).
Overall, Fig.~\ref{fig:g1varcomp} shows that the scaling of the different computed quantities stays unaffected within the KPZ temporal window.
From this analysis, we conclude that Eq.~\eqref{eq:g1-exp} is a reliable approximation of $g^{(1)}$ in the KPZ regime, which amounts to saying that the behavior of $g^{(1)}$ is mainly dominated by phase-phase correlations in the KPZ window. In the following, we finally relate $\langle e^{i\Delta\theta}\rangle$ to the variance of the phase $\mathrm{Var} \left[ \Delta \theta \right]$.

\subsubsection{Spatio-temporal phase maps and calculation of the phase variance}

\begin{figure}[h!]
    \centering
    \includegraphics[scale=0.6]{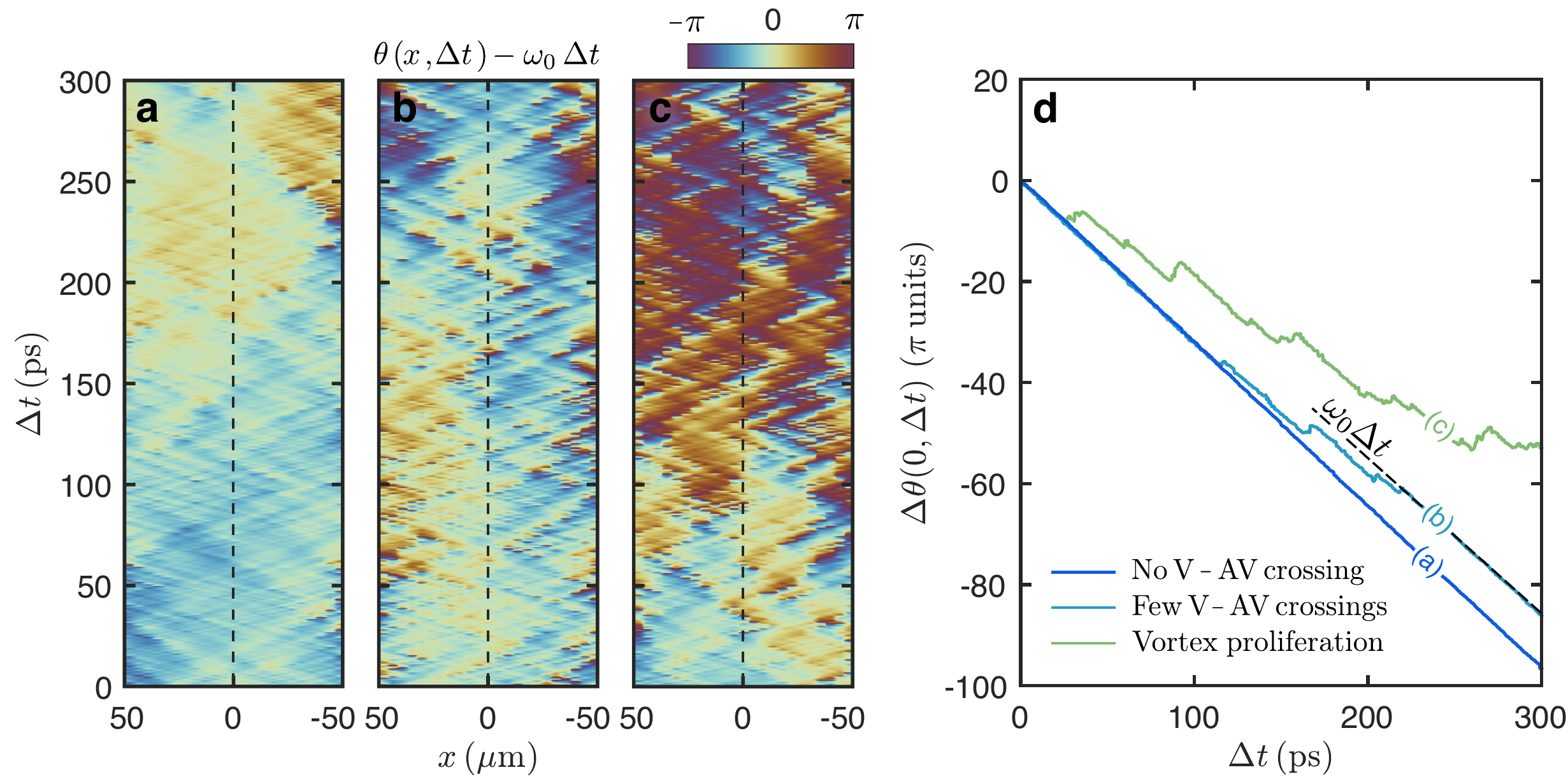}
    \caption{\textbf{a}, \textbf{b} and \textbf{c}. Set of three space-time phase maps generated in the numerical simulations using the parameters given in section~\ref{sec:numerical-simulations}, with (a) no vortices, (b) few vortices, (c) many vortices. The most likely realization is the one depicted in panel (b).
    \textbf{d} Corresponding cuts at $x=0$, showing that the phase trajectory at $x=0$ undergoes jumps whenever it  crosses a vortex-anti-vortex pair.} 
    \label{fig:methods}
\end{figure}

A visual inspection of typical space-time phase maps reveals the presence of space-time vortices. Fig.~\ref{fig:methods} shows three such maps corresponding to three different noise realizations presenting no (a), few (b) or many (c) space-time vortices. By analyzing the distribution of typical vortex distances, we have found that almost all space-time vortices appear as vortex-antivortex (V-AV) pairs (set of two nearby vortices of opposite charge), while the number of single vortices is negligible.
The associated cuts at $x=0$ are shown in Fig.~\ref{fig:methods}d. When no vortices are present, the unwrapped phase at $x=0$ shows a dominant linear behavior in time (dark blue line in Fig.~\ref{fig:methods}d), on top of which KPZ fluctuations develop. When crossing V-AV pairs, the unwrapped phase at $x=0$ undergoes jumps on very short timescales (light blue line in Fig.~\ref{fig:methods}d). Every jump induces a phase shift with respect to vortex-free trajectories, after which the linear behavior is restored. 
The amplitude of these jumps is distributed between 0 and $2\pi$ depending on where the $x=0$ line crosses  the V-AV pair, but dominated by values close to $2\pi$.

Clearly, such phase jumps will have a strong impact on the calculation of the variance of phase fluctuations: vortices will lead to a fast increase of the variance that may hide the KPZ scaling, or even lead to other dynamical regimes such as the ones evidences in Ref.~\cite{21he2017}.
For most of the phase maps generated in the simulations with realistic experimental parameters, we notice that only few jumps occur within the 300~ps time interval under consideration.
More quantitatively, over a set of $10^4$ trajectories, the probability of observing a jump within a 1~ps time interval is around 0.01.
As a consequence, for most realizations (about 75~\%), we are able to find a vortex-free region that extends over a time interval exceeding 100~ps.
We thus use these trajectories without any further processing for evaluating the vortex-free phase variance.
For about 20~\% of the realizations, any 100~ps time window contains at least one or a few V-AV pairs, but that are sparse enough to allow the numerical filtering of the phase jumps and so to evaluate the vortex-free phase variance (in practice, we filter a vortex by adding at its location a second vortex of opposite charge).
In a limited number of cases (5~\% of the trajectories) the jumps are too numerous in any time window of 100~ps to allow computing the vortex-free variance. 
We thus discard these trajectories in the calculation of the phase variance.

The computed vortex-free phase variance is shown in Fig.~\ref{fig:g1varcomp}c (yellow diamonds) and is directly compared with $-2 \, \mathrm{log} \left(\left|\langle e^{i\Delta\theta}\rangle \right| \right)$. 
Both quantities perfectly coincide within the KPZ window, up to the departure from this regime, which shows that the cumulant expansion Eq.~\eqref{eq:g1delta} is valid in this whole range.  
We emphasize that $-2 \, \mathrm{log} \left(\left|\langle e^{i\Delta\theta}\rangle \right| \right)$ is computed over the full duration of all unprocessed trajectories, and thus includes the effect of vortices. Therefore, we conclude that in the low-vortex density regime we explore here, the quantity $-2 \, \mathrm{log} \left(\left|\langle e^{i\Delta\theta}\rangle \right| \right)$ (and thus also $g^{(1)}$) is not sensitive to V-AV pairs.
This analysis fully confirms that $g^{(1)}$ provides a good observable to probe the KPZ scaling of the phase.

\subsubsection{Resilience of KPZ to space-time V-AV pairs}

In the previous Section, we have found that the presence of V-AV pairs weakly affects the temporal scaling of $|g^{(1)}(0,\Delta t)|$, as can be observed in Fig.~\ref{fig:g1varcomp}.
The analysis of the phase jump amplitudes in Fig.4b of the main text provides an explanation for the robustness of the $g^{(1)}$ correlations against space-time V-AV pairs. Indeed, we notice that the jumps are centered around values that are close to multiples of $2\pi$. Hence, such jumps have a negligible impact as they enter in the exponential of the condensate field. This property allows observing the emergence of KPZ universality even in the presence of some space-time  V-AV pairs. 
This property, characteristic of the compact version of the KPZ problem, is even more remarkable if we notice that, in such conditions, Eq.~\eqref{eq:g1delta} cannot be blindly used: in presence of defects every random jump increases considerably the phase-phase correlator bringing it away from the predictions of the KPZ scaling.

\subsubsection{Effect of the condensate linewidth on the long-time coherence}
\label{sec:cond-linewidth}
 
\begin{figure}[h!]
    \centering
    \includegraphics[scale=0.7]{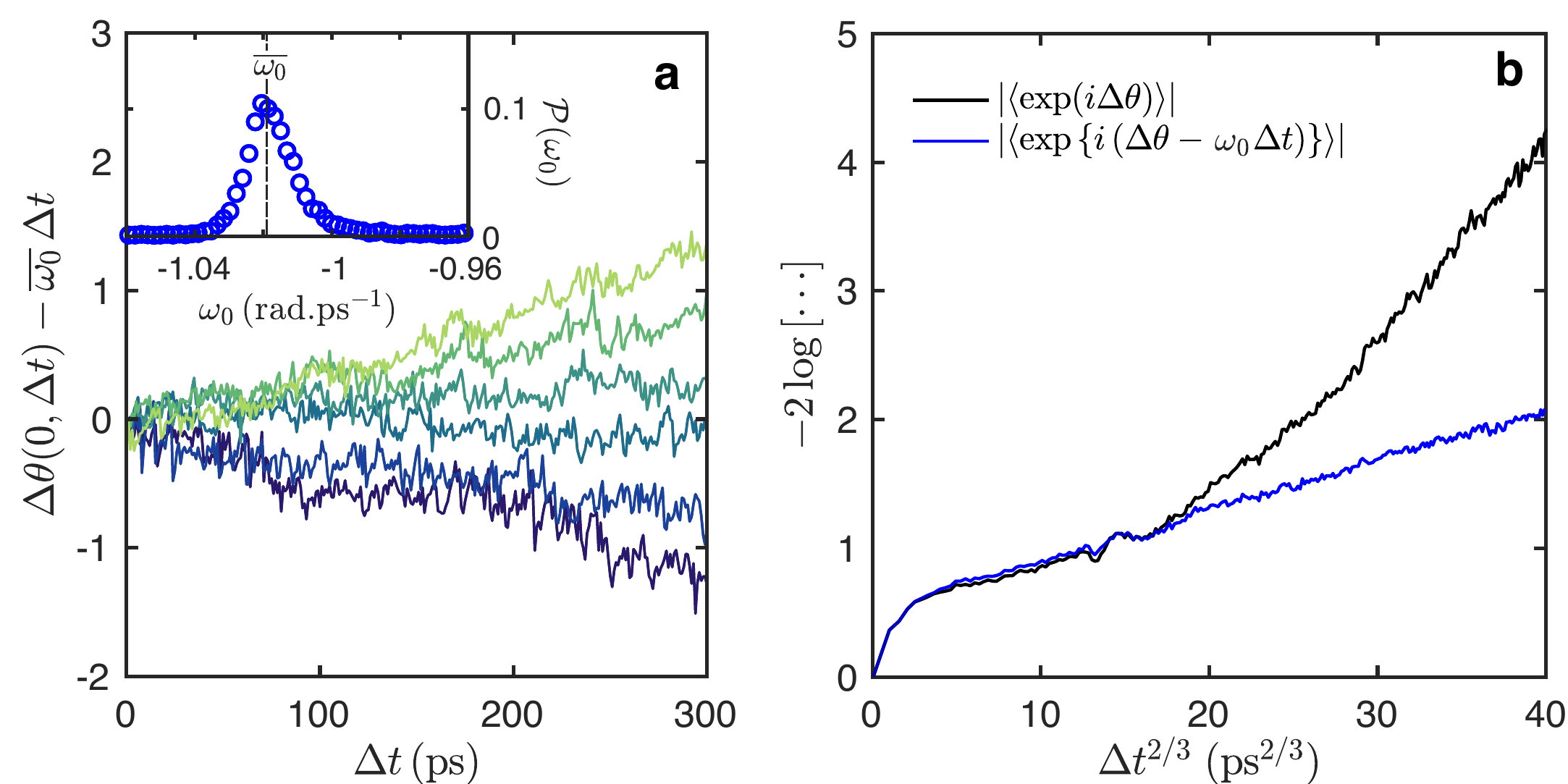}
    \caption{\textbf{a} Set of unwound phase trajectories with no phase jump, after subtracting the ensemble average linear growth $\overline{\omega_{0}}  \Delta t$. Inset: Distribution of the  mean frequency $\omega_0$ of each individual trajectory. The fluctuations of the frequency $\omega_0$ around its ensemble average value $\bar\omega_0$ induces a residual slope in these trajectories. \textbf{b} Correlations $-2\log \left( |\langle e^{i\Delta \theta}\rangle|\right)$ with and without correcting for the frequency fluctuations.}
    \label{fig:localslopes}
\end{figure}

For exciton-polaritons, the mean velocity $\omega_{0}$ of each phase trajectory fluctuates with respect to the ensemble average $\overline{\omega_{0}}$. To illustrate this, we show in Fig.~\ref{fig:localslopes} a set of trajectories with no jumps in the KPZ time window, where we subtracted the average dynamical phase $\langle \Delta\theta(0,\Delta t) \rangle\equiv \overline{\omega_{0}} \Delta t$. One observes a dispersion of the trajectories, which are not on average  constant in time but rather exhibit a residual linear behavior, with slope $\delta \omega_{0} = \omega_{0}-\overline{\omega_{0}}$. This is because the total frequency $\omega_0$ of each individual trajectory varies due to density fluctuations.
Such stochastic variations in the slope of the phase dynamics corresponds to an inhomogeneous spectral broadening. The related distribution $\mathcal{P}(\omega_{0})$ is shown in the inset of Fig.~\ref{fig:localslopes}a. 
  
To determine the influence of this broadening on the KPZ regime, we post-process the phase trajectories so as to subtract for each trajectory its own linear behavior $\omega_0 \Delta t$. We then compute the corresponding correlations $|\langle e^{i\Delta\theta -\omega_0 \Delta t}\rangle|$.
The result is shown in the right panel of  Fig.~\ref{fig:localslopes}, where it is compared with the ``raw''  $|\langle e^{i\Delta\theta}\rangle|$ where the average is performed over the same set of trajectories, but subtracting the average dynamical phase $\overline{\omega_{0}} \Delta t$. This clearly evidences that the KPZ $\Delta t^{2/3}$ scaling holds over much longer times once  the intrinsic slope of each trajectory is properly removed. Otherwise, at large times ($\Delta t > 80$ ps), the residual linear behavior becomes dominant compared to the $\Delta t^{2/3}$ scaling of the fluctuations and introduces faster decay of the coherence. This demonstrates that the departure from the KPZ regime in the raw correlations is induced by inhomogeneous spectral broadening. 

Note that we also performed simulations for higher power densities. As reported in Fig.\ref{fig:SUP_MAT_ComPower}c, we observe that departure from the KPZ regime occurs at shorter time delays when increasing the excitation power, thus suggesting that inhomogeneous broadening increases with excitation power. We point out that the shrinking of the KPZ window obtained in the numerical simulations occurs over a range of excitation powers that is similar to the one observed experimentally.

\subsection{Distribution of phase fluctuations}
\label{sec:distribution}

In this section, we discuss the probability distribution associated to phase fluctuations computed from our numerical simulations. 
We demonstrate that this distribution is well reproduced by the Tracy-Widom (TW) Gaussian Orthogonal Ensemble (GOE) distribution, which is characteristic of the flat KPZ universality subclass.

\subsubsection{The KPZ universality subclasses}

While the KPZ universality class is fully characterized by the critical exponents $\chi$ and $\beta$ for the two-point correlators, the distribution of the height fluctuations of a KPZ interface in 1D allows one to distinguish three universality subclasses. Let us briefly review this result in this section, before detailing our results for the distribution of the phase fluctuations of the exciton-polariton condensates.
\vspace{4pt}
\newline 
\noindent For a classical interface, the height field $h(t) \equiv h(x_{0},t)$ is known to behave at long times and at any given  point $x_{0}$ in space according to
\begin{equation}
h(x_{0},t) \sim v_\infty t + \left(|\Gamma| t \right)^\beta \tilde{h}
\label{eq:ansatz-h} \, ,
\end{equation}
where, $v_{\infty}$ represents the mean velocity of the growing interface \footnote{Note that for flat initial conditions, $v_{\infty}$ is equal to  the microscopic KPZ non-linearity $\lambda$}, and $\tilde h$ is a random variable describing the reduced fluctuations whose distribution is universal.
Much theoretical effort has been dedicated during the last decade to study the properties of the dimensionless random field $\tilde h$, whose distribution $P[\tilde h]$ turns out to be sensitive to the spatial profile of the initial condition $h(x, t=0)$, or equivalently to the global geometry of the interface
(we refer to Ref.~\cite{4takeuchi2018}) for a review). 
Three main possible subclasses emerge. 
For flat initial conditions the reduced KPZ field $\tilde h$ is distributed according to the Tracy-Widom distribution (TW) associated to the largest eigenvalue of random matrices belonging to the Gaussian Orthogonal Ensemble (GOE).
For curved initial conditions, $P[\tilde h]$ is the TW distribution associated to the largest eigenvalue of random matrices belonging to the Gaussian Unitary Ensemble (GUE).
The last subclass is associated to stationary initial condition for the phase profile, for which $P[\tilde h]$ is the Baik-Rains distribution. 

Based on a previous numerical study~\cite{14squizzato2018}, we expect that driven-dissipative condensates, in absence of external confinement, belong to the flat universality subclass. This is supported by the universal plateau value obtained in Sec.\ref{sec:univ-subclass} for the KPZ scaling function, which coincides with the one of the flat subclass.

\newpage

\subsubsection{Analysis of the phase distribution}

\vspace{-4pt}

To treat the condensate phase field in a similar way as classical interface height $h$ (unbounded parameter), one should first unwrap the phase in time at fixed $x=0$. In the KPZ regime, the equivalent of Eq.~\eqref{eq:ansatz-h} then becomes:
\begin{equation}\label{eq:phdecomp}
    \Delta \theta (t_0, \Delta t)
    \equiv
    \theta(t_0 + \Delta t) - \theta(t_0) %\overset{\Delta  t\gg T}
    \ {\sim}  \ \omega_0 \Delta  t + \left( |\Gamma| \Delta  t \right)^{\chi/z} \tilde\theta(\Delta  t) \,,
\end{equation}
with $\omega_0$ the mean frequency associated with the phase dynamics and $t_0$ the reference time for the phase unwinding.
We emphasize that the unwrapping is crucial to study the distribution of the reduced variable $\tilde\theta$. The distribution of the  phase itself has a compact support and cannot be any of the Tracy-Widom or Baik-Rain distributions characteristic of the KPZ realm.
Following the analysis usually performed on a classical interface, a natural strategy to obtain the reduced random variable $\tilde \theta$ from the unwound phase would be to subtract the mean linear behavior and to rescale by $\Delta t^{2/3}$. 
However, as explained in the previous section, the presence of random phase jumps, induced by the proximity of vortices, hinders this strategy, since it disrupts the linear $\Delta t$ evolution. In particular, as visible in Fig. 5 of the main text, these phase jumps induce replicas of the main distribution separated by shifts of value close to $2\pi$. In principle, since the amplitude of the phase jumps are distributed, they could also affect the main central peak. However, as shown below, whereas they do affect the right tail of the central distribution, their effect is negligible on its left tail. 

\begin{figure}[h!]
    \centering
    \includegraphics[scale=1]{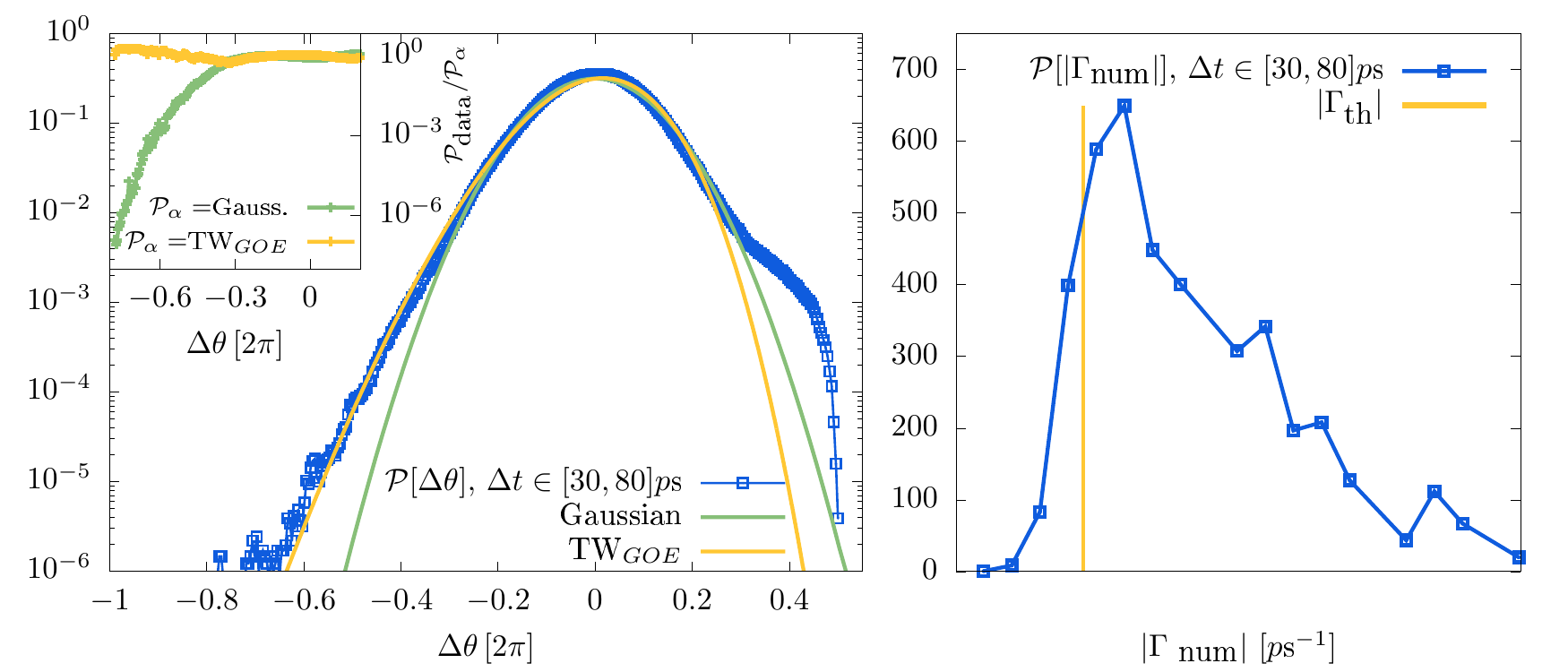}
    \caption{\textbf{a} Distribution of the phase fluctuations obtained by summing up the individual distribution obtained at all delays within the range $30 \,\mathrm{ps} \leq \Delta t \leq 80\,\mathrm{ps}$. The Tracy Widom GOE distribution is shown as a yellow solid line. For comparison, we also plot with green solid line the normal distribution $\mathcal{N}$, centered at $\Delta \theta =0$ with the same variance as the Tracy-Widom distribution. In the inset,  the agreement between the numerical data and either (green) the Normal distribution or (yellow) the Tracy-Widom distribution are compared by showing the ratio $\mathcal{P}[\Delta \theta]/
    \mathcal{P}_\alpha[\Delta \theta] $ with $\alpha=TW_{GOE}, \mathcal{N}$. This clearly demonstrates that TW GOE provides a better description of the numerical data.
    \textbf{b} Distribution of the parameters $|\Gamma_{\textrm{num}}|$ in the same temporal window, extracted from the simulations $via$ \eqref{eq:gamnum}, together with the theoretical value $|\Gamma_{\mathrm{th}}|$ computed using the microscopic parameters of the numerical simulation and \eqref{eq:gamth}. It is important to note that numerically we only have access to the absolute value of $\Gamma$ because it is extracted from the variance of the phase, which is a positive defined quantity.}
    \label{fig:histevol}
\end{figure}

To overcome this difficulty, instead of following the time evolution of the unwrapped phase, as explained in the main text, we rather record the distributions of the phase fluctuations over different realizations at fixed time delays $\Delta t$ lying in the KPZ window. We then normalize each obtained distribution such that it has zero mean and unit variance.
To test with high accuracy the agreement between the phase distribution and Tracy-Widom distribution (in particular to better resolve the tails), we accumulate more statistics by summing the properly normalized fluctuations at all time instants in the KPZ window. The resulting distribution around $\Delta \theta = 0$ is displayed in Fig.~\ref{fig:histevol} (left panel). Strikingly, the left tail of the phase distribution strongly departs from a Gaussian distribution and is reproduced using the Tracy-Widom GOE distribution over more than five decades. Note that the right tail cannot be analysed with such precision because it is affected by the subset of trajectories that contain one or more phase jumps. 
 
Let us emphasize that by normalizing the variance of the distribution at each $\Delta t$ to unity, one can deduce the estimate for the non-universal parameter $|\Gamma|$ as 
 \begin{equation}\label{eq:gamnum}
    \left|\Gamma_{\rm num}\right|=\frac{1}{\Delta t}\left( \frac{\textrm{Var}({\Delta\theta(\Delta t))}} 
    {\textrm{Var}_{TW_{GOE}}}\right)^{3/2} \, .
\end{equation}
The values found for the different time instants can be compared with the theoretical estimate of $|\Gamma|$ given in Eq.~\eqref{eq:gamth}.
%as $ \Gamma_{\rm th}=\frac{1}{2}\lambda A^2, \ A=\frac{D}{\nu}$ with $\lambda$, $\nu$ and $D$ from Eq.~\eqref{eq:bareKPZ} using the mapping from the gGPE to the KPZ equation.
The comparison is shown in the right panel of Fig.~\ref{fig:histevol}, and the agreement is remarkable. This is yet another strong confirmation of the accuracy of the mapping, given that $\Gamma$ is a non-universal constant. 

\end{document}